\documentstyle[12pt,bezier,epsfig,axodraw]{article}
 
\makeatletter
 
\expandafter\ifx\csname mathrm\endcsname\relax\def\mathrm#1{{\rm #1}}\fi
\@ifundefined{mathrm}{\def\mathrm#1{{\rm #1}}}{\relax}
 
\def\@biblabel#1{[#1]\hfill}    
 
\def\refs#1{\let\@refa\@empty
           \@for\@ref:=#1\do{\@refa\def\@refa{,\penalty\@m\ }%
                             \ref{\@ref}}}
\def\refes#1{\let\@refa\@empty
           \@for\@ref:=#1\do{\@refa\def\@refa{,\penalty\@m\ }%
                             \refeqf{\@ref}}}
\def\refeq#1{\mbox{equation~(\ref{#1})}}
\def\refeqs#1{\mbox{equations~\refes{#1}}}
\def\refeqf#1{\mbox{(\ref{#1})}}

\def\reffi#1{\mbox{figure~\ref{#1}}}
\def\reffis#1{\mbox{figures~\refs{#1}}}
\def\reffif#1{\mbox{\refs{#1}}}
\def\refta#1{\mbox{table~\ref{#1}}}

\def\refse#1{\mbox{section~\ref{#1}}}

\def\refcha#1{\mbox{section~\ref{#1}}}
\def\refapp#1{\mbox{appendix~\ref{#1}}}
\def\refapps#1{\mbox{appendices~\ref{#1}}}
\def\refappf#1{\mbox{\ref{#1}}}
 
\def\thefigure{\@arabic\c@figure}
\def\thetable{\@arabic\c@table}
\def\thefootnote{\alph{footnote}}
\def\thefootnote{\@alph\c@footnote}

\makeatother
 
\def\al{\alpha}
\def\be{\beta}
\def\Ga{\Gamma}
\def\ga{\gamma}
\def\de{\delta}
\def\De{\Delta}
\def\eps{\epsilon}
\def\veps{\varepsilon}
\def\la{\lambda}
\def\om{\omega}
\def\si{\sigma}

\def\mathswitchr#1{\relax\ifmmode{\mathrm{#1}}\else$\mathrm{#1}$\fi}
\def\mathswitch#1{\relax\ifmmode#1\else$#1$\fi}

\newcommand{\GeV}{\unskip\,\mathswitchr{GeV}}
\newcommand{\MeV}{\unskip\,\mathswitchr{MeV}}
\newcommand{\TeV}{\unskip\,\mathswitchr{TeV}}

\newcommand{\pba}{\unskip\,\mathswitchr{pb}}

\newcommand{\Pf}{\mathswitch  f}

\newcommand{\Pl}{\mathswitchr l}

\newcommand{\PW}{\mathswitchr W}
\newcommand{\PZ}{\mathswitchr Z}

\newcommand{\PH}{\mathswitchr H}
\newcommand{\Pe}{\mathswitchr e}
\newcommand{\Pmy}{\mathswitch \mu}
\newcommand{\Pmu}{\mathswitch \mu}
\newcommand{\Pta}{\mathswitch \tau}
\newcommand{\Pne}{{\mathswitch {\nu_{\Pe}}}}
\newcommand{\Pnm}{{\mathswitch {\nu_{\mu}}}}
\newcommand{\Pnt}{{\mathswitch {\nu_{\tau}}}}
\newcommand{\Pd}{\mathswitchr d}
\newcommand{\Pu}{\mathswitchr u}
\newcommand{\Ps}{\mathswitchr s}
\newcommand{\Pc}{\mathswitchr c}
\newcommand{\Pb}{\mathswitchr b}
\newcommand{\Pt}{\mathswitchr t}
\newcommand{\Pep}{\mathswitchr {e^+}}
\newcommand{\Pem}{\mathswitchr {e^-}}
\newcommand{\PWp}{\mathswitchr {W^+}}
\newcommand{\PWm}{\mathswitchr {W^-}}
 
\newcommand{\Mf}{\mathswitch {m_\Pf}}

\newcommand{\MW}{\mathswitch {M_\PW}}

\newcommand{\MZ}{\mathswitch {M_\PZ}}
\newcommand{\MH}{\mathswitch {M_\PH}}
\newcommand{\Me}{\mathswitch {m_\Pe}}

\newcommand{\Md}{\mathswitch {m_\Pd}}
\newcommand{\Mu}{\mathswitch {m_\Pu}}
\newcommand{\Ms}{\mathswitch {m_\Ps}}
\newcommand{\Mc}{\mathswitch {m_\Pc}}
\newcommand{\Mb}{\mathswitch {m_\Pb}}
\newcommand{\Mt}{\mathswitch {m_\Pt}}
 
\newcommand{\GW}{\mathswitch {\Gamma_\PW}}

\newcommand{\sw}{\mathswitch {s_\PW}}
\newcommand{\cw}{\mathswitch {c_\PW}}
\newcommand{\NCf}{\mathswitch {N_{\mathrm{C}}^\Pf}}
\newcommand{\Qf}{\mathswitch {Q_\Pf}}
\newcommand{\GF}{\mathswitch {G_\mu}}

\def\da{\Delta\alpha}
\def\dro{\Delta\rho}
\def\cs{cross-section}
\def\css{cross-sections}
\def\eeWW{\mathswitch{\Pep\Pem\to\PWp\PWm}}
\def\eeffff{\mathswitch{\Pep\Pem\to4f}}
\def\Eg{\mathswitch{\langle E_\ga \rangle}}
\def\Mg{\mathswitch{\langle M_\ga \rangle}}
\def\IBA{\mathswitchr{IBA}}
\def\ew{{\mathswitchr{ew}}}
\def\app{\mathswitchr{app}}

\def\born{{\mathswitchr{Born}}}

\def\QCD{{\mathswitchr{QCD}}}

\def\Mhat{\widehat{\cal{M}}}

\def\pep{p_+}
\def\pem{p_-}
\def\pepm{p_\pm}
\def\psep{\ps_+}

\def\pW{k}
\def\pWp{\pW_+}
\def\pWm{\pW_-}
\def\pWpm{\pW_\pm}
\def\psWp{\ks_+}
\def\psWm{\ks_-}
\def\eWp{\veps_+}
\def\eWm{\veps_-}
\def\eWpm{\veps_\pm}
\def\esWp{\es_+}
\def\esWm{\es_-}
\def\he{\kappa}
\def\hep{\kappa_+}
\def\hem{\kappa_-}

\def\hWp{\la_+}
\def\hWm{\la_-}
\def\geeZ{g_{\Pe\Pe\PZ}^{\he}}
\def\fl{f_{\ell}}

\def\als{\alpha_s}
\def\drl{\Delta r_{_L}}
\def\da{\Delta\alpha}
\def\dro{\Delta\rho}
\def\Mfi{m_{\Pf_i}}
\def\Mfj{m_{\Pf'_j}}
\def\LL{\mathswitch{L}}

\def\hpt{\hphantom{2}}
\def\hpm{$\hphantom{-}$}
\def\bebar{\bar\be}
\def\bexp{\be_{\mathrm{exp}}}
\def\bs{\be_{\mathrm{S}}}
\def\bh{\be_{\mathrm{H}}}
\def\ie{i.e.\ }
\def\eg{e.g.\ }

\def\vb{\bar v}

\def\dsidop{\left(\frac{\d\sigma}{\d\Omega}\right)}

\def\dsidc{\frac{\d\sigma}{\d\!\cos\theta}}
\def\dsidct{\d\sigma/(\d\!\cos\theta)}
\def\dsidctp{\d\sigma/(\d\!\cos\theta_+)}

\def\co{\relax}
\def\co{,}
\def\nl{\nonumber\\}
\def\nlc{\co\nonumber\\}

\def\eqskipcorr{\vspace{-\abovedisplayskip}
                \vspace{\abovedisplayshortskip}}
 
\def\beq{\begin{equation}}
\def\eeq{\end{equation}}
\def\beqar{\begin{eqnarray}}
\def\eeqar{\end{eqnarray}}
\def\bma{\begin{displaymath}} 
\def\ema{\end{displaymath}}
\def\barr#1{\begin{array}{#1}}
\def\earr{\end{array}}
\def\bit{\begin{itemize}}
\def\eit{\end{itemize}}
\def\bfi{\begin{figure}}
\def\efi{\end{figure}}
\def\btab{\begin{table}}
\def\etab{\end{table}}
\def\bce{\begin{center}}
\def\ece{\end{center}}
\def\nn{\nonumber}
\def\disp{\displaystyle}
\def\text{\textstyle}
 
\def\d{{\mathrm{d}}}
\def\Li{\mathop{\mathrm{Li}_2}\nolimits}
\def\Re{\mathop{\mathrm{Re}}\nolimits}
\def\Im{\mathop{\mathrm{Im}}\nolimits}
 
\renewcommand{\O}{{\cal{O}}}
\newcommand{\Oa}{\mathswitch{{\cal{O}}(\alpha)}}
\newcommand{\Oaa}{\mathswitch{{\cal{O}}(\alpha^2)}}
\newcommand{\M}{{\cal{M}}}

\def\slash{\mathpalette\make@slash}
\def\make@slash#1#2{\setbox\z@\hbox{$#1#2$}%
  \hbox to 0pt{\hss$#1/$\hss\kern-\wd0}\box0}

\newcommand{\es}{\varepsilon \hspace{-0.48em}/\hspace{0.04em}}
\newcommand{\ks}{k\hspace{-0.52em}/\hspace{0.1em}}
\newcommand{\ps}{p\hspace{-0.42em}/}

\def\@versim#1#2{\lower2.5\p@\vbox{\baselineskip\z@skip\lineskip-.2\p@
    \ialign{$\m@th#1\hfil##\hfil$\crcr#2\crcr\sim\crcr}}}
\newcommand{\lsim}
{\;\raisebox{-.3em}{$\stackrel{\displaystyle <}{\sim}$}\;}
\newcommand{\gsim}
{\;\raisebox{-.3em}{$\stackrel{\displaystyle >}{\sim}$}\;}
 
\def\heavy{\raise.7mm\hbox{\protect\rule{1.1cm}{.5mm}}}
\def\solid{\raise.9mm\hbox{\protect\rule{1.1cm}{.2mm}}}

\newsavebox{\hdwig}
\newsavebox{\dpdwig}
\newsavebox{\dmdwig}
\newsavebox{\wigr}
\newsavebox{\wigu}
\newsavebox{\wigur}
\newsavebox{\wigdr}
\newsavebox{\wigt}
\newsavebox{\blobdz}
\newsavebox{\blobza}
\newsavebox{\Vu}
\newsavebox{\Vr}
\newsavebox{\St}
\newsavebox{\Sr}
\newsavebox{\Fr}
\newsavebox{\Fl}
\newsavebox{\Fu}
\newsavebox{\Vur}
\newsavebox{\Vdr}
\newsavebox{\Vurd}
\newsavebox{\Vdru}
\newsavebox{\Sur}
\newsavebox{\Sdr}
\newsavebox{\Fur}
\newsavebox{\Fdr}
\newsavebox{\Vudr}
\newsavebox{\Fudr}
\newsavebox{\Sudr}
\newsavebox{\Gr}
\newsavebox{\Gudr}
\newsavebox{\Lu}
\newsavebox{\Lr}
\newsavebox{\Lur}
\newsavebox{\Ldr}
\newsavebox{\Ld}
\newsavebox{\Ldl}
\newsavebox{\Lul}
\newsavebox{\Vp}
 
\begin{document}
 
\def\thefootnote{\alph{footnote}}

\begin{center}
{\large \bf WW CROSS-SECTIONS AND DISTRIBUTIONS}
\end{center}
\begin{center}
{\it Conveners:}          
W.~Beenakker and F.A.~Berends  
\end{center}
\begin{center}           
{\it Working group:}      
E.N.~Argyres, D.~Bardin, A.~Denner, S.~Dittmaier, J.~Hoogland, S.~Jadach,   
R.~Kleiss, Y.~Kurihara, D.~Lehner, G.~Montagna, T.~Munehisa, O.~Nicrosini, 
T.~Ohl, G.J.~van~Oldenborgh, C.G.~Papadopoulos, G.~Passarino, F.~Piccinini,
B.~Pietrzyk, T.~Riemann, Y.~Shimizu, M.~Skrzypek           
\end{center}
\vspace*{1cm}
\tableofcontents
\newpage

\section{Introduction}
\label{CHintro}
\subsection{Goals of W-pair production}

The measurements performed at LEP1 have provided us with 
an extremely accurate knowledge of the parameters of
the \PZ\ gauge boson: its mass, partial widths, and total width. Except perhaps
for the $\Pb\bar{\Pb}$ and $\Pc\bar{\Pc}$ decay widths all data are in perfect
agreement with the Standard-Model \cite{SM} (SM) predictions 
\cite{summer95,Ho95}. There even
is first evidence that the contributions of gauge-boson
loops to the gauge-boson self-energies are indeed required \cite{bosons}.
Thus an indirect confirmation of the existence of the triple
gauge-boson couplings (TGC's) has been obtained.

The future measurements of \PW-pair production at LEP2 will 
add two important pieces of information to our knowledge of
the SM. One is a determination of the \PW\ mass, which at present
is only directly measured at hadron colliders. The envisaged
precision of 40--50\MeV\ gives a significant improvement on the
present Tevatron measurements \cite{CDFUA2}. Since the \PW\ mass is one of the
key parameters of the electroweak theory, such an improved
accuracy makes the tests on the SM more stringent.

The second piece of information that \PW-pair production can
provide is the structure of the triple gauge-boson couplings. These
couplings now play a role in the tree-level \cs, in
contrast to LEP1 physics where they only enter through loop corrections. 
So the Yang-Mills character of the TGC's 
can be established by studying \PW-pair differential \css\ at LEP2 
energies. Since these couplings are at the heart
of the non-abelian gauge theories, this information is essential
for a direct confirmation of the SM.

\subsubsection{Measurement of the mass of the W boson}

With the precise measurement of the mass of the \PW\ boson at LEP2
the situation for the electroweak input parameters
changes with respect to LEP1. The common practice at LEP1 \cite{YR95} is to
use for \MW\ a value derived from the Fermi constant \GF, 
which is accurately known from muon decay. The relation to obtain
\MW\ follows from the SM prediction for muon decay
\begin{equation}
   \GF  =  \frac{\alpha \pi}{\sqrt{2}\,\MW^2\,(1-\MW^2/\MZ^2)}\,\frac{1}
           {1- \Delta r},
\end{equation}
where $\Delta r = 0$ at tree level and where $\Delta r$ is \Mt- and
\MH-dependent when loop corrections are included. Thus, \MW\
in LEP1 calculations is \Mt- and \MH-dependent through the
above procedure. At LEP2, where one wants to measure \MW\ and, hence, wants
to treat \MW\ as a fit parameter, the
above relation now primarily acts as a test of the SM. The
above relation predicts for any chosen \MH\ and measured \MW\
a value for \Mt\ that can be used as input for LEP2 loop
calculations and can also be compared with the directly observed
top-quark mass from the Tevatron \cite{FNALmt}.

As to the actual procedure to measure \MW\ from \PW-pair
production, two methods are advocated \cite{Wmassgroup}. A third one, involving
the measurement of the endpoint energy of
a charged lepton originating from a decaying \PW\ boson,
was suggested a few years ago but turns out to be less
promising than other methods.

One procedure requires a measurement of the total \PW-pair
\cs\ close to threshold, where the size of
$\sigma_{\mathrm{tot}}$ is most sensitive to the \PW\ mass. The 
energy proposed is 161\GeV. For this method the theory
should obviously predict $\sigma_{\mathrm{tot}}$ with a sufficient
accuracy, to wit \,$\sim 2\%$.\footnote{Throughout this report the required
theoretical accuracy is taken to be half the expected statistical error.}
Therefore the radiative corrections (RC's) to the
total \cs\ should be under control.

The other method looks at the decay products of the \PW\
bosons, in particular at the four- and two-jet production.
{}From the measured momenta of the decay products one tries
to reconstruct the \PW\ mass. In this reconstruction
a good knowledge of the \PW-pair centre-of-mass energy
is essential. Since there will be an energy loss due
to initial-state radiation (ISR) of primarily photons,
the \PW-pair energy will be different from the laboratory energy of the 
incoming electrons and positrons. So, for this method the ISR should be well 
under control, i.e.~the theoretical error on the average energy loss, 
$\Eg = \int\d E_\ga\, (\d\si/\d E_\ga)\,E_\ga/\si_{\mathrm{tot}}$,
should be less than 30\MeV, which translates into a 
theoretical error of less than 15\MeV\ on the reconstructed \PW\ mass.
This again is an aspect of radiative corrections.

\subsubsection{Test of non-abelian couplings}

Within the SM the triple gauge-boson couplings have the
Yang-Mills (YM) form. Amongst others, this specific form for the TGC's leads to
a proper high-energy behaviour of the \PW-pair \cs\ and is a 
requirement for having a renormalizable theory. Couplings different
from the YM form, called anomalous or non-standard couplings,
will in general lead to a high-energy behaviour of \css\ 
increasing with energy and thereby violating unitarity. At
LEP2 the energy is too low to see such effects and in order 
to establish the presence of anomalous couplings one therefore
has to study in detail the angular distributions of the \PW\ bosons
and their decay products. In particular, the angular distribution
in the \PW-production angle $\theta$ is sensitive to non-standard
couplings. Again, the knowledge of RC's to the tree level SM
predictions is required, since they affect the Born-level angular
distributions \cite{BD94}. 

As elsewhere in this volume a detailed report on non-standard
TGC's is given \cite{ACgroup}, only a few comments will be made here.

When one considers the most general coupling between three gauge bosons
allowed for by Lorentz invariance, assuming the gauge bosons to be coupled to
conserved currents, one ends up with seven
possible couplings for the \PZ\PW\PW\ and $\gamma$\PW\PW\ interaction.
Of these seven there are three which are CP violating and
one which is CP conserving but C and P violating \cite{Ha87}.

In practice it will be impossible to set limits on all these
couplings. Therefore usually some assumptions are made to
reduce the number of 14 couplings \cite{Ha87,Bi93}. For instance, one may
restrict oneself to CP-conserving non-standard couplings
so that 8 couplings are left. Of these the electromagnetic
ones can be reduced further by omitting the C and P
violating one and requiring the strength of the 
electromagnetic coupling to be determined by the charge.
Two possible anomalous electromagnetic couplings remain
and four \PZ\PW\PW\ anomalous couplings. Even with this
reduced number it will be impossible to set experimental
limits on all of them simultaneously.

However, there are theoretical arguments that such a purely
phenomenological approach is also not required. First of all one might use
symmetry arguments, motivated by specific models for the non-standard 
physics, to find relations between the anomalous couplings \cite{Bi93}. 
Alternatively, when one considers the electroweak theory as an effective theory
originating from a field theory that manifests itself at 
higher energies, then also some small anomalous couplings
may be present at lower energies. In such a $SU(2)\times U(1)$ gauge-invariant
framework the non-standard physics, situated at an energy scale $\Lambda$,
decouples at low energies and the anomalous TGC's are suppressed by factors
$(E/\Lambda)^{d-4}$, according to the dimension ($d$) of the 
corresponding operators. This naturally introduces a hierarchy amongst the
anomalous TGC's based on the dimension of the corresponding operators
\cite{AChierarchy}.

{}From the perspective of this report the origin of the
non-standard couplings is not so important, but the
fact that they often modify angular distributions
is relevant. Also SM effects -- like RC's, the finite decay width of the 
\PW\ bosons, and background contributions -- provide deviations from the 
tree-level distributions \cite{BD94,ACgroup}.

\subsection{How to obtain accurate predictions?}

In order to extract the wanted information from \PW-pair
production it is clear from the remarks above that RC's
are needed for total and differential \css\
and, moreover, for the determination of the energy loss. Anticipating 
\,$\sim 10^4$ \PW-pair events, the theoretical accuracy that should be
targeted for the SM predictions is \,$\sim 0.5\%$, although specific final
states, distributions, or observables in fact often do not require such a
precision (like, e.g., $\sigma_{\mathrm{tot}}$ at 161\GeV\ or the energy loss).

Ideally one would like to have the full RC's to the final state
of four fermions, which originate from the two
decaying vector bosons. In practice such a very involved
calculation does not exist and is hopefully not
required in its complete form for the present accuracy. For the discussion
of the LEP2 situation and strategy it is useful to
distinguish three levels of sophistication in the
description of \PW-pair production.

The first level is to consider on-shell \PW-pair
production, \eeWW,
which at tree level is described by three diagrams:
neutrino exchange in the $t$ channel, and $\gamma$ and \PZ\
exchange in the $s$ channel. Here the complete $\O(\alpha)$ 
RC's are known, comprising the virtual one-loop corrections 
and the real-photon bremsstrahlung \cite{Bo88,Fl89}. When one wants to 
divide the $\O(\alpha)$ corrections into
different parts the situation differs from LEP1 \cite{BD94}. 

As to the bremsstrahlung, a gauge-invariant separation
into initial-state radiation, final-state radiation (FSR), and its
interference is not possible like, e.g., in \Pmu-pair
production at LEP1. The reason is that the photon should
couple to all charged particles in a line of the Feynman
diagram. The $t$-channel diagram then makes a separation
into ISR and FSR impossible. However, the leading logarithmic (LL) part of 
ISR is in itself gauge-invariant. This can be combined with the
LL QED virtual corrections so that a LL description
with structure functions for ISR can be given \cite{SFM}.

A separation of the virtual corrections into a
photonic and weak part is also not possible since
charged vector bosons are already present at Born
level, necessitating an interplay between $\ga$- and \PZ-exchange diagrams
in order to preserve $SU(2)$ gauge invariance.

Once one has a description of on-shell \PW-pair
production one can attach to it the on-shell \PW\
decay. Again, the RC's to this decay are known \cite{Al80}--\cite{De90a}. 

The next level of sophistication is to consider off-shell production of 
\PW\ pairs, which then decay into four fermions \cite{Mu86}. 
The \PW\ propagators with energy-dependent widths can be taken into account. 
Although this description of four-fermion production through
virtual \PW\ bosons is a natural extension of the on-shell
evaluation, it is not a gauge-invariant
treatment. In fact there are more diagrams needed to
calculate such a four-fermion process. As to RC's, the
ISR and FSR can be implemented in LL approximation, but the full set of  
virtual corrections have not yet been calculated.

The final level for the study of \PW-pair production
would be a full $\O(\alpha)$-corrected 
evaluation of all possible four-fermion final states.
At tree level there now exist evaluations where,
besides the three off-shell \PW-pair diagrams, all
other diagrams for a specific four-fermion final
state have been included \cite{EGgroup}. On top of that ISR can
be taken into account. Again, one has to be aware
of possible gauge-invariance problems. In particular,
the introduction of energy-dependent widths in the
\PW-propagators will destroy electromagnetic gauge
invariance and may introduce dramatically wrong
\css\ in certain regions of phase space.

As long as a full $\O(\alpha)$-corrected evaluation
of four-fermion production is not available, certain
approximative schemes, like for instance the `pole scheme'
\cite{Ve63}--\cite{Ae94}, may be useful.
This goes beyond the treatments where only ISR through
structure functions is taken into account. Actual
numerical results from a complete `pole-scheme' evaluation
are not yet available.

\section{On-Shell W-Pair Production and W Decay}
\label{CHon}
Although the actual process that will be probed at LEP2 is \eeffff$(\ga,g)$,
we first focus on the production and subsequent decay of on-shell \PW\ bosons,
being basic building blocks in some of the schemes for handling off-shell
\PW\ bosons. In contrast to off-shell \PW-pair production
the on-shell processes are not plagued by the problem of gauge invariance for
unstable particles and a complete set of \Oa\ radiative corrections is 
available. Consequently, they are well suited for studying
the typical sizes of various radiative corrections. Moreover, many
of the important features of the production and decay of off-shell \PW\ bosons
are already contained in the on-shell limit. We indicate explicitly where the 
width of the \PW\ bosons radically changes the on-shell predictions. 

\subsection{Notation and conventions}
\label{SEonnc}
We use the Bj\o{}rken--Drell metric $g_{\mu\nu} =
\mathrm{diag}(+1,-1,-1,-1)$
and fix the totally antisymmetric tensor by $\eps^{0123}=+1$. The
matrix $\ga_5$ is defined as $\ga_5=i\ga^0\ga^1\ga^2\ga^3$ and the
helicity projectors $\om_\pm$, which are used to project on right-
and left-handed massless fermions, as
\beq
\om_\pm = \frac{1}{2}(1\pm\ga_5).
\eeq
 
First we set the conventions for the process
\beq
\Pep(\pep,\hep)+\Pem(\pem,\hem)\to
\PWp(\pWp,\hWp)+\PWm(\pWm,\hWm),
\eeq
where the arguments indicate the momenta and helicities of the incoming
fermions and outgoing bosons ($\he_{i}=\pm \frac{1}{2}$, $\lambda_{i}=
1,0,-1$).
Note that we sometimes use the shorthand version $\he_{i}=\pm$
in certain sub- and superscripts.
In the centre-of-mass (CM) system of the $\Pep\Pem$ pair, which we will
refer to as the laboratory (LAB) system in the following, the momenta
read
\beq
\pepm^\mu = E(1,0,0,\mp1), \qquad
\pWpm^\mu = E(1,\mp\beta\sin\theta,0,\mp\beta\cos\theta),
\eeq
with $E$ denoting the beam energy, $\theta$ the scattering angle
between the $\Pep$ and the $\PWp$,
and $\beta=\sqrt{1-\MW^2/E^2}$  the velocity of the \PW~bosons.
The Mandelstam variables used in the following are given by
\beq \label{Mandelstam}
\barr{lll}
s&=&(\pep+\pem)^{2}=(\pWp+\pWm)^{2}=4E^{2}, \\[1ex]
t&=&(\pep-\pWp)^{2}=(\pem-\pWm)^{2}=
-E^{2}(1+\beta^{2}-2\beta\cos\theta) ,\\[1ex]
u&=&(\pep-\pWm)^{2}=(\pem-\pWp)^{2}=
-E^{2}(1+\beta^{2}+2\beta\cos\theta) .
\earr
\eeq
 
In order to define helicity amplitudes we need to introduce the
corresponding polarization vectors for the \PWp\ and \PWm\ boson%
\footnote{Note that the helicity of the massive \PW~bosons is not
Lorentz-invariant. We define it in the LAB system.}
\beqar
\nn {\eWpm}^{\!\mu}(\pWpm,+1) &=&
\frac{1}{\sqrt{2}}(0,\mp\cos\theta,-i,\pm\sin\theta),\\[1ex]
\nn {\eWpm}^{\!\mu}(\pWpm,-1) &=&
\frac{1}{\sqrt{2}}(0,\mp\cos\theta,+i,\pm\sin\theta),\\[1ex]
    {\eWpm}^{\!\mu}(\pWpm,\phantom{-}0) &=&
\frac{E}{\MW}(\beta,\mp\sin\theta,0,\mp\cos\theta).
\eeqar
 
Because we are neglecting the electron mass, the helicity of the
positron is opposite to the helicity of the electron
\beq
\hem = - \hep = \he.
\eeq
Henceforth we refer to the helicity amplitudes for \PW-pair production as
$\M(\he,\hWp,\hWm,s,t)$.
CP invariance implies the relation:
\beq \label{CPrel}
\M (\he ,\hWp, \hWm,s,t) = \M (\he,-\hWm,-\hWp,s,t).
\eeq
This holds in the SM if we neglect the CP-violating phase in the
quark-mixing matrix. Even in the presence of this CP-violating phase,
the CP breaking occurs at \Oaa\ in SM \PW-pair production and is
additionally suppressed by the smallness of the mixing angles between the 
quarks. Consequently in the SM one effectively has only 12 independent 
helicity-matrix elements instead of 36. In the presence of substantial 
(non-standard) CP violation this number should be increased to 18.
This allows for a decomposition of the matrix elements in terms of an explicit
set of 12 (18) independent basic matrix elements multiplied by purely
kinematical invariant functions (coefficients) \cite{BD94,Bo88,Fl89}.

{}From the helicity amplitudes the differential \css\ for explicit \PW-boson
polarization and various degrees of initial-state polarization can be 
constructed. For example the differential \cs\ for unpolarized electrons, 
positrons, and \PW\ bosons is given by
\beq
\dsidop =
\frac{\beta}{64\pi^{2}s}\sum_{\he,\hWp,\hWm}
\frac{1}{4}\left|\M(\he,\hWp,\hWm,s,t)\right|^{2}.
\eeq
 
For the numerical evaluations we have to fix the input parameters. We use the 
following default set \cite{summer95,PDG94}:
$$\begin{array}{lcllcllcl}
\al &\equiv& \al(0) = 1/137.0359895, \qquad &
\GF & = & \rlap{$1.16639 \times 10^{-5} \GeV^{-2}$,} \\[.3em]
\MZ & = & 91.1884\GeV, &
\MH & = & 300\GeV, &&& \\[.3em]
\Me & = & 0.51099906\MeV,  \qquad &
m_{\mu} & = & 105.658389\MeV,  \qquad &
m_{\tau} & = & 1.7771\GeV, \\[.3em]
\Mu & = & 47.0\MeV, &
\Mc & = & 1.55\GeV, &
    &   & \\[.3em]
\Md & = & 47.0\MeV, &
\Ms & = & 150\MeV, &
\Mb & = & 4.7\GeV.
\end{array}$$
The masses of the light quarks are adjusted in such a way that the
experimentally measured hadronic vacuum polarization \cite{EJ95}
is reproduced. In the actual calculations either these light quark masses are
used or the dispersion-integral result for the hadronic vacuum polarization 
of \cite{EJ95}. The strong coupling constant is calculated according to the
parametrization of \cite{Ma84}, using $\als(\MZ^2)=0.123$ as input.
The \PW-pair threshold region is very sensitive to \MW\ and consequently, if
\MW\ were to be calculated from the other parameters using the muon decay 
width including radiative corrections (like at LEP1), to the masses of the top 
quark and Higgs boson. This is of course not very natural, since we want to 
use \MW\ as a model-independent fit parameter. In view of this we
use $\al$, $\GF$, \MZ\ as input\footnote{We do not eliminate \MZ\ or \GF,
as \MZ\ is needed in the RC's and \GF\ reduces the size of the RC's to the
production and decay of the \PW\ bosons.},
treat \MW\ as free fit parameter,
and calculate \Mt\ from muon decay\footnote{It should be noted that the RC's 
associated with the muon decay make that this parameter set is not 
overcomplete.}. This calculated value of \Mt\ can then be confronted with
the direct bounds from Fermilab (weighted average $\Mt = 180 \pm 12\GeV$ 
\cite{FNALmt}) and the indirect ones from the precision measurements at 
LEP1/SLC ($\Mt = 180^{\,+8\,+17}_{\,-9\,-20}\GeV$ 
\cite{summer95}). In this scheme \Mt\ hence plays the role played by \MW\ 
at LEP1. The above set of parameters and the default \PW\ mass  
\MW=80.26\GeV\ \cite{summer95,CDFUA2} yield
$$\Mt=165.26\GeV.$$ At present the error on this value as a result of the 
160\MeV\ error on \MW\ is roughly 27\GeV. The precise \MW\ measurement at 
LEP2 will lead to a reduction of this error by a factor of four. 
Of course the so-obtained top-quark mass will become
$\als$- and \MH-dependent through the RC's associated with the muon decay.
We come back to that point in \refse{SEonrcho} and only mention here that the
\MH\ dependence of \Mt\ amounts to roughly 33\GeV\ in the range
$60\GeV < \MH < 1\TeV$, which clearly exceeds the above-mentioned 
expected error from \MW\ (see \refta{TAmtcalc}). 
\btab[bt]
\bce
\begin{tabular}{||c|c|c|c||} \hline\hline
 \multicolumn{1}{||r|}{$\MH\ [\mathrm{GeV}] =\ $} & \hpt 60 & 300 & 1000 \\ 
 \hline
 $\MW\ [\mathrm{GeV}]$ & \multicolumn{3}{c||}{$\Mt\
 [\mathrm{GeV}]$} \\ \hline
 80.10 & 119.1 & 137.3 & 154.2 \\ \hline
 80.18 & 133.9 & 151.6 & 168.0 \\ \hline
 80.26 & 148.1 & 165.3 & 181.2 \\ \hline
 80.34 & 161.7 & 178.3 & 193.9 \\ \hline
 80.42 & 174.4 & 190.7 & 206.1 \\ 
 \hline \hline
\end{tabular}
\ece
\caption{Calculated \Mt\ for $\als(\MZ^2)=0.123$ and different Higgs- and 
         \PW-boson masses, using the state-of-the-art calculation described in 
         \refse{SEonrcho}. The theoretical error in \Mt\ is roughly 1--2\GeV.} 
\label{TAmtcalc}
\etab 

Finally, the sine and cosine of the weak mixing angle are defined by
\beq
\cw = \cos\theta_{\PW} = \frac{\MW}{\MZ}, \qquad 
\sw = \sin\theta_{\PW} = \sqrt{1-\cw^2}.
\eeq
 
\subsection{Lowest order}
\label{SEonlo}
 
In the SM there are three lowest-order diagrams
(\reffi{FIeewwdia}), if we omit a Higgs-exchange diagram that is
suppressed by a factor $\Me/\MW$ and thus completely negligible.
\bfi
\unitlength 1pt
\savebox{\wigr}(12,0)[bl]
   {\bezier{20}(0,0)(3, 4)(6,0)
    \bezier{20}(6,0)(9,-4)(12,0)}
\savebox{\Vr}(48,0)[bl]{\multiput(0,0)(12,0){4}{\usebox{\wigr}}}
\savebox{\wigur}(8,6)[bl]
   {\bezier{20}(0,0)(1,4)(4,3)
    \bezier{20}(4,3)(7,2)(8,6)}
\savebox{\Vur}(32,24)[bl]{\multiput(0,0)(8,6){4}{\usebox{\wigur}}}
\savebox{\wigdr}(8,6)[bl]
   {\bezier{20}(0,0)(1,-4)(4,-3)
    \bezier{20}(4,-3)(7,-2)(8,-6)}
\savebox{\Vdr}(32,24)[bl]{\multiput(0,24)(8,-6){4}{\usebox{\wigdr}}}
\savebox{\Vudr}(32,48)[bl]
{\put(00,24){\usebox{\Vur}}
\put(00,00){\usebox{\Vdr}}}
\savebox{\Fu}(0,48)[bl]
{ \put(0,0){\vector(0,1){27}} \put(0,24){\line(0,1){24}} }
\savebox{\Fr}(48,0)[bl]
{ \put(0,0){\vector(1,0){26}} \put(24,0){\line(1,0){24}} }
\savebox{\Fl}(48,0)[bl]
{ \put(48,0){\vector(-1,0){26}} \put(24,0){\line(-1,0){24}} }
\savebox{\Fur}(32,24)[bl]
{ \put(0,0){\vector(4,3){18}} \put(16,12){\line(4,3){16}} }
\savebox{\Fdr}(32,24)[bl]
{ \put(32,0){\vector(-4,3){19}} \put(16,12){\line(-4,3){16}} }
\savebox{\Fudr}(32,48)[bl]
{\put(00,24){\usebox{\Fur}}
\put(00,00){\usebox{\Fdr}}}
\bma
\barr{lll}
\begin{picture}(176,72)
\put(00,60){\makebox(10,10)[bl]{$\Pep$}}
\put(00,5){\makebox(10,10)[bl] {$\Pem$}}
\put(116,60){\makebox(10,10)[bl]{$\PWp$}}
\put(116,5){\makebox(10,10)[bl] {$\PWm$}}
\put(70,33){\makebox(20,10)[bl] {$\Pne$}}
\put(16,60){\usebox{\Fl}}
\put(16,12){\usebox{\Fr}}
\put(64,12){\usebox{\Fu}}
\put(64,60){\circle*{4}}
\put(64,12){\circle*{4}}
\put(64,10){\usebox{\Vr}}
\put(64,58){\usebox{\Vr}}
\end{picture}  &
\begin{picture}(148,72)
\put(00,60){\makebox(10,10)[bl]{$\Pep$}}
\put(00,5){\makebox(10,10)[bl] {$\Pem$}}
\put(134,60){\makebox(10,10)[bl]{$\PWp$}}
\put(134,5){\makebox(10,10)[bl] {$\PWm$}}
\put(62,43){\makebox(20,10)[bl] {$\ga,\PZ$}}
\put(16,36){\usebox{\Fdr}}
\put(16,12){\usebox{\Fur}}
\put(48,36){\circle*{4}}
\put(96,36){\circle*{4}}
\put(48,34){\usebox{\Vr}}
\put(96,36){\usebox{\Vur}}
\put(96,6){\usebox{\Vdr}}
\end{picture}
\earr
\ema
\caption{Lowest-order diagrams for \eeWW.}
\label{FIeewwdia}
\efi
The $t$-channel diagram involving the $\Pne$ exchange
only contributes for left-handed electrons.
The $s$-channel diagrams, containing the non-abelian triple
gauge-boson couplings, contribute for both helicities of the electron.
The corresponding matrix element reads
\beq \label{M0eeWW}
\M_{\born}(\he,\hWp,\hWm,s,t) =
\frac{e^{2}}{2\sw^2}\,\frac{1}{t}\, \M^{\he}_1\,\de_{\he-}
+ e^2
\biggl[\frac{1}{s}-\frac{\cw}{\sw}\,\geeZ\,\frac{1}{s-\MZ^2}\biggr]\,
2\,(\M^{\he}_{3}-\M^{\he}_{2}),
\eeq
with $\de_{\he-} = 1$ for left-handed electrons and $\de_{\he-}=0$
for right-handed electrons, and
\beqar \label{MeeWW}
{\M}^{\he}_{1} & = &
\vb(\pep)\,\esWp(\pWp,\hWp)\,(\psWp - \psep)\,\esWm(\pWm,\hWm)
\,\om_{\he}\,u(\pem) \nlc[1ex]
{\M}^{\he}_{2} & = &
\vb(\pep)\,\frac{\psWp-\psWm}{2}\,[\eWp(\pWp,\hWp)\cdot\eWm(\pWm,\hWm)]
\,\om_{\he}\,u(\pem) \nlc[2ex]
{\M}^{\he}_{3} & = &
\vb(\pep) \left(\esWp(\pWp,\hWp)\,[\eWm(\pWm,\hWm)\cdot\pWp] 
-\esWm(\pWm,\hWm)\,[\eWp(\pWp,\hWp)\cdot\pWm]\right)\om_{\he}\,u(\pem).
\hphantom{AA}
\eeqar
After inserting the explicit form of the \PZ-boson--fermion couplings
\beq
\geeZ = \frac{\sw}{\cw} - \de_{\he-}\frac{1}{2\sw\cw},
\eeq
we can organize the lowest-order amplitude into
two gauge-invariant subsets:
\beq \label{M0eeWWgi}
\M_{\born}(\he,\hWp,\hWm,s,t) =
\frac{e^{2}}{2\sw^2}\,\M_I(\he,\hWp,\hWm,s,t)\,\delta_{\he-}
+e^{2} \M_{Q}(\he,\hWp,\hWm,s,t),
\eeq
where \eqskipcorr
\beqar \label{MIMQ}
\M_{I}(\he,\hWp,\hWm,s,t) &=&
\frac{1}{t}\,\M^{\he}_{1}
+\frac{1}{s-\MZ^2}\,2\,(\M^{\he}_{3}-\M^{\he}_{2}) \nlc
\M_{Q}(\he,\hWp,\hWm,s,t) &=&
\biggl[\frac{1}{s}-\frac{1}{s-\MZ^2}\biggr]\,2\,(\M^{\he}_{3}-\M^{\he}_{2}).
\eeqar

The gauge invariance of the two contributions $\M_I$ and $\M_Q$
can be simply inferred from the fact that they are accompanied by
different coupling constants, one of which involving the
electromagnetic coupling constant $e$, the other
the charged-current coupling constant $e/(\sqrt{2}\sw)$.
Whereas $\M_I$ corresponds to the pure $SU(2)$ contribution,
the parity-conserving contribution $\M_Q$ is a result of the
symmetry-breaking mechanism.
 
The lowest-order \cs\ determines the essential features of \PW-pair
production.
The threshold behaviour is important for the determination of the \PW~mass 
from the measurement of the \cs\ in a single energy
point very close to threshold \cite{Wmassgroup}, i.e.~at $\sqrt{s}=161\GeV$.  
For small $\beta $ the matrix elements behave as
\beq
\M^{\he}_{2},\M^{\he}_{3}\propto\beta, \qquad
\M^{\he}_{1}\propto 1
\eeq
for fixed scattering angles.
Consequently, the $s$-channel matrix elements vanish at thres\-hold%
\footnote{This holds for arbitrary CP-conserving $s$-channel
contributions in the limit of vanishing electron mass \cite{Ha87}.}
and the $t$-channel graph dominates in the threshold region.
For $\beta \ll 1$ the differential \cs\ for unpolarized beams
and \PW~bosons is given by \cite{BD94}
\beq \label{dsigthr}
\dsidop_{\born}\approx
\frac{\al^{2}}{s}\,\frac{1}{4\sw^4}\,\be\,
\biggl[1 +
4\be\cos\theta\,\frac{3\cw^2-1}{4\cw^2-1}
+\O(\beta^{2})\biggr],
\eeq
where the leading term $\propto\beta$ originates from the $t$-channel
diagram only. Note that the leading term is angular-independent.
In the total \cs\
\beq \label{sigthr}
\sigma_{\born}\approx
\frac{\pi \al ^{2}}{s}\,\frac{1}{4\sw^4}\,4\beta
+\O(\beta ^{3}),
\eeq
all terms $\propto\beta^{2}$ 
drop out and the $s$-channel and the $s$--$t$-interference
contributions are proportional to $\beta^3$. This is the consequence of CP 
conservation, fermion-helicity conservation in the initial state, and the 
orthogonality of different partial waves \cite{BD94}.
Hence in the threshold region the $t$ channel is dominant and
the \cs\ for \eeWW\ is not very sensitive to the triple gauge-boson
couplings.

In \refta{TAsiww} we give the integrated \cs\ for different
centre-of-mass energies and different polarizations
of the electrons (+,$-$). Positrons are assumed to be unpolarized.
\btab
\arraycolsep 6pt
$$
\begin{array}{||r|r|r|r||} \hline\hline
\multicolumn{1}{||c|}{ \,\sqrt{s}\ [\mathrm{GeV}]\,} &
\multicolumn{1}{c|}{\si_\born} &
\multicolumn{1}{c|}{\si^-_\born} &
\multicolumn{1}{c||}{\si^+_\born} \\
\hline
 161.0 \quad& 3.812 & 7.622 & 0.002 \\
\hline
 175.0 \quad& 15.959 & 31.716 & 0.202  \\
\hline
 184.0 \quad& 17.427 & 34.567 & 0.287 \\
\hline
 190.0 \quad& 17.762 & 35.203 & 0.321 \\
\hline
 205.0 \quad& 17.609 & 34.867 & 0.350 \\
\hline\hline
\end{array}
$$
\caption{Integrated lowest-order \cs\ in pb for different polarizations of the 
electrons and different centre-of-mass energies.}
\label{TAsiww}
\etab
Using right-handed electrons one could study a pure triple-gauge-%
coupling process, but this would require longitudinally polarized electron 
beams, the prospect of which looks rather unfavourable for LEP2.
Furthermore, for all energies the \cs\ for right-handed electrons is
suppressed by two orders of magnitude compared with the dominant
left-handed mode, mainly because there is no $t$-channel contribution.
Therefore essentially only the latter can be investigated at LEP2.
With transverse beam polarization, however, one could obtain information
on the right-handed matrix element via its interference with the
left-handed one.
 
As stated in \refse{SEonnc}, for on-shell \PW-pair production and
unpolarized electrons and positrons there are nine independent
helicity-matrix elements or six if CP is conserved. These yield
nine or six independent observables. Taking into account the
decay of the \PW~bosons there are many more observables:
81 or, if CP is conserved, 36 products of the various helicity-matrix elements
\cite{Ha87}.
Because the $V-A$ structure of the \PW~decays is well established,
these observables can be extracted in a model-independent
way from the data for the five-fold differential \cs\
$\d\si/(\d\!\cos\theta\,\d\!\cos\theta_1\,\d\phi_1\,\d\!\cos\theta_2\,
\d\phi_2)$,
where $\theta_{1,2}$ and $\phi_{1,2}$ represent the decay angles of the two
\PW~bosons. These observables may serve to put limits on anomalous 
couplings \cite{ACgroup,Bi93}.
Note that the five-fold differential \cs\ requires the identification of the
charge of at least one of the decaying \PW\ bosons, as otherwise the 
information on the sign of $\cos\theta$ would be lost. This is possible for 
hadronic--leptonic or leptonic--leptonic events. If no charge identification is
possible only production-forward--backward-symmetric
observables would be left. As argued above, in those observables
for instance the $s$-channel
contribution is suppressed by $\be^2$ in the threshold region\footnote{If CP is
violated there exists an anomalous gauge-boson coupling that does not yield a
suppressed $s$-channel contribution \cite{BD94,Ha87}.}.
Consequently, if one cannot identify the jet charges the purely
hadronic events will not be of much use for studies of the gauge couplings
at LEP2. Of course this does not concern the \PW-mass
determination as it does not rely on $s$-channel contributions.
 
\subsection{Radiative corrections}
\label{SEonrc}
\unitlength 1pt
 
As has been argued in \refcha{CHintro}, the SM theoretical predictions for
\PW-pair production should have an uncertainty of about 0.5\% (2\% at 
$\sqrt{s}=161\GeV$) in order to
obtain reasonable limits on the structure of the gauge-boson
self-couplings, and to determine the \PW-boson
mass with the envisaged precision of roughly 40--50\MeV.
In this context radiative corrections are indispensable.
 
Much effort has been made in recent years to obtain such precise
theoretical predictions for \PW-pair production.
In the following we discuss the existing results for the
virtual and real electroweak corrections in the on-shell case.
In addition we discuss the quality of an improved Born approximation (IBA)
that contains all familiar, LEP1-like leading corrections for the \PW-pair
production \cs\ at LEP2 energies. Such a discussion is in particular relevant
as the present off-shell LEP2 Monte Carlos often make use of such an 
approximation.
 
\subsubsection {Radiative electroweak \Oa\ corrections}
\label{SEonrcal}

The $\O(\al)$ radiative corrections can be naturally divided into
three classes, the virtual, soft-photonic, and hard-photonic
contributions. Since the process $\Pep\Pem\to \PWp\PWm$ involves the
charged current in lowest order, the corresponding radiative corrections
cannot be separated on the basis of Feynman diagrams into
electromagnetic and weak contributions in a gauge-invariant way. 
Like we have already observed in \refeq{MIMQ}, $SU(2)$ gauge invariance 
requires an interplay between the $\ga$- and \PZ-exchange diagrams.
 
The complete radiative electroweak \Oa\ corrections have been calculated 
independently by two groups \cite{Bo88,Fl89}%
\footnote{The process $\eeWW\gamma$ has also been calculated in 
\cite{WWgam}.}. For the unpolarized case they 
have been checked and found to agree within a couple of per-mil (i.e.~within
the integration error). 

The virtual corrections contain infra-red (IR) divergences, which result from
virtual photons exchanged between external charged particles. They are 
compensated for by adding the \cs\ for
the process $\eeWW\ga$. If the energy of the emitted photon is small
compared with the detector resolution (soft photons), this process cannot be
distinguished experimentally from the non-radiative \PW-pair production
process. In practical experiments the soft-photon approximation is in general
not sufficient and one has to include the radiation of hard photons, too.
When adding these real-photon effects to the contribution of the
virtual corrections, not only the IR singularities but also the large
Sudakov double logarithms $\log^{2}(s/\Me^{2})$ drop out.
 
 Still there are various sources of potentially large \Oa\ corrections left
at LEP2 energies. First of all there are large QED corrections of the
form $(\al/\pi)\log(Q^2/\Me^2)$ with $Q^2 \gg \Me^2$, originating from
collinear photon radiation off the electron or positron (see \refapp{app}).
They form a gauge-invariant subset of QED corrections and amount to roughly 
6\% at LEP2, not taking into account possible enhancements from the 
corresponding coefficients (like, e.g., close to thresholds). 
 
{}From the renormalization two sets of potentially large fermionic (formally
weak) corrections arise. The first set is associated with the charge
renormalization at zero momentum transfer, where the relevant scale is
set by the fermion masses entering the vacuum polarization.
In high-energy experiments, however, the running charge should be
evaluated at scales much larger than the masses of the light fermions
$\fl = \{\Pe,\Pmy,\Pta,\Pu,\Pd,\Pc,\Ps,\Pb\}$. This leads to large
logarithmic (`mass singular') contributions of the form $(\al/\pi)
\log(Q^2/m_{\fl}^2)$ with $Q^2 \gg m_{\fl}^2$, which can amount to a
shift in $\al$ of 8\% at LEP2 energies.
Related to the top quark, corrections $\propto \Mt^2/\MW^2$ will
occur. They show up as universal corrections via the renormalization of
the \PW\ and \PZ~masses (or equivalently $\sw^2$) if the corresponding
renormalization scales are small compared with the mass splitting in
the $(\Pt,\Pb)$ isospin doublet.
 
Finally the long-range electromagnetic interaction between
the slowly moving \PW~bosons leads to the so-called Coulomb singularity
\cite{CS}.
This singularity yields an \Oa\ correction factor $\al\pi/(2\beta)$,
resulting in an \Oa-corrected \cs\  that does not vanish at threshold
for left-handed electrons. The right-handed \cs\ remains
suppressed by at least $\be^2$.
This correction factor exhibits the fact that the
free-particle approximation is inadequate near the \PW-pair production
threshold in the presence of the long-range Coulomb interaction. We
want to stress here that the Coulomb singularity is changed
substantially by effects that effectively truncate the range of the
interaction, like the off-shellness and the decay of the \PW~bosons.
This will be treated in \refse{SEofcoul} and will lead to the
conclusion that higher-order Coulombic corrections to the total \cs\ are not 
important.

The sensitivity of the total unpolarized \cs\ to $\als$ and the unknown 
mass of the Higgs boson is illustrated in \refta{mhalstab}.
\btab
\bce
\begin{tabular}{||c|c|c|c||c|c|c||} \hline\hline
 \multicolumn{1}{||r|}{$\MH\ [\mathrm{GeV}] =\ $} & 300 & 300
 & 300 & \hpt 60 & 300 & 1000 \\ \hline
 \multicolumn{1}{||c|}{$\als =\ $} & 0.117 & 0.123
 & 0.129 & 0.123 & 0.123 & 0.123 \\ \hline
 \multicolumn{1}{||r|}{$\Mt\ [\mathrm{GeV}] =\ $} & 164.80 & 165.26 & 165.73
 & 148.14 & 165.26 & 181.20 \\ \hline
 $\sqrt{s}\ [\mathrm{GeV}]$ & \multicolumn{6}{c||}{$\sigma\
 [\mathrm{pb}]$} \\ \hline
 \hpt 161.0 & \hpt 2.473 & \hpt 2.472 & \hpt 2.470 & \hpt 2.514 &
 \hpt 2.472 & \hpt 2.460 \\ \hline
 \hpt 175.0 & 14.471 & 14.465 & 14.459 & 14.548 & 14.465 & 14.422 \\ \hline
 \hpt 184.0 & 16.619 & 16.613 & 16.607 & 16.671 & 16.613 & 16.568 \\ \hline
 \hpt 190.0 & 17.263 & 17.257 & 17.251 & 17.301 & 17.257 & 17.213 \\ \hline
 \hpt 205.0 & 17.683 & 17.677 & 17.671 & 17.698 & 17.677 & 17.637 \\ 
 \hline \hline
\end{tabular}
\ece
\caption{Integrated unpolarized \cs\ including radiative electroweak \Oa\
corrections for different values for the Higgs-boson mass and 
$\als$, at various centre-of-mass energies. The theoretical error in
\Mt\ is roughly 1--2\GeV.} 
\label{mhalstab}
\etab
The dependence on $\als$, originating from the calculation of \Mt, is 
completely negligible ($\lsim 0.1\%$). Compared with the lowest-order \css\
of \refta{TAsiww}, a variation of \MH\
between 60 and 1000\GeV, however, influences the total \cs\ by around
0.5\% at the three highest energy points, by about 0.8\% at $\sqrt{s}=175\GeV$,
and by 1.4\% at threshold. These numbers are lowered by about 0.5\% if the 
radiative corrections are calculated in the so-called \GF\ representation,
which absorbs the universal \Mt\ and \MH\ effects present
in the \PW\ wave-function factors (see \refse{SEgwrc}). In this representation
a variation of \MH\ between 300 and 1000\GeV\ has a negligible impact on the
\cs. Keeping in mind that we would like to reach
a theoretical accuracy of 2\% (0.5\%) at $\sqrt{s}=161\GeV$ 
($\sqrt{s}\geq 175\GeV$), it should be clear that the Higgs-mass dependence 
constitutes a major uncertainty. A more detailed investigation \cite{BvO96} 
revealed that this Higgs-boson-mass dependence is the result of the Yukawa 
interaction between the two slowly moving \PW\ bosons (mediated by the Higgs 
boson). As such the effect is largest close to threshold and for the lightest 
Higgs masses, which yield the shortest range of the interaction. For instance,
upon increasing the lower Higgs-mass bound from 60 to 90\GeV, the resulting
uncertainty is reduced by roughly a factor of two%
\footnote{At the start of the 161\GeV\ run no significant change in the \MH\
bound is expected. So, only after the higher-energy LEP2 runs have taken place
the improved knowledge on \MH\ can be used for an a posteriori reduction of 
the \MH\ dependence of the 161\GeV\ run.}.

In \reffi{totthr} we display
the influence of the full \Oa\ corrections on the total unpolarized
\cs\ near the \PW-pair production threshold.

\bfi
 \begin{center}
  \setlength{\unitlength}{1cm}
  \hspace*{-0.2cm}
  \begin{picture}(15.5,8.7)
  \put (-1.3,-2.5){\includegraphics{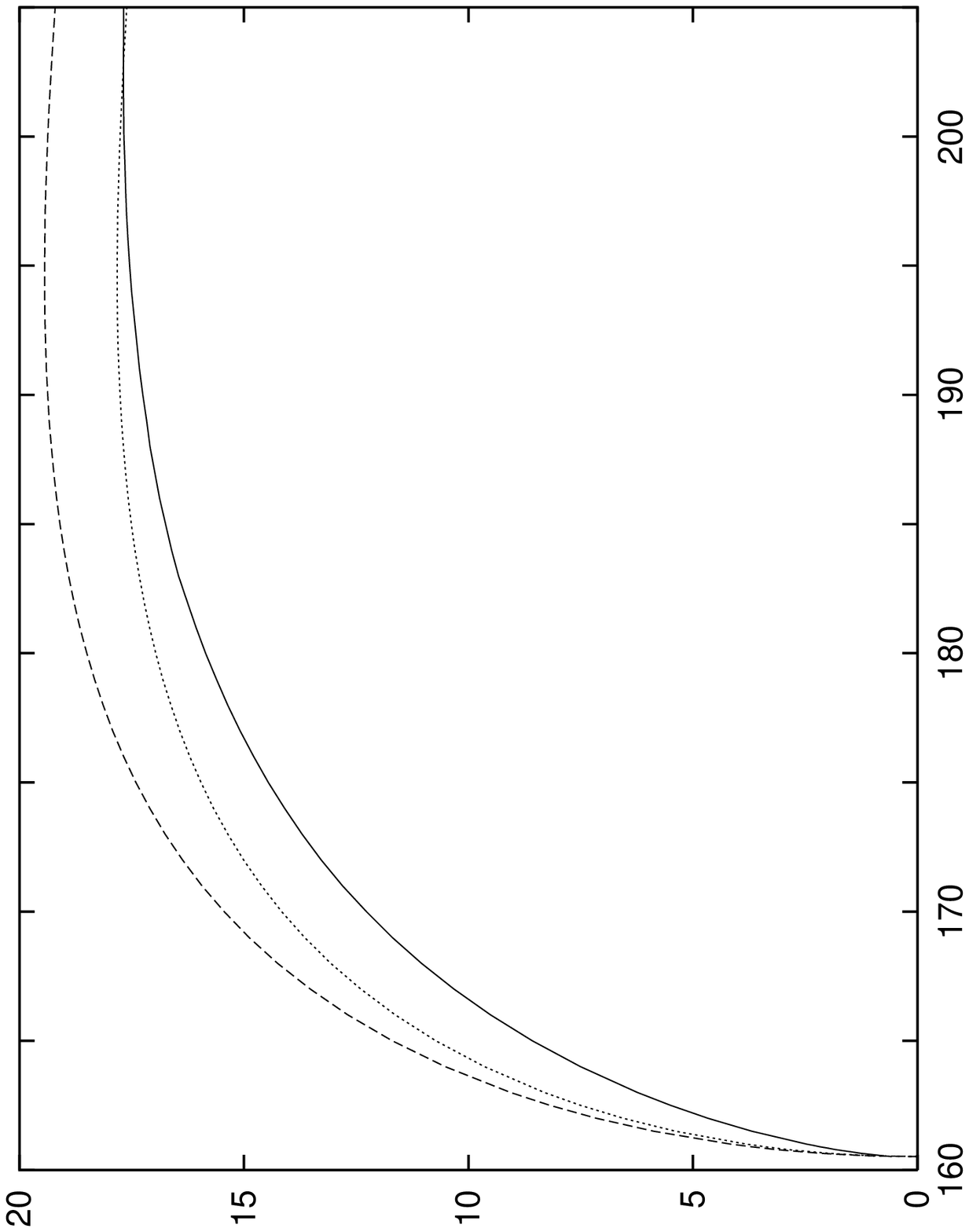}}
  \put (4,1.3){\includegraphics{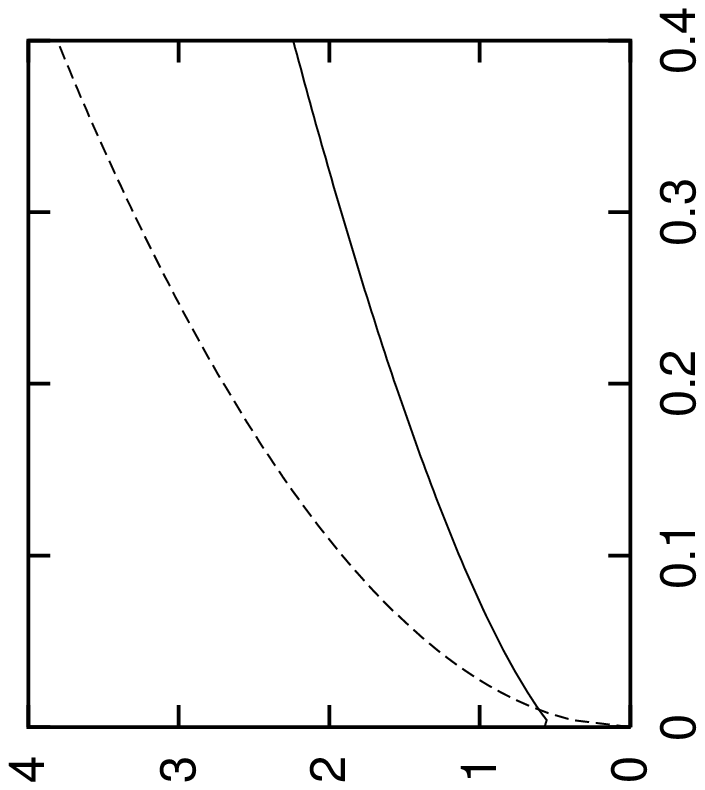}}
  \put (11.7,1.6){\mbox{\small $\sqrt{s}-2\MW$}}
  \put (10.9,-0.5){\mbox{\small $\sqrt{s}\ \ [\GeV]$}}
  \put (0.4,8.5){\mbox{\small $\sigma$}}
  \put (0.17,7.8){\mbox{\small $[{\rm pb}]$}}
  \end{picture}
 \end{center}
 \caption{Unpolarized total \cs\ in the threshold region. The dotted curve 
          corresponds to Born, the dashed one to \GF-Born, and the solid one 
          to the results including full radiative electroweak \Oa\ 
          corrections.} 
\label{totthr}
\efi
As has been stated before, it should be noted that, unlike at LEP1, the 
presence of the charged current at lowest order in \eeWW\ complicates the 
separation between QED and weak corrections. So, in order to have a cleaner 
look at the QED corrections we
present in addition to the Born \cs\ also the lowest-order \cs\ in the
so-called \GF\ representation, i.e.~the Born \cs\ with $\al$ replaced by
$\sqrt{2}\,\GF\MW^2\sw^2/\pi$, which already contains an important part of
the leading weak effects discussed above\footnote{As the $t$ channel is 
dominant at LEP2 energies, the \GF-Born describes the leading weak corrections 
reasonably well for the default Higgs mass \MH=300\GeV.} 
(see also \refse{SEonrcho}). We will refer to this \cs\ as \GF-Born.
As a result of the steep drop with decreasing centre-of-mass
energy of the \PW-pair \cs\  close to threshold,
large and predominantly soft QED effects can be observed.
Compared with \GF-Born the \Oa\ corrections amount to more than
$-25\%$ in the direct vicinity of the threshold ($\sqrt{s}<165\GeV$),
apart from the region very close to threshold where the positive
Coulomb-singularity contribution takes over (see close-up in
\reffi{totthr}), and still $-17\%$ ($-11\%$) at $\sqrt{s}=175\GeV$ (190\GeV).
The size of these effects necessitates the inclusion of higher-order
QED corrections in order to end up with an acceptable theoretical
uncertainty. A discussion of these higher-order QED corrections can be
found in \refse{SEonrcho}. The finite \PW\ width has a drastic impact on 
the effects related to the Coulomb singularity (see \refse{SEofcoul}), but the
large soft QED effects are merely smoothened and stay sizeable.
 
In \reffi{dcu190} the effect of the \Oa\ corrections on the unpolarized
differential \cs\ $\dsidctp$ is shown for
$\sqrt{s}=190\GeV$. Here $\theta_+$ stands for the polar angle of the $\PWp$\
boson with respect to the incoming positron\footnote{In the presence of 
hard-photon radiation in general $\theta_+ \neq \theta_-$.}.
\bfi
 \begin{center}
  \setlength{\unitlength}{1cm}
  \hspace*{-0.2cm}
  \begin{picture}(15.5,8.7)
  \put (-1.3,-2.5){\includegraphics{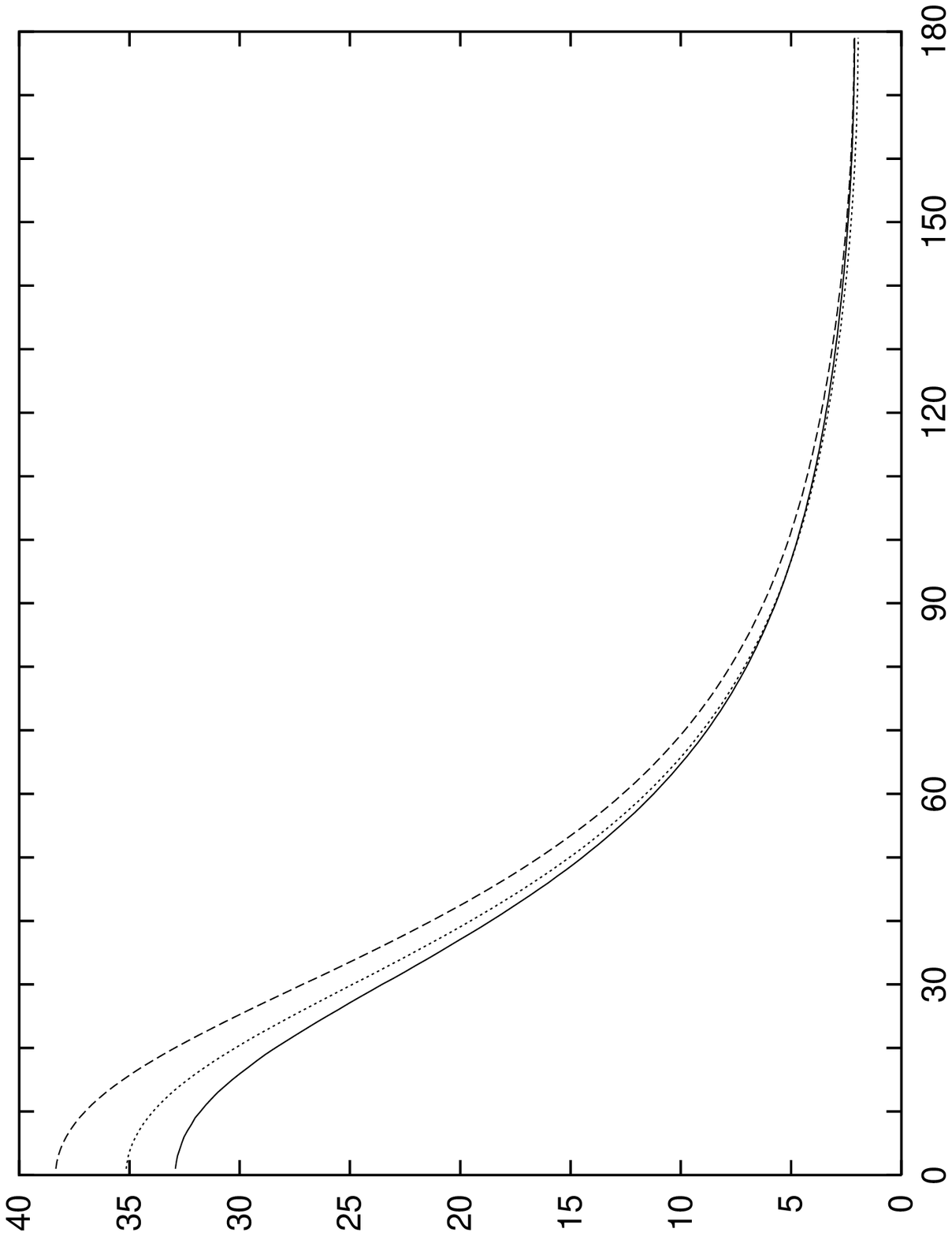}}
  \put (13.2,-0.3){\mbox{\small $\theta\ [\vphantom{a}^o]$}}
  \put (-0.4,8.1){\mbox{\small $\disp\dsidc$}}
  \put (-0.2,7.1){\mbox{\small $[{\rm pb}]$}}
  \end{picture}
 \end{center}
 \caption{Unpolarized differential \cs\ $\dsidctp$ at 
          $\protect\sqrt{s}=190\GeV$. The same signature as adopted in
          \protect\reffi{totthr}.} 
\label{dcu190}
\efi
Apart from the expected normalization effects that are already observed
in the total \cs, also a distortion of the distribution occurs. As this
is exactly the type of signature one might expect from anomalous
gauge-boson couplings, this underlines again the importance of having a
profound knowledge of the SM radiative corrections. The origin
of the distortion can be traced back to hard initial-state photonic
corrections. Hard-photon emissions boost the centre-of-mass system of the
\PWp\PWm\ pair. As a result of that, events that are forward in the 
centre-of-mass system of the produced \PW\ bosons can show up as backward
events in the LAB system and vice versa. Since the cross-section in the
backward direction is substantially lower than in the forward direction,
the net effect of this redistribution (migration) of events will be a
distortion of the differential distribution with respect to the lowest-order
one. Of course these boost effects are much more
pronounced at high energies \cite{BD94}.

\subsubsection{Approximations in the LEP2 region}
\label{SEonap}

The complete analytic results for the electroweak \Oa\ corrections are
very lengthy and complicated, resulting in huge and rather slow
computer programs. Moreover, the formulae are completely untransparent.
In view of this, simple approximative expressions are
desirable. Apart from providing fast computer programs, which are
useful for many applications, simple transparent formulae should reveal
the physical content and the origin of the dominant radiative
corrections. Furthermore, if these approximative expressions represent the 
exact corrections adequately, one might consider implementing them in the
existing LEP2 Monte Carlos.
 
Owing to the lack of a calculation of the complete $\Oa$ corrections,
the present LEP2 Monte Carlos for off-shell \PW-pair production include
only the known leading universal corrections. In order to assess the
theoretical uncertainty inherent in this approach, the on-shell case
can be used as guideline. The size of the non-leading $\Oa$ corrections 
in this case should provide a reasonable estimate for the corresponding 
left-out non-leading corrections in the off-shell case.

We start out by investigating the structure of the matrix
element for $\Pep\Pem\to\PWp\PWm$.
Whereas it involves only three different tensor structures in lowest order,
at $\Oa$
twelve independent tensor structures occur, each of which is associated
with an independent invariant function, which can be considered as
an $s$- and $t$-dependent effective coupling.
The dominant radiative corrections, as \eg those that are related
to UV, IR, or mass singularities, in general have factorization
properties
and are at \Oa\ restricted to those invariant functions that
appear already at lowest order. Therefore the contributions of
the other invariant functions should be relatively small.
Indeed, a numerical analysis reveals that in a suitably chosen
representation for the basic set of independent matrix elements
only the three Born-like invariant functions plus one extra right-handed piece,
related to $\M^+_I$, are relevant for a sufficiently good 
approximation \cite{Di92,Fl92}.
This results in the following approximation for the matrix
element:
\beq \label{Mapp}
\M^{\he}_{\app}  =
\M^{\he}_{I} F_I^{\he} + \M^{\he}_{Q} F_Q^{\he},
\eeq
with $\M_I$ and $\M_Q$ defined in \refeq{MIMQ}. The term involving $F_I^+$ 
is needed for right-handed electrons
because of contributions $\propto \be^2\cos\theta$, originating from
the interference of $\M_\born$ with particular 1-loop matrix elements, which 
are not present at lowest order.
Neglecting the other invariant functions in the basis given in \cite{Di92}
introduces errors well below the per-cent level (see \refta{wwibata}).
This is of course only true for observables where $\M^\he_\app$ is not
suppressed or absent.
All this demonstrates that improved Born approximations are possible.
Note, however, that in contrast to the situation at LEP1 the invariant
functions $F_{I,Q}^{\he}$ are both energy- and angular-dependent.
 
In order to construct an improved Born approximation (IBA) one has to specify
simple expressions for the invariant functions $F_{I,Q}^\he$, which
reproduce the corresponding exact expressions with sufficient accuracy.
In the LEP2 energy region the following
expressions can be used as a reasonable ansatz \cite{Di92}
\beqar \label{FIapp}
\nn F_I^\he &=& \left[
2\sqrt{2}\,\GF\,\MW^2+ \frac{4\pi\al}{2\sw^2}\,\frac{\pi\al}{4\beta}\,
(1-\beta^2)^2\right]\delta_{\he-} \nlc[1ex]
F_Q^\he &=& \left[
4\pi\al(s)+ 4\pi\al\,\frac{\pi\al}{4\beta}\,
(1-\beta^2)^2 \right].
\eeqar
 
The terms containing $\GF$ and $\al(s)$ incorporate all the
leading universal corrections associated with the running of $\al$
and the corrections $\propto \al\Mt^2/\MW^2$ associated with the 
$\rho$ parameter (see also \refse{SEonrcho}).
As these are linked to the renormalization of the electric charge at zero
momentum transfer and of the weak mixing angle,
they only contribute to the structures already present at lowest order.
The corresponding leading \Oa~contributions can be recovered from
\refeq{FIapp} by substituting
\beq
\al(s) \to \al[1+\da(s)], \qquad
2\sqrt{2}\,\GF\,\MW^2 \to \frac{4\pi\al}{2\sw^2}\,
\biggl[1+\da(\MW^2) - \frac{\cw^2}{\sw^2}\,\dro\biggr],
\eeq
with
\beq
\da(s) = \frac{\al}{3\pi}
\sum_{\Pf\ne\Pt}\NCf\Qf^2\log\biggl(\frac{s}{\Mf^2}\biggr), \qquad
\dro = \frac{3\al\Mt^2}{16\pi\sw^2\MW^2}.
\eeq
 
The $1/\beta$ term describes the effect of the Coulomb singularity,
which for $\beta\ll1$ yields a simple correction factor
to the lowest-order \cs:
\beq
\de\si_{\mathrm{Coul}} = \frac{\pi\al}{2\be}\, \si_\born.
\eeq
The factor $(1-\be^2)^2$ is introduced by hand to switch off the Coulomb 
singularity fast enough. In this way one avoids corrections that are too large
away from threshold, where the Coulomb singularity should not play a role 
anymore. As has been studied in \cite{Di92}, heavy-mass contributions 
$\propto \log(\Mt)$ and $\log(\MH)$ that 
are not covered by \refeq{FIapp} have a negligible impact
on the approximation. Since the above IBA analysis has been performed for
the default Higgs-boson mass $\MH=300\GeV$, 
the large light-Higgs-boson corrections close to threshold, shown
in \refta{mhalstab}, are absent. By adding a simple approximation for
these corrections \cite{BvO96} to the IBA, however, the full \MH\ dependence 
of the exact \Oa\ corrections can be reproduced.

In addition to the contributions described so far, one has to include
the leading logarithmic QED corrections. These can be calculated using
the structure-function method as described in \refapp{app}.
They comprise all contributions $\propto (\al/\pi)\log(\Me^2/Q^2)$.
In the following numerical analysis the scale $Q^2=s$ has been used and, in
order to extract the effect of the non-leading virtual corrections,
hard-photon radiation is left out.
 
As we want to compare the approximation with an $\Oa$ calculation we
do not use the square of the matrix element defined in \refeqs{Mapp} and
\refeqf{FIapp} for the numerical analysis,
but only the square of the terms involving $\GF$ and $\al(s)$ and
the interference of these terms with the others. Moreover, in this
interference $\GF$ and $\al(s)$ are replaced by the corresponding
lowest-order expressions in terms of $\sw^2$ and $\al(0)$.
Nevertheless, the so-obtained approximation still contains higher-order
contributions through the squares of $\al(s)$ and $\GF$. To allow for a 
meaningful comparison these have been included in the numbers for the exact
one-loop results given in \refta{wwibata} in the
same way (see \cite{Di92} for more details).

\btab[p]
\begin{center}
\begin{tabular}{||c||c|c||c|c||c|c||} \hline\hline
 & \multicolumn{2}{|c||}{unpolarized} 
 & \multicolumn{2}{|c||}{right-handed} 
 & \multicolumn{2}{|c||}{left-handed}\\ \hline 
\multicolumn{7}{||c||}{$\sqrt{s}=161\GeV$}\\
\hline
  total &  1.45 &  0.00 & $-$1.56 & $-$0.01 &  1.45 &  0.00\\
    10$^\circ$  &  1.63 &  0.00 & \hpm 4.41 & \hpm 0.00 &  1.63 &  0.00\\
    90$^\circ$  &  1.44 &  0.00 & $-$1.57 & $-$0.01 &  1.44 &  0.00\\
   170$^\circ$  &  1.26 &  0.00 & $-$7.52 & \hpm 0.00 &  1.26 &  0.00\\
\hline
\multicolumn{7}{||c||}{$\sqrt{s}=165\GeV$}\\
\hline
  total &  1.27 &  0.02 & $-$2.09 & $-$0.01 &  1.28 &  0.02\\
    10$^\circ$  &  1.67 &  0.00 & \hpm 0.49 & \hpm 0.00 &  1.67 &  0.00\\
    90$^\circ$  &  1.17 &  0.02 & $-$2.09 & $-$0.02 &  1.18 &  0.02\\
   170$^\circ$  &  0.75 &  0.00 & $-$4.64 & \hpm 0.00 &  0.77 &  0.00\\
\hline
\multicolumn{7}{||c||}{$\sqrt{s}=175\GeV$}\\
\hline
  total &  1.26 &  0.03 & $-$2.58 & $-$0.02 &  1.28 &  0.03\\
    10$^\circ$  &  1.71 &  0.00 & \hpm 0.18 & \hpm 0.00 &  1.71 &  0.00\\
    90$^\circ$  &  1.03 &  0.05 & $-$2.59 & $-$0.03 &  1.06 &  0.05\\
   170$^\circ$  &  0.59 &  0.00 & $-$5.30 & \hpm 0.00 &  0.69 &  0.00\\
\hline
\multicolumn{7}{||c||}{$\sqrt{s}=184\GeV$}\\
\hline
  total &  1.02 &  0.04 & $-$2.80 & $-$0.03 &  1.06 &  0.04\\
    10$^\circ$  &  1.57 &  0.00 & \hpm 2.17 & $-$0.01 &  1.57 &  0.00\\
    90$^\circ$  &  0.67 &  0.08 & $-$2.82 & $-$0.05 &  0.72 &  0.08\\
   170$^\circ$  &  0.10 &  0.00 & $-$7.72 & \hpm 0.01 &  0.32 &  0.00\\
\hline
\multicolumn{7}{||c||}{$\sqrt{s}=190\GeV$}\\
\hline
  total &  1.24 &  0.03 & $-$2.91 & $-$0.04 &  1.28 &  0.03\\
    10$^\circ$  &  1.67 &  0.00 & \hpm 0.59 & $-$0.01 &  1.67 &  0.00\\
    90$^\circ$  &  0.95 &  0.06 & $-$2.92 & $-$0.06 &  1.01 &  0.06\\
   170$^\circ$  &  0.58 &  0.00 & $-$6.32 & \hpm 0.00 &  0.83 &  0.00\\
\hline
\multicolumn{7}{||c||}{$\sqrt{s}=205\GeV$}\\
\hline
  total &  1.60 &  0.00 & $-$3.11 & $-$0.09 &  1.65 &  0.00\\
    10$^\circ$  &  1.77 &  0.00 & $-$1.68 & $-$0.01 &  1.77 &  0.00\\
    90$^\circ$  &  1.55 &  0.00 & $-$3.14 & $-$0.12 &  1.64 &  0.00\\
   170$^\circ$  &  1.61 &  0.00 & $-$4.37 & \hpm 0.00 &  1.94 &  0.00\\ 
\hline\hline
\end{tabular}
\end{center}
\vspace*{-2pt}
\caption{Quality of the IBA (first column per polarization) and the
         form-factor approximation \protect\refeqf{Mapp} (second column per
         polarization), given in per cent relative to Born. Results 
         are given for the total \cs\ (total) and the differential \cs\ at
         10, 90, and $170^\circ$.}
\label{wwibata}
\etab

In \refta{wwibata} we show the difference between the 
exact and approximated virtual and soft 
\Oa\ radiative corrections to the total and differential \cs\ for 
right-handed, left-handed, and unpolarized electrons. The positrons are assumed
to be unpolarized. In the LEP2 energy region the relative difference between 
the exact result and the approximation can be as large as 1--2\% for 
left-handed or unpolarized electrons, and reach 3--8\% for 
right-handed electrons. As there is no obvious reason
why these remaining non-leading corrections should be smaller in the case of 
off-shell \PW\ bosons or \eeffff, their inclusion in the LEP2 data analysis 
seems to be unavoidable.

\subsubsection {Higher-order corrections}
\label{SEonrcho}
 
In \refse{SEonrcal} we have encountered large \Oa\ corrections.
A short description of the way to include the corresponding
higher-order corrections is hence in place.

In two distinct ways the higher-order corrections enter the calculation of the
distributions for \PW-pair production. First of all there is the calculation 
of \Mt\ from \MW\ and the input parameters, either to be used in the 
calculation of the \PW-pair RC's or to be confronted with the Tevatron data.
As \Mt\ enters the relation between $\al$, \GF, \MZ, and \MW, resulting 
from the muon decay width, at the 1-loop 
level, the highest precision possible is required for this relation. As this 
relation relies on calculations performed at $Q^2\approx 0$ for the muon decay
width and at the subtraction points $Q^2=\MW^2,\MZ^2$ for the renormalization 
procedure, we can use the state-of-the-art calculation developed for the LEP1
analysis \cite{YR95}. This yields 
\begin{eqnarray}
\GF &=& \frac{\al\pi}{\sqrt{2}\sw^2\MW^2}\,\rho_{c}.
\label{gmu}
\end{eqnarray}
Usually $\rho_{c}$ is written in the form
\begin{equation}
\rho_{c} = \frac{1}{1-\Delta r},
\label{rhoc}
\end{equation}
where $\Delta r$ contains all the one-loop corrections to the muon decay width 
with the inclusion and the proper arrangement of the higher-order terms.
Next we subdivide $\rho_{c}$ as introduced
in \refeq{rhoc} into a {\it leading} term, $\Delta r_{_L}$, and 
{\it remainder} terms, $\Delta r_{\mathrm{rem}}$, as follows:
\begin{equation}
\rho_{c} =
\frac{1}{1-\Delta r} = \frac{1}{1- \Delta r_{_L}- \Delta r_{\mathrm{rem}}}
= \frac{1} {\left(1 - \Delta \al \right)
\left( 1 +{\displaystyle \frac {\cw^2}{\sw^2}} \Delta \rho_{_X} \right)
- \Delta r_{\mathrm{rem}} },
\label{romus}
\end{equation}
with
\begin{eqnarray}
\Delta r_{\mathrm{rem}} &=&
\Delta r^\al
+ \Delta r^{\al \als}
+ \frac{\cw^2}{\sw^2} \Delta {\bar \rho}_{_X}
- \Delta \al.
\label{newdr}
\end{eqnarray}
This contains all the terms known at present: the complete one-loop
${\cal O}(\al )$ corrections $\Delta r^\al$ (two-, three-, four-point 
functions) and complete two-loop ${\cal O} (\al\als)$ insertions 
$\Delta r^{\al \als}$ to two-point functions,
from which the leading ${\cal O}(\al )$ and ${\cal O}(\al\als)$
terms are subtracted.
The Born version of \refeq{gmu}, i.e.~$\rho_c = 1$, relates \GF\ directly to 
$\al$, \MW, and \MZ, hence leaving no room for using \GF\ as additional
independent input parameter.
However, having introduced radiative corrections to the Born relation we can 
solve \refeq{gmu} in terms of the top quark mass and, in turn, this 
particular value for \Mt\ will be used throughout the rest of the calculation. 
One should notice that this procedure is not free of ambiguities since
$\Delta r$ is also a function of \MH, but with much smaller sensitivity
due to the well known screening. So in the end we should also allow for
some variation in \MH. Moreover, $\Delta r$ is also a function of $\als$ 
through higher-order corrections, for instance ${\O}(\al\als)$, and, as a
consequence, also some variation in the strong coupling constant should be 
allowed.

The leading terms $\Delta\al$ and $\Delta {\bar \rho}_{X}$ appearing in
\refeq{newdr} are given by
\begin{eqnarray}
\Delta {\bar \rho}_{X} &=&
\Delta {\bar \rho}^{\al} +
\Delta {\bar \rho}^{\al \als} + {\bar X}
 \nonumber \\
 &=&  \frac{3 \al}{16 \pi \sw^2 \cw^2} \;
\frac{\Mt^2}{\MZ^2} \left[ 1 - \frac{2}{3} \left(1 + \frac{\pi^2}{3}\right)
\frac {\als(\Mt^2)}{\pi}\right] + {\bar X},\nn\\
\Delta\al &=& 1-\frac{\al}{\al(\MW^2)},\nn\\ 
\al(s) &=& \frac{\al}{1+\Sigma^{\mathrm{f}\neq \Pt}_\ga(s)/s},
\label{drobar}
\end{eqnarray}
where $\Sigma^{\mathrm{f}\neq \Pt}_\ga(s)$ is the renormalized \Oa\
transverse photon self-energy originating from fermion loops,
excluding top-quark loops.

The term ${\bar X}$ in \refeq{drobar} is a next-to-leading order term, whose
proper treatment is rather important:
\begin{equation}
{\bar X} = \Re\left[ \frac{ \Pi_\PZ(\MZ^2)}{\MZ^2}
      - \frac{ \Pi_\PW(\MW^2)}{\MW^2}
      \right]^{\mathrm{1loop}}_{\overline {\mathrm{MS}}}
      - \Delta \rho^{\al},
\label{xbar}
\end{equation}
where $\Pi_{\mathrm{V}}$ denotes the unrenormalized transverse self-energy of 
the V gauge boson. The ultra-violet (UV) divergences are removed according to 
the $\overline {\mathrm{MS}}$ renormalization scheme with $\mu=\MZ$.

In contrast to $\Delta {\bar \rho}_{X}$, the leading contribution 
$\Delta \rho_X$ appearing in \refeq{romus} is normalized by \GF\
rather than by $\al/(\sw^2\MW^2)$, as is required by the resummation
proposed in \cite{resum}:
\begin{eqnarray}
\Delta \rho_{X} &=& \Delta \rho^{\al}
     + \Delta \rho^{\al^2} + \Delta \rho^{\al \als}
     + \Delta \rho^{\al \al^2_s} + X    \nonumber \\
 &=& N_{\mathrm{C}}^\Pt\, x_t \left[ 1 
     + x_t\, \Delta \rho^{(2)} \left(\frac{\Mt^2}{\MH^2}\right)
     + c_1 \,\frac{\als(\Mt^2)}{\pi}
     + c_2 \left(\frac{\als(\Mt^2)}{\pi}\right)^2 \right] + X,
\label{rholead}
\end{eqnarray}
where $N_{\mathrm{C}}^\Pt=3$ and
\begin{eqnarray}
x_t &=& \frac{\GF}{\sqrt{2}} \,\frac{\Mt^2}{8 \pi^2},
\label{calt} \\
X &=& 2 \sw^2 \cw^2 \,\frac{\GF \MZ^2}
{\sqrt{2} \pi \al} \,{\bar X}.
\label{defx}
\end{eqnarray}
The coefficients $c_1$ and $c_2$ describe
the first- and second-order QCD corrections for the leading
$x_t$ contribution to $\Delta \rho$, calculated
in \cite{aasse,afmt}. Correspondingly:
\begin{eqnarray}
c_1 &=& - \frac{2}{3} \left(1+ \frac{\pi^2}{3}\right),
\label{c1qcd} \\
c_2 &=& - \pi^2 \left( 2.564571 - 0.180981\;n_f\right),
\label{c2qcd}
\end{eqnarray}
with $n_f$ the total number of flavours ($n_f=6$).
The function $\Delta \rho^{(2)}(\Mt^2/\MH^2)$ describes the leading
two-loop electroweak $x_t$ correction to $\Delta \rho$,
calculated first in the $\MH=0$ approximation in \cite{vanhoog}
and later in \cite{barb} for an arbitrary relation between \MH\
and \Mt.

It should be noted that the higher-order radiative corrections discussed above
are usually
calculated in the limit of a heavy top mass, i.e.~usually only the leading
part of the corrections is under control. This often raises
the question of what effect can be estimated from the sub-leading terms, since
$\Mt \approx 2\,\MZ$. As far as QCD ${\O}(\als)$ and ${\O}(\als^2)$
corrections are concerned the sub-leading terms are well under control. 
In the pure electroweak sector, however, there has been, so far, no calculation
of higher-order sub-leading terms at an arbitrary scale. The only available
piece of calculation concerns the $\rho$ parameter at $Q^2=0$ \cite{DeYR95},
therefore relevant only for $\nu_{\mu}$--$e$ scattering. If however one is
willing to extrapolate the $Q^2=0$ result to a higher scale, by
assuming that the ratio of leading to next-to-leading corrections is 
representative for the corresponding theoretical uncertainty, then the answer 
is next-to-leading $\approx$ leading.

Even before considering the actual impact of the higher-order terms on 
\Mt, we should mention at this point that the way in which the non-leading 
terms can be treated and the exact form of
the {\it leading--remainder} splitting give rise to several possible options
in the actual implementation of radiative corrections. This in turn becomes
a source of theoretical uncertainty. For instance, for $\Delta r$ we can 
introduce the decomposition into leading and remainder. Since we
know how to proceed with all objects in the leading approximation, the only
ambiguity is due to the treatment of the remainders. Clearly, after
the splitting $\Delta r = \drl + \Delta r_{\mathrm{rem}}$ there are in 
principle several possible ways of handling the remainder:
\begin{equation}
\frac{1}{1 - \Delta r} = \frac{1}{1- \drl- \Delta r_{\mathrm{rem}}}
\ ,\ 
\frac{1}{1 - \drl}\,\left( 1 + \frac{\Delta r_{\mathrm{rem}}}{1 - \drl}\right)
\ ,\  
\frac{1 + \Delta r_{\mathrm{rem}}}{1 - \drl} 
\ ,\ 
\frac{1}{1 - \drl} + \Delta r_{\mathrm{rem}}.
\label{ifacr}
\end{equation}
Actually, these options differ among themselves, but the difference can be
related to the choice of the scale in the remainder term.
A complete evaluation of the sub-leading ${\O}(\GF^2 \MZ^2 \Mt^2)$
corrections would greatly reduce the associated uncertainty.
In conclusion, we observe that a natural and familiar language 
for the basic ingredients of the physical observables is that of effective 
couplings.
However, it should be stressed that the above formulae belong to a specific
realization of this language and other realizations could also be used.

Now we can assess the influence of the higher-order corrections on the 
calculation of \Mt\ from \refeq{gmu}. Using for instance $\MW = 80.26\GeV$ and
$\MZ= 91.1884\GeV$ we find
\begin{equation}
\Mt = 165^{+16}_{-18}\,(\MH,\als)\GeV,
\end{equation}
the central value of which corresponds to $\MH = 300\GeV$ and 
$\als(\MZ^2) = 0.123$. The errors 
are derived by varying \MH\ and $\als$ in the range $60\GeV < \MH
< 1\TeV$ and $\als = 0.123 \pm 0.006$. It should be noted in this respect 
that the total variation induced by \MH\ alone (at $\als = 0.123$) is about 
$33\GeV$. The following is observed for the higher-order corrections: 

\begin{itemize}

\item by neglecting the ${\O}(\al^2)$ term in $\Delta r$
we find for the same input parameters (and $\MH = 300\GeV$, $\als = 0.123$)
a shift in \Mt\ of $-1.9\GeV$. 

\item If instead we switch off the ${\O}(\al\als^2)$ correction the 
corresponding shift will be $-1.5\GeV$. 

\item If finally both the ${\O}(\al\als)$ and ${\O}(\al\als^2)$ 
corrections are neglected we find a shift in \Mt\ of $-10.5\GeV$. Here
by ${\O(\al\als)}$ the full result is implied and not
only the leading part of it.

\end{itemize}

Based on the above
observations, the remaining theoretical uncertainty in the calculation of \Mt\
from missing higher-order corrections
and sub-leading $\Oaa$ corrections to $\Delta r$ is estimated to be 
roughly 1--2\GeV.

The second way the higher-order corrections enter the calculations for \PW-pair
production is through the process itself, so through self-energies, vertices, 
etc. The known weak higher-order effects comprise the running of $\al$ [see 
\refeq{drobar}] and the complete $\O(\al\als)$ corrections to the
gauge-boson self-energies \cite{aasse}. Other higher-order calculations, as 
those for $\Delta\rho$ at ${\O}(\al^2)$ or ${\O}(\al\als^2)$, have been
performed in the limit where \Mt\ is the largest scale. The QCD corrections
associated with the gauge-boson vertex corrections have, to our
knowledge, not been calculated yet, but they are at most logarithmic in the 
top mass. If one takes into account the leading weak effects by replacing 
$e^2/(2\sw^2)$ by $2\sqrt{2}\,\GF\MW^2$ in the $\M_I$ part of \refeq{MIMQ} and
$\al$ by $\al(s)$ in the $\M_Q$ part, the remaining unknown higher-order weak
effects are expected to be negligible compared with the required theoretical
accuracy.
 
As pointed out in the previous subsection, the virtual and real
corrections reveal the presence of large logarithmic QED effects of the
form $\al \LL/\pi \equiv (\al/\pi)\,\log(Q^2/\Me^2)$
with $Q^2 \gg \Me^2$. They arise when photons
or light fermions are radiated off in the direction of incoming or
outgoing light particles. In the case of \PW-pair production the only
terms of this sort originate from initial-state photon emission.
Radiation of photons from the final-state \PW~bosons can only lead to
sizeable corrections if the energy of the \PW~bosons is much larger
than their mass.
The leading large logarithmic corrections can be calculated by
using the so-called structure-function method \cite{SFM} in leading-log
(LL) approximation, \ie only taking along the terms $\propto (\al
\LL/\pi)^n$. This procedure also allows the inclusion of
soft-photon effects to all orders by means of exponentiation and is discussed
in detail in \refapp{app}. 

\btab
\begin{center}
\begin{tabular}{||c|c|c|c|c|c||} \hline\hline
 $ \sqrt{s} $ & Born & \multicolumn{2}{|c|}{ $+$ h.o.t. for $Q^2=s$ } 
 & \multicolumn{2}{|c||}{ $+$ h.o.t. for $Q^2=s\,\frac{1-\beta}{1+\beta}$ }\\
 \cline{3-6}
 $ [\mathrm{GeV}] $ & $+$ \Oa & \Oaa\ LL & exp. LL & \Oaa\ LL & exp. LL \\[1ex]
 \hline
 \hpt 161.0 & \hpt 2.472 $\pm$
 0.001 & \hpt 3.103  & \hpt 3.003 & \hpt 3.095  & \hpt 2.998 \\ \hline
 \hpt 165.0 & \hpt 8.581 $\pm$  0.003
 & \hpt 9.079 & \hpt 9.049 & \hpt 9.061 & \hpt 9.033 \\ \hline
 \hpt 170.0 & 12.270 $\pm$  0.004 & 12.583 & 12.585 & 12.567 & 12.568 \\ \hline
 \hpt 175.0 & 14.465 $\pm$  0.004 & 14.654 & 14.670 & 14.642 & 14.656 \\ \hline
 \hpt 184.0 & 16.613 $\pm$  0.005 & 16.668 & 16.693 & 16.663 & 16.686 \\ \hline
 \hpt 190.0 & 17.257 $\pm$  0.006 & 17.259 & 17.286 & 17.259 & 17.283 \\ \hline
 \hpt 205.0 & 17.677 $\pm$  0.006 & 17.613 & 17.638 & 17.620 & 17.641 \\ \hline
\hline
\end{tabular}
\end{center}
\caption{Unpolarized total \cs, given in pb, including radiative electroweak 
         \Oa\ corrections and higher-order terms in the leading-log 
         approximation. These higher-order terms are given with and without
         soft-photon exponentiation for two different `natural' scale choices.}
\label{lltab}
\etab

In \refta{lltab} the higher-order effects related to the large logarithmic 
QED corrections are displayed. The \Oaa\ LL entry contains,
in addition to the full \Oa\ result, the contribution from \Oaa\ LL
corrections using $\hat{\sigma}_0=\sigma_{\mathrm{Born}}$ in the
convolution \refeqf{LLint}. The exp.~LL entry contains on top of that the
exponentiation of soft-photon effects. Compared with the full \Oa\ results 
we observe large \Oaa\ LL effects near threshold, \eg 25\% at 
$\sqrt{s}=161\GeV$, and moderate ones when sufficiently
far above threshold, \ie $<1\%$ for energies above roughly 175\GeV. 
The additional soft-photon exponentiation is only sizeable near threshold,
\eg $-4\%$ at $\sqrt{s}=161\GeV$.
Note that finite-\PW-width effects
will smoothen the threshold behaviour and hence reduce the size of
the \Oaa\ LL effects, nevertheless they will stay sizeable near
threshold. The dependence of the higher-order LL corrections [beyond \Oa]
on the scale choice $Q^2$ is negligible, since all natural scales are 
roughly equal close to threshold. When comparing the popular scale choice 
$Q^2=s$ with 
$Q^2 = s(1-\beta)/(1+\beta)$, motivated by the behaviour of the total \cs\ 
near threshold and at high energies \cite{BD94}, the differences are at the 
0.1\% level (0.2--0.3\% at $\sqrt{s}=161\GeV$). 

An additional improvement of the theoretical predictions can be obtained by 
using an improved Born \cs\ in the convolution \refeqf{LLint}, taking into 
account corrections related to $\GF$, $\al(s)$, \MH\ etc.%
\footnote{As the finite decay width of the \PW\ bosons will have a substantial
impact on the Coulomb singularity, a LL analysis involving this Coulomb
singularity only makes sense for off-shell \PW\ bosons.}
 
\subsection{The width of the W boson}
\label{SEgw}
Evidently the width of the \PW~boson is a crucial ingredient for 
the (off-shell) \PW-pair production \cs, especially\ in the threshold region.
Moreover, the branching ratios enter the \css\ for definite
fermions in the final state. As at present the width of the \PW\ boson is 
experimentally poorly known, we need a precise theoretical calculation instead
in order to obtain adequate theoretical predictions for off-shell \PW-pair
production.
 
\subsubsection{The W-boson width in lowest order}
\label{SEgwlo}
The \PW~width is dominated by decays into fermion--antifermion pairs.
In lowest order the partial width for the decay of a \PW~boson into
two fermions with masses $\Mfi$ and $\Mfj$
($i,j$ denote the generation index and $\Pf,\Pf'$ stand for $\Pu,\Pd$
or $\nu,\Pl$) is given by
\beqar \label{G0Wff}
\Gamma_{\PW\Pf_{i}\Pf'_{j}}^{\raisebox{1mm}{$\scriptstyle \mathrm{Born}$}} &=&
\NCf\,\frac{\al}{6}\,
\frac{\MW}{2\sw^{2}}\, |V_{ij}|^{2}
\left[1-\frac{\Mfi^{2}+\Mfj^{2}}{2\MW^2}-\frac{(\Mfi^{2}-
\Mfj^{2})^{2}}{2\MW^{4}}\right]  \nl[1ex]
&&\qquad\times \frac{\sqrt{\Bigl(\MW^2-(\Mfi+\Mfj)^2\Bigr)
\Bigl(\MW^2-(\Mfi-\Mfj)^2\Bigr)}}{\MW^2} .
\eeqar
For leptonic decays the mixing matrix is diagonal ($V_{ij}=\delta_{ij}$)
and the colour factor $\NCf$ equals one.
For decays into quarks there
is a non-trivial quark mixing matrix and ${\NCf=3}$.
For the quark mixing matrix we have used $s_{12}=0.221$, $s_{23}=0.04$, and
$s_{13}=0.004$ \cite{PDG94}. 
The total width is obtained
as a sum over the partial fermionic decay widths with
$\Mfi+\Mfj< \MW$
\beq
\Gamma_\PW^{\raisebox{1mm}{$\scriptstyle \mathrm{Born}$}}=
     \sum_{i,j}\Gamma^{\raisebox{1mm}{$\scriptstyle 
     \mathrm{Born}$}}_{\PW\Pu_{i}\Pd_{j}}
     +\sum _{i}\Gamma^{\raisebox{1mm}{$\scriptstyle 
     \mathrm{Born}$}}_{\PW\nu_{i}\Pl_{i}}.
\eeq
As the quark masses are small compared with \MW, the fermion-mass effects
are small for the \PW~decay.
If we neglect them we obtain the simple formulae
\beq \label{G0Wff0}
 \Gamma_{\PW\Pf_{i}\Pf'_{j}}^{\raisebox{1mm}{$\scriptstyle \mathrm{Born}$}} =
   \NCf\,\frac{\al}{6}\,\frac{\MW}{2\sw^{2}}\, |V_{ij}|^{2}  , \qquad
 \Gamma_{\PW}^{\raisebox{1mm}{$\scriptstyle \mathrm{Born}$}}
   \approx\frac{3\al}{2}\,\frac{\MW}{2\sw^{2}}.
\eeq
 
\subsubsection{Higher-order corrections to the W-boson width}
\label{SEgwrc}
The electroweak and QCD radiative
corrections for decays into massless fermions
($\Mf\ll \MW$) have been calculated in \cite{Al80}--\cite{Ba86}.
The full one-loop electroweak and QCD corrections, together with the
complete photonic and gluonic bremsstrahlung, were evaluated for
arbitrary fermion masses in \cite{De90a}.
The various calculations are in good agreement.
\begin{table}
$$
\tabcolsep 6pt
\begin{array}{||c|c|c|c||c|c|c||} \hline\hline
 \,\MH\ [\mathrm{GeV}]\,        
 &     300 &     300 &     300 &      60 &     300 &    1000 \\ 
 \hline
 \,\als\,    
 &   0.117 &   0.123 &   0.129 &   0.123 &   0.123 &   0.123 \\ 
 \hline
 \,\Mt\ [\mathrm{GeV}]\,        
 &  164.80 &  165.26 &  165.73 &  148.14 &  165.26 &  181.20 \\ 
 \hline\hline
 \,\Gamma_\PW^{\raisebox{1mm}{$\scriptstyle \mathrm{Born}$}}\ [\mathrm{GeV}]\, 
 &  1.9490 &  1.9490 &  1.9490 &  1.9490 &  1.9490 &  1.9490 \\ 
 \hline
 \,\overline{\Gamma}_\PW^{\born}\ [\mathrm{GeV}]\,  
 &  2.0354 &  2.0354 &  2.0354 &  2.0354 &  2.0354 &  2.0354 \\ 
 \hline
 \,\Gamma_\PW\ [\mathrm{GeV}]\, 
 &  2.0642 &  2.0663 &  2.0684 &  2.0681 &  2.0663 &  2.0639 \\ 
 \hline
 \,\overline{\Gamma}_\PW\ [\mathrm{GeV}]\,  
 &  2.0791 &  2.0817 &  2.0844 &  2.0813 &  2.0817 &  2.0819 \\
 \hline\hline
 \,\delta_{\mathrm{ew}}\,  
 & \hpm 0.03416 & \hpm 0.03398 & \hpm 0.03380 & \hpm 0.03491 & \hpm 0.03398 
 & \hpm 0.03275 \\ 
 \hline
 \,\overline{\delta}_{\mathrm{ew}}\, 
 & $-$0.00347 & $-$0.00347 & $-$0.00347 & $-$0.00369 & $-$0.00347 
 & $-$0.00341 \\ 
 \hline
 \,\delta_{\mathrm{QCD}}\,  
 & \hpm 0.02495 & \hpm 0.02623 & \hpm 0.02751 & \hpm 0.02623 & \hpm 0.02623 
 & \hpm 0.02623 \\ 
 \hline\hline
\end{array}
$$
\caption{Higgs-mass and $\als$ dependence of the \PW-boson width}
\label{FIgw}
\end{table}
The relative corrections to the total \PW-boson decay width
are given in \refta{FIgw}. The electroweak corrections in the on-shell 
scheme $\de^\ew\equiv\GW/\Ga^{\raisebox{1mm}{$\scriptstyle \mathrm{Born}$}}_\PW
-1-\de^\QCD$
are predominantly originating from the running of $\al$ and the 
corrections $\propto \al\,\Mt^{2}/\MW^{2}$ to the $\rho$ parameter.
These corrections can be easily accounted for
by parametrizing the lowest-order width in terms of $\GF$
and $\MW$ instead of $\al$ and $\sw^2$.
The width in this \GF\ parametrization $\overline{\Gamma}_\PW$ is
related to the width in the on-shell parametrization $\GW$ by
\beqar
\overline{\Ga}_\PW &=& 
  \frac{\GW-\Ga^{\raisebox{1mm}{$\scriptstyle \mathrm{Born}$}}_\PW 
  \De r^{\mbox{\scriptsize1-loop}}}{1-\De r}
  = \Ga^{\raisebox{1mm}{$\scriptstyle \mathrm{Born}$}}_\PW
  \frac{1+\de^\ew+\de^\QCD-\De r^{\mbox{\scriptsize1-loop}}}{1-\De r}
  \nonumber \\[1ex]
                   &=& \overline{\Ga}_\PW^\born (1+\de^\ew+\de^\QCD
  -\De r^{\mbox{\scriptsize1-loop}})
  \equiv \overline{\Ga}_\PW^\born (1+\overline{\de}^\ew+\de^\QCD).
\eeqar
As can be seen from \refta{FIgw}
the electroweak corrections with respect to the parametrization
with $\GF$, $\bar{\de}^\ew$,
depend in a negligible way on \MH, and remain below 0.5\% for the total width.
The QCD corrections $\de^\QCD$ are practically constant and equal to
$2\als(\MW^2)/(3\pi)$, their value for zero fermion masses.
For the numerical evaluation we use $\als(\MW^2)=0.123$ (\ie equal to the 
default input value). The difference
between \GW\ and the more precise $\overline{\Ga}_\PW$ is caused by 
missing higher-order terms related to $\De r$.
 
We obtain the following improved Born approximation
for the total and partial widths
\cite{Dehab}
\beqar \label{GWIBA}
{\Gamma}_{\PW\nu_i\Pl_i} &\approx&
\overline{\Gamma}_{\PW\nu_i\Pl_i}^\born =
\frac{\GF \MW^3}{6\sqrt{2}\pi}  \nlc[1ex]
\Gamma_{\PW\Pu_{i}\Pd_{j}} &\approx&
\overline{\Gamma}_{\PW\Pu_{i}\Pd_{j}}^\born
\left(1+\frac{\als(\MW^2)}{\pi}\right) =
\frac{\GF \MW^3}{2\sqrt{2}\pi}\, |{V_{ij}}|^{2}
\left(1+\frac{\als(\MW^2)}{\pi}\right) \nlc[1ex]
{\Gamma}_\PW &\approx&
\overline{\Gamma}_\PW^\born \left(1+\frac{2\als(\MW^2)}{3\pi}\right)
= \frac{3\GF \MW^3}{2\sqrt{2}\pi}
\left(1+\frac{2\als(\MW^2)}{3\pi}\right) .
\eeqar
 
In \refta{TAgw} we compare the improved Born approximation (\IBA)
\begin{table}
\begin{center}
\tabcolsep 6pt
\begin{tabular}{||l|c|c|c|c||l||} \hline\hline
& \multicolumn{1}{|c|}{Born}      & \multicolumn{1}{c|}{complete}
& \multicolumn{1}{c|}{complete}   & \multicolumn{1}{c||}{$\IBA$}
& \multicolumn{1}{c||}{Branching} \\
& \multicolumn{1}{c|}{$\Mf\ne0$}& \multicolumn{1}{c|}{$\Mf\ne0$}
& \multicolumn{1}{c|}{$\Mf=0$}  & \multicolumn{1}{c||}{($\Mf=0$)}
& \multicolumn{1}{c||}{ratio} \\
\hline\hline
 $\Gamma(\PW\to \Pe \Pne      )$
 &   0.2262 &   0.2255 &   0.2255 &   0.2262 &   0.1083 \\
 \hline
 $\Gamma(\PW\to \Pmu\Pnm      )$
 &   0.2262 &   0.2255 &   0.2255 &   0.2262 &   0.1083 \\
 \hline
 $\Gamma(\PW\to \Pta\Pnt      )$
 &   0.2261 &   0.2253 &   0.2255 &   0.2262 &   0.1082 \\
 \hline
 $\Gamma(\PW\to\mathrm{lep.})$
 &   0.6785 &   0.6763 &   0.6765 &   0.6787 &   0.3249 \\
 \hline
 $\Gamma(\PW\to \Pu \Pd       )$   
 &   0.6455 &   0.6684 &   0.6684 &   0.6708 &   0.3211 \\
 \hline
 $\Gamma(\PW\to \Pu \Ps       ) \times 10 $   
 &   0.3315 &   0.3432 &   0.3432 &   0.3444 &   0.0165 \\
 \hline
 $\Gamma(\PW\to \Pu \Pb       ) \times 10^4 $   
 &   0.1080 &   0.1122 &   0.1124 &   0.1128 &   0.000005 \\
 \hline
 $\Gamma(\PW\to \Pc \Pd       ) \times 10 $   
 &   0.3312 &   0.3431 &   0.3432 &   0.3444 &   0.0165 \\
 \hline
 $\Gamma(\PW\to \Pc \Ps       )$   
 &   0.6441 &   0.6672 &   0.6673 &   0.6697 &   0.3205 \\
 \hline
 $\Gamma(\PW\to \Pc \Pb       ) \times 10^2 $   
 &   0.1080 &   0.1121 &   0.1124 &   0.1128 &   0.0005 \\
 \hline
 $\Gamma(\PW\to\mathrm{had.}) $
 &  1.3569 &  1.4054 &  1.4055 &  1.4104 &  0.6751 \\
 \hline
 $\Gamma(\PW\to\mathrm{all})$    
 &  2.0354 &  2.0817 &  2.0820 &  2.0891 & \\
 \hline\hline
\end{tabular}
\end{center}
\caption{Partial and total \protect\PW-decay widths 
         $\overline{\Gamma}_{\PW}$ in different
         approximations given in GeV.} 
\label{TAgw}
\end{table}
for the partial and total widths, given by \refeq{GWIBA}, with the 
lowest-order widths,
the widths including the complete first-order and leading higher-order
corrections for finite fermion masses, and the same for vanishing
fermion masses, all in the \GF\ parametrization.
The effects of the fermion masses, which are of the order
$\Mf^{2}/\MW^2$, are below 0.3\%.
Consequently the exact numerical values for the masses of
the external fermions are irrelevant.
The \IBA\ reproduces the exact
results within 0.4\% (0.6\% for the decays into a \Pb-quark).
The branching ratios for the individual decay channels derived from 
\refeq{GWIBA}, which depend only on $\al_{s}$ and $V_{ij}$,
agree numerically within 0.1\% with those obtained from the full
one-loop results.

\section{Off-Shell W-Pair Production}
\label{CHof}
\subsection{Lowest order: an introduction}
\label{SEofintro}

So far we have only considered the production of stable \PW\ bosons. 
This is, however, only an approximation and in particular in the threshold 
region it is not sufficient.
Rather, one has to describe the \PW\ bosons as resonances, with a finite
width so as to avoid singularities inside the physical phase space,
and analyse their presence through their decay products:
\beq
\Pep+\Pem \to \PWp+\PWm \to \Pf_1 + \bar\Pf_2 + \Pf_3 + \bar \Pf_4.
\label{WWofshell}
\eeq
Process \refeqf{WWofshell} involves two resonant \PW\ bosons (doubly-resonant)
and can be viewed as a very natural first step beyond the on-shell limit. 
In lowest order this process is represented by the three Feynman diagrams 
shown in \reffi{FIee4fdia}. However, the full four-fermion process 
does not only proceed through the three doubly-resonant diagrams.
\begin{figure}[t]
\unitlength 1pt
\savebox{\hdwig}(20,0){
  \setlength{\unitlength}{1.075pt}
  \begin{picture}(20,0)
    \bezier {20}(0,0)(2.5,2.5)(5,0)
    \bezier {20}(10,0)(12.5,2.5)(15,0)
    \bezier {20}(5,0)(7.5,-2.5)(10,0)
    \bezier {20}(15,0)(17.5,-2.5)(20,0)
   \end{picture}
  \setlength{\unitlength}{1pt} }
\savebox{\dpdwig}(14.4,0){
  \setlength{\unitlength}{0.9625pt}
  \begin{picture}(16,0)
  \bezier{20}(0,0)(0,4)(4,4)
  \bezier{20}(4,4)(8,4)(8,8)
  \bezier{20}(8,8)(8,12)(12,12)
  \bezier{20}(12,12)(16,12)(16,16)
  \end{picture}
  \setlength{\unitlength}{1pt} }
\savebox{\dmdwig}(14.4,0){
  \setlength{\unitlength}{0.9625pt}
  \begin{picture}(16,0)
  \bezier{20}(0,0)(0,-4)(4,-4)
  \bezier{20}(4,-4)(8,-4)(8,-8)
  \bezier{20}(8,-8)(8,-12)(12,-12)
  \bezier{20}(12,-12)(16,-12)(16,-16)
  \end{picture}
  \setlength{\unitlength}{1pt} }
\begin{center}
\begin{picture}(386,97)(0,3)
 \put (0,0){ \begin{picture}(164,103)(0,-20)
  \put (20,0){\vector(1,0){23}}
  \put (43,0){\line(1,0){17}}
  \put (60,63){\vector(-1,0){23}}
  \put (37,63){\line(-1,0){17}}
  \put (61.5,1.5){\vector(0,1){33}}
  \put (61.5,34.5){\line(0,1){27}}
  \put (0,-3){\mbox{$\Pem$}}
  \put (0,60){\mbox{$\Pep$}}
  \put (68.5,28.5){\mbox{$\Pne$}}
  \put (61.5,0){\circle*{4}}
  \put (61.5,63){\circle*{4}}
  \put (61.5,0){\usebox{\hdwig}}
  \put (83,0){\usebox{\hdwig}}
  \put (61.5,63){\usebox{\hdwig}}
  \put (83,63){\usebox{\hdwig}}
  \put (78,-14){\mbox{$\PWm$}}
  \put (78,69){\mbox{$\PWp$}}
  \put (104.5,0){\circle*{4}}
  \put (104.5,63){\circle*{4}}
  \put (104.5,63){\vector(2,1){23}}
  \put (127.5,74.5){\line(2,1){17}}
  \put (144.5,43){\vector(-2,1){23}}
  \put (121.5,54.5){\line(-2,1){17}}
  \put (104.5,0){\vector(2,1){23}}
  \put (127.5,11.5){\line(2,1){17}}
  \put (144.5,-20){\vector(-2,1){23}}
  \put (121.5,-8.5){\line(-2,1){17}}
  \put (150,80){\mbox{$f_1$}}
  \put (150,40){\mbox{$\bar f_2$}}
  \put (150,17){\mbox{$f_3$}}
  \put (150,-23){\mbox{$\bar f_4$}}
  \end{picture} }
 \put (201,0.5){ \begin{picture}(173,102)(0,-20)
  \put (20,0){\vector(1,1){17}}
  \put (37,17){\line(1,1){13}}
  \put (50,32){\vector(-1,1){17}}
  \put (33,49){\line(-1,1){13}}
  \put (0,-3.5){\mbox{$\Pem$}}
  \put (0,59.5){\mbox{$\Pep$}}
  \put (51.5,31){\circle*{4}}
  \put (51.5,31){\usebox{\hdwig}}
  \put (62,17){\mbox{$\gamma,\,\PZ$}}
  \put (73,31){\usebox{\hdwig}}
  \put (94.5,31){\circle*{4}}
  \put (94.5,31.0){\usebox{\dpdwig}}
  \put (94.5,31.0){\usebox{\dmdwig}}
  \put (109.9,46.4){\usebox{\dpdwig}}
  \put (109.9,15.6){\usebox{\dmdwig}}
  \put (98,-1){\mbox{$\PWm$}}
  \put (98,55){\mbox{$\PWp$}}
  \put (125.3,0){\circle*{4}}
  \put (125.3,61.8){\circle*{4}}
  \put (125.3,61.8){\vector(2,1){23}}
  \put (148.3,73.3){\line(2,1){17}}
  \put (165.3,41.8){\vector(-2,1){23}}
  \put (142.3,53.3){\line(-2,1){17}}
  \put (125.3,0.2){\vector(2,1){23}}
  \put (148.3,11.7){\line(2,1){17}}
  \put (165.3,-20){\vector(-2,1){23}}
  \put (142.5,-8.5){\line(-2,1){17}}
  \put (169.3,80){\mbox{$f_1$}}
  \put (169.3,40){\mbox{$\bar f_2$}}
  \put (169.3,17){\mbox{$f_3$}}
  \put (169.3,-23){\mbox{$\bar f_4$}}
  \end{picture} }
\end{picture}
\end{center}
\caption{Lowest-order diagrams for $\protect\eeWW\to4f$}
\label{FIee4fdia}
\end{figure}
There are also contributions from other diagrams 
with the same initial and final states, but different intermediate states.
Classifications of four-fermion production processes and of the contributing 
diagrams are given in \cite{Ba94a,Be94}. In \refta{CCnum} we give the number 
of diagrams contributing for final states that can be 
reached by \PW-pair intermediate states.
These so-called charged-current processes are sometimes referred to as
{\tt CCn}, with {\tt n} denoting the number of contributing diagrams [\eg
{\tt CC3} denotes process \refeqf{WWofshell}].
\begin{table}[t]
  \begin{center}
    \begin{tabular}{||c|c|c|c|c|c||}
      \hline \hline
      & \raisebox{0.pt}[2.5ex][0.0ex]{${\bar \Pd} \Pu$}
      & ${\bar \Ps} \Pc$ & ${\bar \Pe} \nu_{\Pe}$ &
      ${\bar \mu} \Pnm$ & ${\bar \tau} \Pnt$   \\
      \hline
      $\Pd {\bar \Pu}$ & {\it  43}& {\bf 11} &  20 & {\bf 10} & {\bf 10} \\
      \hline
      $\Pe {\bar \nu}_{\Pe}$ & 20 &  20 &{\it 56}&  18 &  18 \\
      \hline
      $\mu {\bar \nu}_{\mu}$ & {\bf 10} & {\bf 10} &  18 & {\it 19} &
      {\bf 9}  \\
      \hline \hline
    \end{tabular}
  \end{center}
  \caption{Number of Feynman diagrams for \PW-pair produced four-fermion
           final states.}
  \label{CCnum}
\end{table}
The simplest case ({\bf boldface} numbers in \refta{CCnum}) is
fully covered by doubly- and singly-resonant diagrams as
given in \reffis{FIee4fdia} and~\reffif{FIbackdia}.
\bfi
\unitlength 1pt
\savebox{\wigur}(12,4)[bl]
   {\bezier{20}(0,0)(1.7,5)(6,2)
    \bezier{20}(6,2)(10.3,-1)(12,4)}
\savebox{\Vur}(36,12)[bl]{\multiput(0,0)(12,4){3}{\usebox{\wigur}}}
\savebox{\wigr}(12,0)[bl]
   {\bezier{20}(0,0)(3, 4)(6,0)
    \bezier{20}(6,0)(9,-4)(12,0)}
\savebox{\Vr}(36,0)[bl]{\multiput(0,0)(12,0){3}{\usebox{\wigr}}}
\bma
\barr{l}
\begin{picture}(204,80)
\put(48,36){\circle*{4}}
\put(84,36){\circle*{4}}
\put(120,24){\circle*{4}}
\put(156,36){\circle*{4}}
\put(138,39){\makebox(0,0){$\PW$}}
\put(30,24){\vector(3,2){3}} \put(12,12){\line(3,2){36}}
\put(30,48){\vector(-3,2){3}} \put(12,60){\line(3,-2){36}}
\put(48,34){\usebox{\Vr}}
\put(55,20){\mbox{$\gamma,\,\PZ$}}
\put(138,54){\vector(3,1){3}} \put(84,36){\line(3,1){108}}
\put(102,30){\vector(-3,1){3}} \put(84,36){\line(3,-1){108}}
\put(156,12){\vector(-3,1){3}}
\put(120,24){\usebox{\Vur}}
\put(174,30){\vector(-3,1){3}} \put(156,36){\line(3,-1){36}}
\put(174,42){\vector(3,1){3}} \put(156,36){\line(3,1){36}}
\end{picture}
\earr
\ema
\caption{Example of a singly-resonant diagram.}
\label{FIbackdia}
\efi
Additional graphs of the types shown in \reffi{FIenuedia}
must be taken into account if electrons or electron-neutrinos are
produced (roman numbers in \refta{CCnum}).
\begin{figure}[t]
\begin{picture}(400,130)(0,0)
  \put(60,90){\line(-3,1){50}}
  \put(60,90){\line(3,1){50}}
  \put(33,99){\vector(3,-1){3}}
  \put(72,94){\vector(3,1){3}}
  \put(96,102){\vector(3,1){3}}
  \put(60,90){\circle*{4}}
  \put(60,90){\Photon(0,0)(0,-60) 2 5}
  \put(60,30){\line(-3,-1){50}}
  \put(60,30){\line(3,-1){50}}
  \put(36,22){\vector(-3,-1){3}}
  \put(87,21){\vector(-3,1){3}}
  \put(60,30){\circle*{4}}
  \put(85,98.3){\Photon(0,0)(23,-23) 2 3}
  \put(85,98.3){\circle*{4}}
  \put(108,75.3){\circle*{4}}
  \put(108,75.3){\line(3,1){30}}
  \put(108,75.3){\line(3,-1){30}}
  \put(123,80.3){\vector(3,1){3}}
  \put(123,70.3){\vector(-3,1){3}}
  \put(180,110){\line(1,0){100}}
  \put(204,110){\vector(1,0){3}}
  \put(254,110){\vector(1,0){3}}
  \put(230,110){\circle*{4}}
  \put(230,110){\Photon(0,0)(0,-33) 2 3}
  \put(230,77){\circle*{4}}
  \put(230,77){\line(0,-1){33}}
  \put(230,77){\line(1,0){50}}
  \put(254,77){\vector(1,0){3}}
  \put(230,61){\vector(0,1){3}}
  \put(230,44){\line(1,0){50}}
  \put(254,44){\vector(-1,0){3}}
  \put(230,44){\circle*{4}}
  \put(230,44){\Photon(0,0)(0,-33) 2 3}
  \put(230,11){\circle*{4}}
  \put(180,11){\line(1,0){100}}
  \put(204,11){\vector(-1,0){3}}
  \put(254,11){\vector(-1,0){3}}
  \put(322,110){\line(1,0){100}}
  \put(346,110){\vector(1,0){3}}
  \put(396,110){\vector(1,0){3}}
  \put(372,110){\circle*{4}}
  \put(372,110){\Photon(0,0)(0,-50) 2 5}
  \put(372,60){\circle*{4}}
  \put(372,60){\Photon(0,0)(0,-50) 2 5}
  \put(372,60){\Photon(0,0)(30,0) 2 3}
  \put(402,60){\circle*{4}}
  \put(402,60){\line(3,1){30}}
  \put(402,60){\line(3,-1){30}}
  \put(417,65){\vector(3,1){3}}
  \put(417,55){\vector(-3,1){3}}
  \put(322,11){\line(1,0){100}}
  \put(372,11){\circle*{4}}
  \put(346,11){\vector(-1,0){3}}
  \put(396,11){\vector(-1,0){3}}
\end{picture}
\caption{Examples of additional diagrams for final states with
         electrons and electron-neutrinos.}
\label{FIenuedia}
\end{figure}
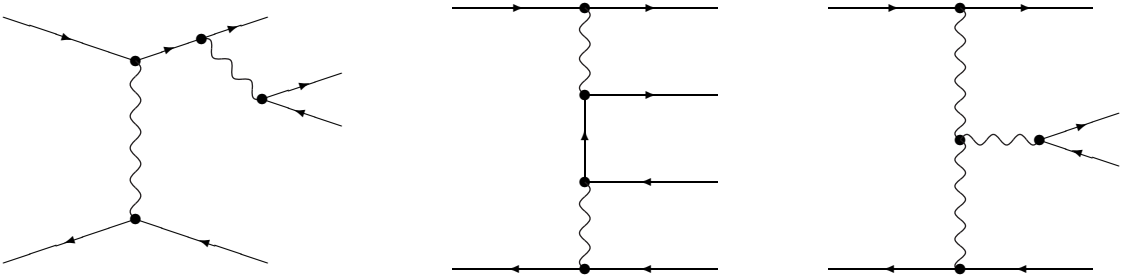
If the produced final state consists of particle--antiparticle pairs, the final
state can also be obtained through intermediate \PZ-pair production, 
leading to extra Feynman diagrams ({\it italic} numbers in \refta{CCnum}). 
Finally, it should be mentioned that QCD diagrams, involving an intermediate 
gluon, contribute in the case of final states consisting of two 
quark--antiquark pairs.

\subsection{Semi-analytical approach}
\label{SEoflosa}

\begin{figure}[bth]
 \begin{center}
  \setlength{\unitlength}{1cm}
  \hspace*{-0.2cm}
  \begin{picture}(11,11)
  \put (-2.3,-3.6){\includegraphics{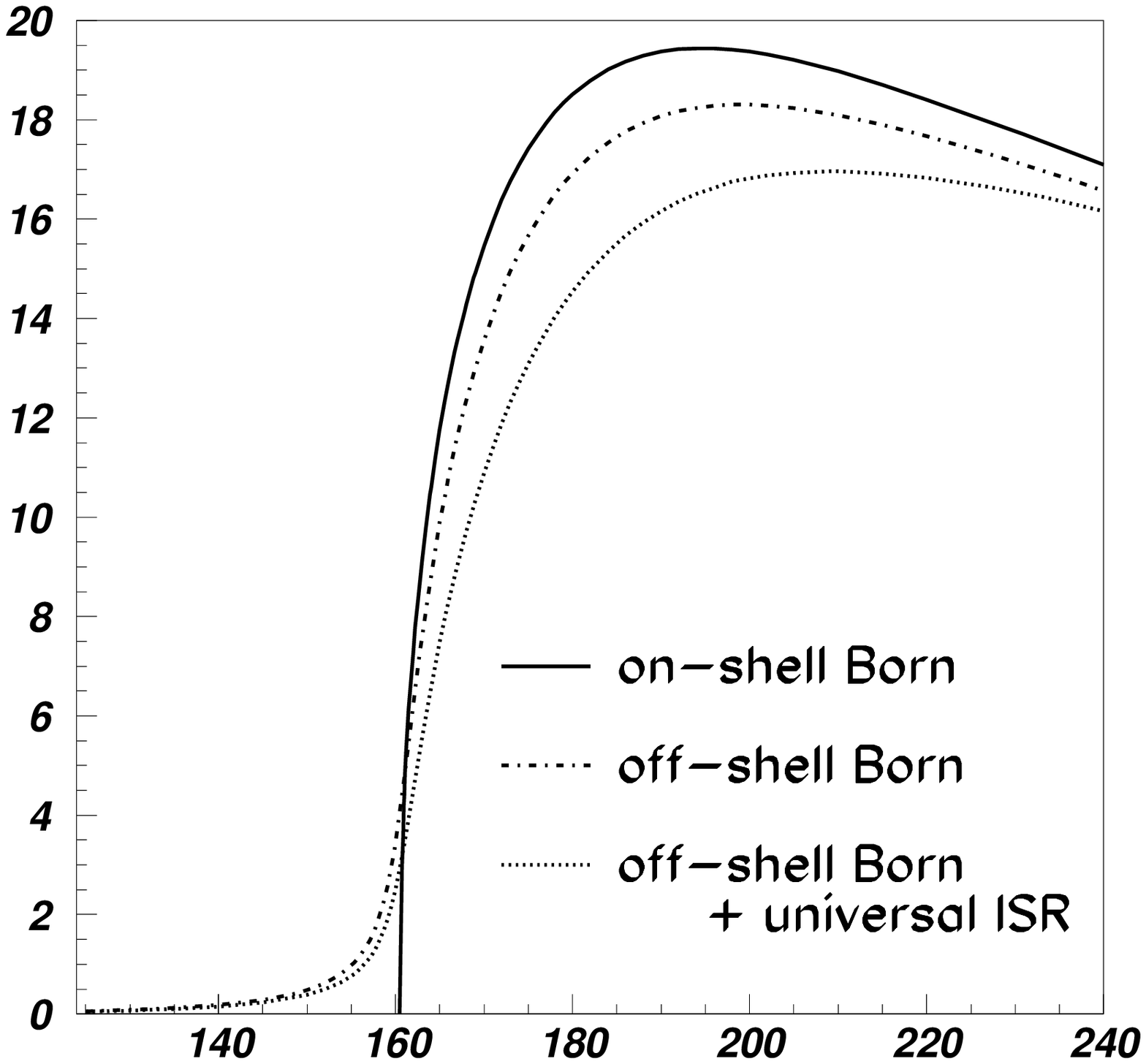}}
  \put (9.8,-0.5){\mbox{\boldmath $\sqrt{s}\ \ [\GeV]$}}
  \put (-1.8,9.6){\mbox{\boldmath $\sigma^{\tt\bf CC3}$}}
  \put (-1.8,8.7){\mbox{\boldmath $[{\rm pb}]$}}
\end{picture}
\end{center}
\caption{The inclusive {\tt CC3} \cs.}
\label{CC3xs}
\end{figure}

We will now, in a first step of sophistication beyond on-shell
\PW-pair production, introduce a finite \PW\ width and perform the {\tt CC3} 
calculation.
In the following a semi-analytical  method will be emphasized, whereas
Monte Carlo methods are treated elsewhere \cite{EGgroup}. The starting 
point is 
\begin{eqnarray}
  \sigma^{\tt CC3}(s) & = &
  \int\limits_0^s \d s_+ \, \rho_\PW(s_+)
  \int\limits_0^{(\sqrt{s} - \sqrt{s_+})^2} \d s_- \, \rho_\PW(s_-)
  \;\; \sigma^{\tt CC3}_0(s;s_+,s_-),
  \label{sigww}
\end{eqnarray}
with $s_+ = \pWp^2$\ and $s_-= \pWm^2$ the virtualities of the internal \PW\
bosons.
The Breit-Wigner densities
\begin{eqnarray}
  \rho_\PW(s_{\pm}) & = & \frac{1}{\pi} \frac {\MW \GW}
   {|s_{\pm} - \MW^2 + i \MW\GW |^2}
  \times {\mathrm {BR}}
   \label{rhow}
\end{eqnarray}
contain the finite width of the \PW\ boson, the coupling
constants of its decay to fermions, and the corresponding branching
ratio. 
In the limit of stable \PW\ bosons, the on-shell \cs\ is
recovered via 
\begin{equation}
  \rho_\PW(s_{\pm}) \stackrel {\GW \rightarrow 0} {\longrightarrow}
  \delta(s_{\pm} - \MW^2) \times {\mathrm {BR}}.
  \label{normal}
\end{equation}
The two-fold differential \cs\ contains terms corresponding to
the Feynman diagrams of \reffi{FIee4fdia} and their
interferences.
In the notation of \cite{Ba95} it may be described by three terms, 
\begin{eqnarray}
 \sigma^{\tt CC3}_0(s;s_+,s_-) & = &
 \frac{\left(\GF\MW^2 \right)^2}{8 \pi s}\, 
 \Biggl[  {\cal C}^s_{\tt CC3} \, {\cal G}^{33}_{\tt CC3}
        + {\cal C}^{st}_{\tt CC3} \, {\cal G}^{3f}_{\tt CC3}
        + {\cal C}^t_{\tt CC3} \, {\cal G}^{ff}_{\tt CC3} \Biggr].
  \label{sigww0}
\end{eqnarray}
The couplings ${\cal C}^{s, st,t}_{\tt CC3}$ contain the weak mixing
angle, the \PZ--\Pe\ vector and axial-vector coupling, the triple-gauge-boson
couplings, and the $s$-channel \PZ\ and $\gamma$ propagators. 
The kinematical functions ${\cal G}^{33,3f,ff}_{\tt CC3}$ are known 
analytically \cite{Mu86}:
\begin{eqnarray}
  {\cal G}^{ff}_{\tt CC3}(s;s_+,s_-) & = & \frac{1}{48}
    \left[ \rule[-.1cm]{0cm}{.5cm} \: \lambda(s,s_+,s_-) +
           12\, s\,\left( s_+ + s_- \right) - 48\, s_+\, s_- \right.
    \nl & & \left. \hphantom{\frac{1}{48}a} \rule[-.1cm]{0cm}{.5cm}
           + 24\, s_+\, s_- \left(s - s_+ - s_-\right)\,
           {\cal L}(s;s_+,s_-)\right],
    \label{cc3gt} \\ \nl
  {\cal G}^{33}_{\tt CC3}(s;s_+,s_-) & = & \frac{\lambda(s,s_+,s_-)}{192}
    \left[ \rule[-.1cm]{0cm}{.5cm} \lambda(s,s_+,s_-) 
           + 12\,\left(s\, s_+ + s\, s_- + s_+\, s_- \right)\right],
    \label{cc3gs} \\ \nl
  {\cal G}^{3f}_{\tt CC3}(s;s_+,s_-) & = & \frac{1}{48}
    \left\{ \rule[-.2cm]{0cm}{.7cm} (s-s_+-s_-)
            \left[ \rule[-.1cm]{0cm}{.5cm}
                   \lambda(s,s_+,s_-) 
                   + 12\, s\, (s\, s_+ + s\, s_- + s_+\,s_- ) \right]
    \right.
    \nl & &  \left. \hphantom{\frac{1}{48}A} \rule[-.2cm]{0cm}{.7cm}
                   - 24\, s_+\, s_-\, \left( s\, s_+ + s\, s_- 
                   + s_+\, s_- \right)\,{\cal L}(s;s_+,s_-) \right\},
\label{cc3gst}
\end{eqnarray}
where
\begin{eqnarray}
  \lambda(s,s_+,s_-) & = &
  s^2 + s_+^2 + s_-^2 - 2\, s\, s_+ - 2\, s\, s_- - 2\, s_+\, s_-,
  \label{lambda} \\[1ex]
  {\cal L}(s;s_+,s_-) & = & \frac{1}{\sqrt{\lambda(s,s_+,s_-)}} \,
  \log\left(\frac{s-s_+-s_-+\sqrt{\lambda(s,s_+,s_-)}}
                 {s-s_+-s_--\sqrt{\lambda(s,s_+,s_-)}}
  \right).
\end{eqnarray}
The numerical importance of the \PW-boson off-shellness in
\refeq{sigww} is displayed in \reffi{CC3xs}. 
Clearly, the effect of the finite \PW\ width is comparable to the
one from universal initial-state radiation, defined in \refse{SEofrcpc}. 

The second level of sophistication of \PW-pair production is
reached by performing calculations for complete sets of Feynman
diagrams for specific final states.
The simplest case, the {\tt CC11} class of processes, corresponds to
four-fermion final states without electrons or electron-neutrinos and
without particle--antiparticle pairs.
The corresponding gauge-invariant set of Feynman
diagrams consists of the three doubly-resonant diagrams of
\reffi{FIee4fdia} plus at most eight singly-resonant diagrams of
the type shown in \reffi{FIbackdia}. This class of processes is still 
calculable in the semi-analytical approach and is of specific interest 
for the \PW-mass determination and the TGC studies.
The corresponding \cs\ is given by \cite{Ba95}
\begin{equation}
  \label{CC11xsec}
  \sigma^{\tt CC11}(s) \; = \; \int \d s_+ \int \d s_- \;\;
    \frac{\sqrt{\lambda(s,s_+,s_-)}}{\pi s^2} \,
    \sum_{k=1}^{15} \frac{\d^2 \sigma_k(s,s_+,s_-)}{\d s_+\, \d s_-}
\end{equation}
with
\begin{equation}
  \label{CC11diff}
  \frac{\d^2 \sigma_k}{\d s_+\, \d s_-} \; = \;
    {\cal C}_k \, {\cal G}_k(s,s_+,s_-)~.
\end{equation}
Initial- and final-state couplings, as well as intermediate \PW, \PZ,
and $\gamma$\ propagators are collected in the ${\cal C}_k$'s. The kinematical
details are contained in the ${\cal G}_k$'s. Because of symmetry properties,
under the exchanges of the three virtualities, only six of these kinematical
functions are independent. Apart from the three kinematical functions
of the {\tt CC3} process, three more enter as a result of the singly-resonant
diagrams. Explicit expressions can be found in \cite{Ba95}.

A similar semi-analytical analysis of the remaining four-fermion final states
has not been performed so far, although this would be desirable.

\subsection{Lowest order: the gauge-invariance issue}
\label{SEoflogi}
 
The discussion in the previous subsections has avoided the question of gauge 
invariance which arises when going from on-shell \PW-pair production to the 
off-shell case. There are two sources of gauge non-invariance one has to be 
aware of.

First of all there is the issue of gauge non-invariance as a result of 
incomplete sets of contributions. Consider to this end again the process
\beq
\Pep+\Pem \to \PWp+\PWm \to \Pf_1 + \bar\Pf_2 + \Pf_3 + \bar \Pf_4,
\eeq
represented by the three Feynman diagrams shown in \reffi{FIee4fdia}.
Forgetting about the finite \PW\ width the corresponding matrix element can be
written in the following form:
\beqar \label{MEofs}
\Mhat_{\born}^{\he} &=&
\left[\delta_{\he -} \,\frac{e^2}{2\sw^{2}}\, \M_{I}^{\he,\rho\si}
+e^2 \M_{Q}^{\he,\rho\si}\right]    \nl &&
\times\,\frac{1}{\pWp^2-\MW^2}
\times\frac{e}{\sqrt{2}\sw}\,\bar u_1\ga_\rho\,\omega_{-}\, v_2
\times\frac{1}{\pWm^2-\MW^2}
\times\frac{e}{\sqrt{2}\sw}\,\bar u_3\ga_\si\,\omega_{-}\, v_4,
\eeqar
where
\beqar
\nn \M_{Q}^{\he,\rho\si}&=&
\frac{2\MZ^2}{s(s-\MZ^2)}  \,\bar{v}_{+}
\left[\ks_{+}g^{\rho\si} -\ga^{\rho} k_{+}^\si +\ga^{\si} k_{-}^\rho\right]
\omega_{\he}\, u_-, \\
\M_{I}^{\he,\rho\si}&=& \frac{1}{t} \,\bar{v}_{+}
\ga^{\rho}(\ks_{+}-\ps_{+})\ga^{\si} \omega_{\he}\, u_-
-\frac{s}{\MZ^2} \,\M_Q^{\he,\rho\si}
\eeqar
are directly related to the corresponding on-shell quantities $\M_I$
and $\M_Q$ defined in \refeq{MIMQ}.
Here $\pWpm$ are the momenta of the $\PW^\pm$~bosons as reconstructed
from the decay products,
and $s$ and $t$ the usual Mandelstam variables for
$\Pep\Pem\to\PWp\PWm$. 
We have assumed that all fermion masses are negligible.
In a renormalizable gauge such as a $R_\xi$ gauge one would otherwise
have to consider in addition diagrams involving one or two
unphysical charged Higgs bosons instead of the \PW~bosons.
 
Whereas for on-shell \PW~bosons the
contributions of $\M_I$ and $\M_Q$ are separately gauge-invariant,
for off-shell \PW~bosons not even $\Mhat_\born$ is gauge-invariant.
This can be illustrated by considering the doubly-resonant diagrams
in an axial gauge \cite{BD94}. To this end we replace the
conventional 't~Hooft--Feynman gauge-fixing term for the \PW~boson by
$-(n^\mu W^+_\mu)(n^\nu W^-_\nu)$. This modifies the \PW-boson
propagators in the diagrams of \reffi{FIee4fdia} and,
since this gauge-fixing term does
not eliminate the mixing between the \PW-boson field and the
corresponding unphysical Higgs field, leads to mixing propagators
between those fields, giving rise to additional Feynman diagrams.
Combining all relevant diagrams yields the following extra
contribution to $\Mhat_{\born}^{\he}$ in the axial gauge:
\beqar \label{axialgauge}
  \lefteqn{ \left.\Mhat_{\born}^{\he}\right|_{\mathrm{axial~gauge}}
    -\left.\Mhat_{\born}^{\he}\right|_{\mbox{\scriptsize 
    't~Hooft--Feynman~gauge}} 
  = e^2\left[\delta_{\he -} \,\frac{1}{2\sw^{2}}
    - \frac{\MZ^2}{s}\right] \frac{1}{s-\MZ^2}\, \bar{v}_{+}
    \ga_{\mu}\, \omega_{\he}\, u_-} \nl && \hspace*{-1.4cm}
  \qquad\qquad\times \left[ \frac{g^{\mu\rho}n^\si}{(\pWm^2-\MW^2)(n\cdot\pWm)}
    -\frac{g^{\mu\si}n^\rho}{(\pWp^2-\MW^2)(n\cdot\pWp)} \right]
    \times\frac{e}{\sqrt{2}\sw}\, \bar u_1\ga_\rho\,\omega_{-}\, v_2
    \times\frac{e}{\sqrt{2}\sw}\, \bar u_3\ga_\si\,\omega_{-}\, v_4.
\hspace{1cm}
\eeqar
Note that these gauge-dependent terms involve either a pole in
$(\pWp^2-\MW^2)$ or $(\pWm^2-\MW^2)$ but not both, \ie they are only
singly-resonant. These terms are exactly cancelled by the
gauge-dependent contributions of eight singly-resonant diagrams contributing
to the same final state with the topology shown in \reffi{FIbackdia}.
 
The sum of the doubly-resonant diagrams and the singly-resonant ones
with the topology shown in \reffi{FIbackdia} is in general gauge-invariant.
This follows directly from the fact that for final states with four
different fermions and no electrons or positrons those diagrams are the only
ones that contribute. As a consequence the non-resonant diagrams are not
needed to cancel the gauge-dependent terms of the doubly-resonant
diagrams. Among the non-resonant diagrams and the singly-resonant ones
that do not have the topology shown in \reffi{FIbackdia} further
gauge-invariant subsets can be identified by considering other four-fermion
final states.

Thus, in general all graphs that contribute to a given final state
have to be taken into account and
one is lead to consider the complete process \eeffff,
including all resonant and non-resonant graphs, in order to
obtain a manifestly gauge-independent result.

Simple estimates indicate that all 
non-doubly-resonant contributions are typically suppressed by a factor 
$\GW/\MW\approx2.5\%$ on the \cs\ level for each non-resonant \PW\ propagator%
\footnote{For differential distributions that do not involve an explicit 
phase, like \eg $\si_{\mathrm{tot}}$ or $\dsidct$, there will be no 
interference between doubly- and singly-resonant diagrams. Consequently the 
non-doubly-resonant contributions are suppressed by 
$\GW^2/\MW^2\approx0.1\%$ in those cases \cite{Ae93}.}.
This is confirmed by explicit calculations \cite{EGgroup}.
For instance, using covariant ($R_\xi$) gauges the universal 
non-doubly-resonant graphs that occur for all final states (see 
\reffi{FIbackdia}) contribute less than $0.15\%$ to the total \cs\ for 
$175\GeV < \sqrt{s} < 205\GeV$, and 0.3\% at 161\GeV. 
The contribution of the non-doubly-resonant $t$-channel photon-exchange 
graphs, which only occur when there are electrons in the final state, depends 
very much on the angular cut imposed on the outgoing electrons; for $10^\circ$ 
they contribute at the per-cent level, \eg $\sim 4\%$ at $\sqrt{s}=190\GeV$.
Although all these non-doubly-resonant contributions are suppressed, they 
should nevertheless be taken into account to reach a theoretical accuracy of 
$\sim 0.5\%$.  
Finally, the QCD graphs have been shown not to 
interfere with the electroweak graphs to any sizable extent, and can thus be 
computed using standard (Monte Carlo) QCD programs.
 
Even when considering the complete set of graphs contributing to a given 
final state, there is still a more fundamental gauge-invariance problem to be 
solved. The resonant graphs discussed above involve poles at 
$\pWpm^2=\MW^2$.%
\footnote{There are similar poles associated with
diagrams containing internal \PZ\ propagators. These correspond to \PZ-pair 
production and are here considered as background to \PW-pair production.}
These have to be cured by introducing the finite width in one way or another,
while at the same time preserving gauge independence and, through a proper
energy dependence, unitarity. In field theory, such widths arise naturally 
from the imaginary parts of higher-order
diagrams describing the boson self-energies, resummed to all orders.
This procedure has been used with great success in the past: indeed,
the $\PZ$ resonance can be described to very high numerical accuracy.
However, in doing a Dyson summation
of self-energy graphs, we are singling out only a very limited subset
of all the possible higher-order diagrams. It is therefore not
surprising that one often ends up with a result that retains
some gauge dependence.

For example it is very tempting to systematically
replace $1/(q^2-M^2)$ by $1/(q^2-M^2+iM\Gamma)$, also for $q^2<0$. Here 
$\Gamma$ denotes the physical width of the particle with mass $M$ and momentum
$q$. This is the so-called `fixed-width scheme'.
As in general the resonant diagrams are not gauge-invariant by themselves, this
substitution will again destroy gauge invariance. 
Moreover, the `fixed-width scheme' has no
physical motivation. In perturbation theory the propagator for space-like 
momenta does not develop an imaginary part. Consequently, unitarity is violated
in this scheme. To improve on the latter the constant width could be replaced 
by a running one. This can, however, not cure the gauge non-invariance problem.
At this point one might ask oneself the legitimate question whether 
the gauge breaking occurring in the `fixed-width scheme' is numerically
relevant or, like the gauge breaking in the LEP1 analyses, negligible for all 
practical purposes.
Of course, such a statement can only be made on the basis of an explicit 
comparison with truly gauge-invariant schemes in the full LEP2 energy range.
We will come back to that point later on, after having defined a scheme that
is gauge-invariant and reliable at LEP2. 

Below we will list a few ways to come to a gauge-invariant result and
discuss their validity.

One way to sidestep the gauge non-invariance problem is by simply multiplying 
the full matrix element by $[q^2-M^2]/[q^2-M^2+iM\Gamma(q^2)]$, which is 
evidently gauge-invariant \cite{Baur92,Ku95}. In this 
way the pole at $q^2=M^2$ is softened into a resonance, at the expense of 
mistreating the non-resonant parts.
It should be noted that when the doubly-resonant diagrams are not dominant,
like at energies at and below the \PW-pair production threshold, this 
so-called `fudge-factor scheme' can lead to large 
deviations \cite{Baur95}.

The second possibility is the so-called `pole scheme' \cite{Ve63}--\cite{Ae94}.
In this scheme one decomposes the
complete amplitude, consisting of contributions from doubly-resonant
diagrams $R_{+-}$ (corresponding in lowest order to $\Mhat_\born$),
singly-resonant diagrams $R_+$, $R_-$, and
non-resonant diagrams $N$, according to their poles as follows:
\beqar \label{poledec}
\M &=&
       \frac{R_{+-}(\pWp^2,\pWm^2,\theta)}{(\pWp^2-\MW^2)(\pWm^2-\MW^2)}
      +\frac{R_{+}(\pWp^2,\pWm^2,\theta)}{\pWp^2-\MW^2}
      +\frac{R_{-}(\pWp^2,\pWm^2,\theta)}{\pWm^2-\MW^2}
      +N(\pWp^2,\pWm^2,\theta) \nn \\
&=& \frac{R_{+-}(\MW^2,\MW^2,\theta)}{(\pWp^2-\MW^2)(\pWm^2-\MW^2)} \nn\\
&& {} +\frac{1}{\pWp^2-\MW^2}\biggl[
       \frac{R_{+-}(\MW^2,\pWm^2,\theta)-R_{+-}(\MW^2,\MW^2,\theta)}
       {\pWm^2-\MW^2} + R_{+}(\MW^2,\pWm^2,\theta) \biggr] \nn\\
&& {} +\frac{1}{\pWm^2-\MW^2}\biggl[
       \frac{R_{+-}(\pWp^2,\MW^2,\theta)-R_{+-}(\MW^2,\MW^2,\theta)}
       {\pWp^2-\MW^2} + R_{-}(\pWp^2,\MW^2,\theta) \biggr] \nn\\
     && {}+\biggl[
     \frac{R_{+-}(\pWp^2,\pWm^2,\theta) +R_{+-}(\MW^2,\MW^2,\theta)
     -R_{+-}(\MW^2,\pWm^2,\theta) -R_{+-}(\pWp^2,\MW^2,\theta)}
     {(\pWp^2-\MW^2)(\pWm^2-\MW^2)} \nn\\
&& {} +\frac{R_{+}(\pWp^2,\pWm^2,\theta)
            -R_{+}(\MW^2,\pWm^2,\theta)}{\pWp^2-\MW^2}  \nn \\
&& {} +\frac{R_{-}(\pWp^2,\pWm^2,\theta)
            -R_{-}(\pWp^2,\MW^2,\theta)}{\pWm^2-\MW^2}
      +N(\pWp^2,\pWm^2,\theta) \biggr].
\eeqar
Here $\theta$ stands generically for all angular variables
which should be defined in such a way that their integration boundaries are
independent of $\pWp^2$ and $\pWm^2$.
Otherwise these angular variables would introduce
an additional dependence on $\pWp^2$ and $\pWm^2$, and the
correct pole terms could only be extracted after the angular
integrations had been performed. This would complicate the
pole decomposition and, in particular, would not be suited for
a Monte-Carlo generator. Appropriate variables are
\eg the angles in the $\Pep\Pem$-CM system but not
the Mandelstam variables.
In \refeq{poledec} $\M$ is decomposed into gauge-invariant
subsets originating from double-pole terms, single-pole terms, and
non-pole terms (with repect to \MW).
Introducing now the finite width only in the pole factors and not in
the finite constant residues in brackets does not destroy
gauge invariance. 
We note that different ways of introducing the finite width, e.g.~constant or
running, differ only by terms that are of higher-order
and/or that are not of the double-pole type%
\footnote{There is no unique prescription for the propagator including the
finite width. Instead of the constant term $\MW\GW$ one can also use
a width depending on the invariant mass $\pWpm^2$.
This corresponds to different definitions of the (renormalized)
\PW~mass, which is fixed by the pole of the propagator.
A popular choice is $k^2\GW\!/\!\MW$, which amounts to a shift in
\MW\ by $\GW^2/(2\MW)\approx 26\MeV$ relative to the propagator
with the constant-width term.}. 
The same holds for different choices of the angular variables~$\theta$ 
(see \refse{SEofrcps}).

A drawback of the `pole scheme' is the fact that it is not defined below
the \PW-pair production threshold and that it yields unreliable results just
above this threshold (see \refse{SEofrcps}). 

Apart from yielding gauge-invariant results, the `pole-scheme' decomposition
also constitutes a systematic expansion according to the degree of 
resonance, \ie in powers of $\GW/(\MW\be)$. Here the enhancement factor 
$1/\be$ represents the influence of the nearby threshold on the expansion.
Sufficiently far above the \PW-pair threshold and after imposing appropriate
angular and invariant mass cuts, in order to reduce \PZ-pair and $t$-channel
photon-exchange backgrounds,
the \cs\ for off-shell \PW-pair production is dominated by
(or may even by defined by)
the double-pole terms $R_{+-}(\MW^2,\MW^2,\theta)$.
At least at lowest order these may be related to
on-shell \PW-pair production in the following way:
\beq \label{Mfactor}
R_{+-}(\MW^2,\MW^2,\theta)  = \sum_{\hWp,\hWm}
\M_{\Pep\Pem\to\PWp\PWm}^{\hWp,\hWm}
\times \M_{\PWp\to\Pf_1\bar\Pf_2}^{\hWp}
\times \M_{\PWm\to\Pf_3\bar\Pf_4}^{\hWm},
\eeq
where $\M_{\Pep\Pem\to\PWp\PWm}$, $\M_{\PWp\to\Pf_1\bar\Pf_2}$,
$\M_{\PWm\to\Pf_3\bar\Pf_4}$ denote the matrix elements for the
production of two on-shell \PW~bosons and their subsequent
decay into fermion--antifermion pairs. 
As such the `pole scheme' is a 
natural starting point for the systematic evaluation of higher-order
corrections \cite{BD94,Ae94} (see also \refse{SEofrcps}).

As a third method, one may determine the minimal set of Feynman diagrams that 
is necessary to compensate for the gauge violation caused by the self-energy 
graphs, and try to include these \cite{Si91,BeEE50092}.
This is obviously the theoretically most satisfying solution, but
it may cause an increase in the complexity of the matrix elements
and a consequent slowing down of the numerical calculations.
For the vector bosons, the lowest-order widths are given by the imaginary 
parts of the fermion loops in the one-loop self-energies.  It is therefore 
natural to include the other possible one-particle-irreducible fermionic 
one-loop corrections \cite{Baur95,Pa95,BHF1}. For the process \eeffff\ this
amounts to adding the fermionic triple-gauge-boson vertex corrections.  
The complete set of fermionic contributions form a gauge-independent subset
and obey all Ward identities exactly, even with resummed propagators 
\cite{BHF2}. 
This implies that the high-energy and collinear limits are properly behaved. 
In contrast to all other schemes mentioned above, the `fermion-loop scheme' 
recommended here does not modify the theory by hand but selects 
an appropriate set of higher-order contributions to restore gauge invariance.
To solve the problem of gauge invariance related to the width, 
we in fact only have to consider the imaginary parts of these fermionic
contributions%
\footnote{As the Ward identities are linear, we can separate the real and 
imaginary parts.}.
  
The `fermion-loop scheme' should work properly for all tree-level
calculations involving resonant \PW\ bosons and \PZ\ bosons or other
particles decaying exclusively into fermions. This also includes, for 
instance, the hard-photon process $\eeffff + \ga$, which requires in addition
to fermionic vertex corrections also fermionic box corrections.
For resonating particles decaying also into bosons, such as the top quark, 
or for calculating RC's to \eeffff, which also involves bosonic corrections,
the `fermion-loop scheme' is not really suited. 

Although the latter scheme is well-justified in standard perturbation 
theory, it should be stressed that any working scheme is arbitrary to
a greater or lesser extent: since the Dyson summation must necessarily
be taken to all orders of perturbation theory, and we are not able
to compute the complete set of {\it all\/} Feynman diagrams to {\it all\/}
orders, the various schemes differ even if they lead to formally 
gauge-invariant results. In \cite{PP95} another technique, the so-called
`pinch technique', has been introduced in order to construct a 
gauge-parameter-independent Dyson summation. Even if the `pinch technique' 
yields gauge-independent results, it still contains some arbitrariness in the 
sense that one still has the freedom to shift gauge-independent parts that 
fulfill the Ward identities from the vertex corrections to the self-energies. 
For instance, it has been demonstrated in \cite{DDW94} that the 
`background-field method' can be used to construct an infinite variety of
such shifts, all representing (theoretically) equally well-justified schemes 
for resumming self-energies.

Now it is a numerical question how much the predictions of different schemes 
differ. In \cite{BHF1} a detailed study has been given for the process
$\Pep\Pem \to \Pem\bar{\nu}_\Pe \Pu \bar{\Pd}$, a process that is highly 
sensitive to $U(1)$ electromagnetic gauge violation. In this process
the electron may emit a virtual photon, whose $k^2$ can be as small as 
$\Me^2$: with a total centre-of-mass energy of $\sqrt{s}$ available, 
we have a mass ratio of $s/\Me^2 = {\mathcal{O}}(10^{11})$, large enough
to amplify even a tiny gauge violation in a disastrous way\footnote{This was 
noted already in \cite{BW70}, and investigated further in 
\cite{Ku95}.}. In \refta{TAU1schemes} we give the \cs\ corresponding to the
$t$-channel photon-exchange diagrams, responsible for the amplification of 
the gauge-breaking terms in the collinear limit. The results are given for two
values of the minimum electron scattering angle $\theta_{\rm{min}}$, displaying
the effect of cutting away the dangerous collinear limit. It is clear that a
naive introduction of a running width without a proper inclusion of fermionic
corrections to the three-vector-boson vertex, which breaks $U(1)$ 
electromagnetic gauge invariance, leads to completely unreliable results. 
The above-described gauge-invariant methods as well as the $U(1)$-preserving 
`fixed-width scheme' numerically deviate by much less than $\GW/\MW$. Hence, 
a naive running width is not suited for LEP2, whereas a constant width, 
although $SU(2)\times U(1)$ gauge breaking, might
constitute a workable approach. As in the collinear limit $k^2\to 0$ the 
gauge-breaking terms originating from a naive running width are proportional 
to the dominant lowest-order graphs, it is possible to multiply
the $\gamma\PW\PW$ Yang-Mills vertex with a simple factor to successfully
restore the $U(1)$ gauge invariance. This factor is, however, certainly not
universal. It will depend on the way the running width is introduced and 
on the process under investigation. Moreover, such a simple factor breaks
unitarity and at high energies the full expression from the  
fermion loops is required for having a proper energy dependence.
\begin{table}[htb]
\begin{center}
\begin{tabular}{||l|r|r||}\hline\hline
Scheme                    & \multicolumn{2}{c||}{$\sigma$ [pb]} \\
\cline{2-3}
                          & $\theta_{\rm{min}} = 0^\circ$ & 
                                             $\theta_{\rm{min}} = 10^\circ$ \\
\hline
Fixed width                               & .08887(8)   & .01660(3) \\
Running width, no correction              & 60738(176)  & .01713(3) \\
Fudge factor, with running width          & .08892(8)   & .01671(3) \\
Pole scheme, with running width           & .08921(8)   & .01666(3) \\
fermion-loop scheme                       & .08896(8)   & .01661(3) \\
\hline\hline
\end{tabular}
\end{center}
\caption[]{Cross-section in different schemes for the $t$-channel 
           photon-exchange diagrams of 
           $\Pep\Pem \to \Pem\bar{\nu}_\Pe \Pu \bar{\Pd}$.
           All schemes were computed using the same sample, so the differences
           are much more significant than the integration error suggests.}
\label{TAU1schemes}
\end{table}

\subsection{Radiative corrections}
\label{SEofrc}
So far no complete treatment for the \Oa\ corrections to off-shell
\PW-pair production is available.
Essentially only the initial-state 
photonic corrections, the final-state Coulomb correction, and the full hard
process \eeffff$+\gamma$ have been treated so far%
\footnote{An evaluation of all resummed one-particle-irreducible fermionic 
\Oa\ corrections, in the context of the `fermion-loop scheme', is in progress
\cite{BHF2}.}. 
These are discussed in the following. The leading weak effects are normally 
taken into account through dressed lowest-order matrix elements, using \GF\ 
and $\al(s)$.%
\footnote{The uncertainty associated with different theoretical definitions 
of $\sw^2$, \ie $\sw^2=1-\MW^2/\MZ^2$ or 
$\sw^2=\pi\,\al(4\MW^2)/(\sqrt{2}\,\GF\,\MW^2)$, 
are found to be below 0.1\%.}
In addition we describe a general strategy for the calculation
of corrections beyond the lowest order using the `pole scheme'.

\subsubsection{Initial-state radiation}
\label{SEofrcpc}
Most of the published calculations for corrections to off-shell \PW-pair 
production that have been done so far cover only part of the photonic 
corrections, mainly because these are easily treatable and constitute a large 
part of the RC's. 
{}From the discussion of the on-shell process we know that 
initial-state radiation (ISR) yields large corrections originating from the 
leading collinear logarithms. As will be discussed in \refapp{app} these can
be easily obtained by applying the structure-function method. Just in the same
way as in the on-shell case the corrections to the off-shell \cs\ are
calculated by convoluting the lowest-order \cs\ (in a certain gauge-invariant
scheme) with the appropriate structure functions. In \refapp{app} a detailed
analysis is given of the theoretical uncertainties associated with the 
leading-log procedure.

In the present-day Monte Carlos different ways of implementing the leading 
logarithmic corrections have been adopted \cite{EGgroup}. One method involves
solving the evolution equations for the structure functions numerically using 
techniques known from parton-shower algorithms. In this way soft-photon 
exponentiation and resummation of the leading logarithms from multiple
hard-photon emission are automatically taken into account. Photons are 
generated according to the matrix elements in the collinear limit, in this way
allowing for a finite $p_T$ kick to the photon. The second method involves a
fully inclusive treatment of the radiated photons by folding the improved
lowest-order \cs\ with leading-log structure functions. 
So, no explicit photons are
generated. The most recent development is a kind of merger of an explicit
$\eeffff\gamma$ Monte Carlo folded with structure functions, allowing for a
consistent definition of observable and unobservable photon radiation (see 
\refse{SEofph}).

Another approach is to include the complete ISR. To this end one has to
define it in a way that preserves the $U(1)$ electromagnetic gauge
invariance. This is non-trivial because of the
presence of the $t$-channel diagram which involves a non-conserved
charge flow in the initial state. One technique to circumvent this
problem is the so-called current-splitting technique \cite{Ba93},
which amounts to
splitting the electrically neutral neutrino in the $t$-channel diagram
into two oppositely flowing leptons each with charge one. One of them
is attributed to the initial state to build a continous flow of charge,
the other is attributed to the final state to do the same there.
The modified \Pne~propagator
leads to additional real and virtual initial-state photonic diagrams
which render the ISR gauge-invariant.
In this way one obtains the usual universal $s$-channel ISR known
from LEP1 plus additional non-universal contributions arising from
the $t$-channel diagrams.
The latter are non-factorizing with respect to the lowest-order
\cs, but are screened by a factor $\pWp^2\pWm^2/s^2$, which automatically 
guarantees a unitary behaviour at high energies. Moreover they turn out to be
numerically small at LEP2 energies. The universal ISR contains all leading 
logarithms and can be supplemented by the known universal higher-order 
terms \cite{Be87}.

It should be noted that both the leading-log and the current-splitting method
leave out non-leading photonic corrections, for instance associated with 
radiation of photons off the intermediate \PW\ bosons. For on-shell \PW\
bosons we give in \refta{Egtab} a comparison of exact and leading-log 
evaluations of the quantity $\int \d E_\ga\,E_\ga\,(\d\si/\d E_\ga)$, needed 
for the average energy loss \Eg.%
\footnote{The (lowest-order) average energy loss \Eg\ can be obtained by 
normalizing to the lowest-order \cs\ of \refta{TAsiww}. Dressing this 
lowest-order \cs\ by LL structure functions, running couplings etc., only 
makes sense when the energy-weighted \cs\ appearing in the numerator is 
treated in the same way!}    
\btab
\begin{center}
\begin{tabular}{||c|c|c|c|c||} \hline\hline
 \raisebox{-4mm}{$\sqrt{s}$} & \raisebox{-4mm}{$\int \d E_\ga\,E_\ga\,
 \frac{\textstyle \d\si}{\textstyle \d E_\ga}$} & 
 \multicolumn{3}{|c||}{ \raisebox{-2mm}{leading-log results} }\\
 \cline{3-5}
 & & $Q^2=s$ & $Q^2=s$ & $Q^2=Q_0^2$\\
 \raisebox{2mm}{$[ \mathrm{GeV} ]$} &
 \raisebox{2mm}{$[ \mathrm{GeV}\cdot\mathrm{pb} ]$} & \LL & $\LL-1$ 
 & $\LL-1$\\[1ex]
 \hline
 \hpt 161.0 & \,0.1377 $\pm$ 0.0001 & \,0.1436 & \,0.1379 & \,0.1375\\ \hline
 \hpt 165.0 & \hpt 3.619 $\pm$ 0.002 & \hpt 3.792 & \hpt 3.642 & \hpt 3.607\\ 
 \hline
 \hpt 170.0 & 10.120 $\pm$ 0.006 & 10.651 & 10.232 & 10.089 \\ \hline
 \hpt 175.0 & 17.437 $\pm$ 0.011 & 18.431 & 17.708 & 17.403 \\ \hline
 \hpt 184.0 & 30.882 $\pm$ 0.029 & 32.873 & 31.589 & 30.902 \\ \hline
 \hpt 190.0 & 39.562 $\pm$ 0.037 & 42.224 & 40.578 & 39.595 \\ \hline
 \hpt 205.0 & 59.126 $\pm$ 0.076 & 63.581 & 61.117 & 59.322 \\ \hline\hline
\end{tabular}
\end{center}
\caption{The quantity $\int \d E_\ga\,E_\ga\,(\d\si/\d E_\ga)$ needed for the 
         average energy loss. The exact on-shell result from hard-photon 
         radiation is given as well as the corresponding leading-log 
         approximation, using $\LL=\log(Q^2/\Me^2)$ or $\LL-1$.} 
\label{Egtab}
\etab
It is clear that a sufficiently accurate leading-log 
evaluation should be based on $\LL\!-\!1$ rather than \LL. 
Furthermore, the often-used scale choice $Q^2=s$
leads to deviations of about 3\% at 190\GeV, which would translate into an
error of \,$\sim 60\MeV$ on $\Eg$. The scale choice 
$Q_0^2 = 4\MW E/(1+\beta)$, however, agrees with the exact result at the 
0.3\% level for all LEP2 energies, leading to errors below 10\MeV\ in $\Eg$.
As there is no obvious reason why these non-leading terms should be smaller 
in the off-shell case, some care has to be taken with the choice of a suitable 
leading-log scale.

\subsubsection{The Coulomb singularity}
\label{SEofcoul}
Another potentially large photonic correction, not associated with the initial
state, is due to the Coulomb singularity in the threshold region, which
has been discussed for off-shell \PW-pair production in 
\cite{Fa93c}--\cite{CFK95}. It originates from the IR limit and
at one-loop it emerges from a single IR-singular scalar three-point
function and a related IR-singular scalar four-point function, the
IR-singular part of which is just a scaled version of the one contained in
the three-point function (see \reffi{FIcouldia}).
\bfi
\unitlength 1pt
\savebox{\wigur}(12,7)[bl]
   {\bezier {20}(0.0,0.0)(1.5,5.5)(6.0,3.5)
    \bezier {20}(6.0,3.5)(10.5,1.5)(12.0,7.0)}
\savebox{\Vur}(72,42)[bl]{\multiput(0,0)(12,7){6}{\usebox{\wigur}}}
\savebox{\wigur}(60,24)[bl]
   {\bezier {20}(0.0,0.0)(1.4,5.2)(6.0,2.4)
    \bezier {20}(6.0,2.4)(10.6,0.6)(12.0,4.8)}
\savebox{\Vur}(60,24)[bl]{\multiput(0,0)(12,4.8){5}{\usebox{\wigur}}}
\savebox{\wigdr}(60,24)[bl]
   {\bezier {20}(0.0,0.0)(1.4,-5.2)(6.0,-2.4)
    \bezier {20}(6.0,-2.4)(10.6,-0.6)(12.0,-4.8)}
\savebox{\Vdr}(60,24)[bl]{\multiput(0,0)(12,-4.8){5}{\usebox{\wigdr}}}
\savebox{\wigr}(12,0)[bl]
   {\bezier{20}(0,0)(3, 4)(6,0)
    \bezier{20}(6,0)(9,-4)(12,0)}
\savebox{\Vr}(36,0)[bl]{\multiput(0,0)(12,0){3}{\usebox{\wigr}}}
\savebox{\wigu}(0,12)[bl]
   {\bezier{20}(0,0)( 4,3)(0,6)
    \bezier{20}(0,6)(-4,9)(0,12)}
\savebox{\Vu}(0,24)[bl]{\multiput(0,0)(0,12){2}{\usebox{\wigu}}}
\savebox{\Fu}(0,48)[bl]
{ \put(0,0){\vector(0,1){27}} \put(0,24){\line(0,1){24}} }
\savebox{\Fr}(36,0)[bl]
{ \put(0,0){\vector(1,0){20}} \put(18,0){\line(1,0){18}} }
\savebox{\Fl}(36,0)[bl]
{ \put(36,0){\vector(-1,0){20}} \put(18,0){\line(-1,0){18}} }
\savebox{\Fur}(36,12)[bl]
{ \put(0,0){\vector(3,1){20}} \put(18,6){\line(3,1){18}} }
\savebox{\Fdr}(36,12)[bl]
{ \put(36,0){\vector(-3,1){21}} \put(18,6){\line(-3,1){18}} }
\savebox{\Lur}(36,12)[bl]
{ \put(0,0){\line(3,1){20}} \put(18,6){\line(3,1){18}} }
\savebox{\Ldr}(36,12)[bl]
{ \put(36,0){\line(-3,1){21}} \put(18,6){\line(-3,1){18}} }
\bma
\barr{lllll}
\begin{picture}(180,80)
\put(120,36){$\gamma$}
\put(48,36){\circle*{4}}
\put(84,36){\circle*{4}}
\put(114,48){\circle*{4}}
\put(114,24){\circle*{4}}
\put(144,12){\circle*{4}}
\put(144,60){\circle*{4}}
\put(12,60){\line(3,-2){36}}
\put(12,12){\line(3, 2){36}}
\put(30,24){\vector( 3, 2){3}}
\put(30,48){\vector(-3,2){3}}
\put(48,34){\usebox{\Vr}}
\put(84,36){\usebox{\Vur}}
\put(84,12){\usebox{\Vdr}}
\put(114,24){\usebox{\Vu}}
\put(144,60){\usebox{\Fur}}
\put(144,48){\usebox{\Fdr}}
\put(144,12){\usebox{\Fur}}
\put(144,00){\usebox{\Fdr}}
\end{picture}
&\qquad&
\begin{picture}(120,80)
\savebox{\Vu}(0,48)[bl]{\multiput(0,0)(0,12){4}{\usebox{\wigu}}}
\savebox{\Vr}(60,0)[bl]{\multiput(0,0)(12,0){5}{\usebox{\wigr}}}
\put(84,36){$\gamma$}
\put(48,12){\circle*{4}}
\put(48,60){\circle*{4}}
\put(78,60){\circle*{4}}
\put(78,12){\circle*{4}}
\put(108,12){\circle*{4}}
\put(108,60){\circle*{4}}
\put(12,12){\usebox{\Fr}}
\put(12,60){\usebox{\Fl}}
\put(48,12){\usebox{\Fu}}
\put(48,10){\usebox{\Vr}}
\put(48,58){\usebox{\Vr}}
\put(108,60){\usebox{\Fur}}
\put(108,48){\usebox{\Fdr}}
\put(108,12){\usebox{\Fur}}
\put(108,00){\usebox{\Fdr}}
\put(78,12){\usebox{\Vu}}
\end{picture}
\earr
\ema
\caption{Diagrams that contribute to the Coulomb singularity.}
\label{FIcouldia}
\efi
The
gauge invariance of the coefficients of these two scalar functions
allows us to include the width in a gauge-invariant way.
Thus one obtains a gauge-invariant correction factor to the
lowest-order \cs\ resulting from the doubly-resonant diagrams:
\newcommand{\sigborn}{\sigma_{\mathrm{Born}}}
\beq   \label{CS}
  \sigma_{\mathrm{Coul}} = \sigma_{\born}^{\tt CC3}\,\frac{\alpha\pi}{2\bebar}
    \left[ 1 - \frac{2}{\pi}\,\arctan\left( \frac{|\be_M+\Delta|^2-\bebar^2}
    {2\bebar\Im\be_M}\right) \right],
\eeq
with
\beqar \label{shorth}
  \bebar &=& \frac{1}{s}\,\sqrt{s^2-2s(\pWp^2+\pWm^2)+(\pWp^2-\pWm^2)^2}
       \nlc[1ex]
  \be_M &=& \sqrt{1-4M^2/s},
     \qquad M^2 = \MW^2 - i\MW\GW -i\epsilon  \nlc[1ex]
  \De &=& \frac{|\pWp^2-\pWm^2|}{s}  ,
\eeqar
and $-\pi/2 < \arctan{y} < \pi/2$. Here $\bebar$ is the average
velocity of the \PW~bosons in their centre-of-mass system.
Equation~(\ref{CS}) only refers to the Coulomb singularity; the
finite remnants of the above mentioned three-point function,
containing also the IR divergence, have been left out.
{}From \refeq{CS} we obtain the usual on-shell Coulomb singularity
for stable \PW\ bosons and $\bebar \neq 0$ by first going on-shell,
\ie $\Delta=0$ and $\bebar^2=\Re\be_M^2=\be^2$,
and subsequently taking the limit $\GW \to 0$. Note that in this
limit the otherwise negligible $i\eps$ in $M^2$ becomes relevant.
 
In contrast to the on-shell case there are various effects present
in \refeq{CS} that
effectively truncate the range of the Coulomb interaction.
The presence of a Coulomb singularity requires that $\bebar$ be at least
of the same order of magnitude as $\Delta$ and
$|\be_M|$, which is bounded by the \PW\ width
according to $|\be_M| \gsim \sqrt{\GW/\MW}$.
This confirms the intuitive argument \cite{Fa93c} that
the Coulomb singularity is modified
substantially by finite-width effects if the characteristic time
of the Coulomb interaction ($t_{\mathrm{Coul}} \sim 1/[\bebar^2\MW]$)
is of the same order as or larger than the typical decay time
($t_{\tau} \sim 1/\GW$) of the \PW\ bosons.
While the effect of the finite \PW~width is contained in $\be_M$, the
off-shellness shows up in $\Delta$ and in the difference
$|\be_M^2-\bebar^2|$, which effectively involves
$|\pWp^2+\pWm^2-2\MW^2|/s$.
This is in agreement with the argument \cite{Coulomb} that
the off-shellness of the virtual \PW~bosons
will affect the Coulomb singularity, if the typical times
($t_{\mathrm{OFS}}^\pm
\sim {1/|\pW_{\pm,0}-\sqrt{\MW^2+\vec{\pW}^2_\pm}|}
\sim {1/|\sqrt{\pWpm^2}-\MW|}$)
for which off-shell $\PW^{\pm}$ bosons with four-momenta $\pWpm$
exist is of the same order 
as or smaller than $t_{\mathrm{Coul}}$. It should be noted 
that as a result of the convolution with the Breit-Wigner functions from
the \PW\ resonances, the quantity $\Delta$ plays a negligible role in an 
adequate description of the Coulomb phenomenon \cite{Coulomb}. Consequently the
at first sight deviating formulae of \cite{Coulomb} and \cite{CFK95}
constitute equally well-justified representations of this phenomenon.
 
In the case of the total {\tt CC3} \cs, the Coulomb singularity gives rise to 
a correction factor which reaches its maximal value of $\sim5.7\%$ at the 
nominal threshold and drops smoothly below and above threshold. 
While it amounts to
2.4\% at $\sqrt{s}=176\GeV$ it is only 1.8\% at $190\GeV$.
The corresponding effect on the reconstructed \PW~mass, resulting from
the pronounced energy dependence, is a
shift at the level of 5--10\MeV. Because of the fact that the Coulomb 
singularity is screened by the finite width of the \PW~bosons and 
the off-shell effects, higher-order Coulomb corrections are not important 
for off-shell \PW-pair production \cite{Coulomb}--\cite{KS95}.
Moreover, bound states of the two \PW~bosons do not have the
time to form because of the finite-width effects. The typical
time scale needed for the formation of a bound state
$t_{\mathrm{form}} \sim 1/(\al^2\MW)\approx 234\GeV^{-1}$ is much larger 
than the typical decay time $t_\tau = 1/\GW\approx 0.5\GeV^{-1}$.

\subsubsection{The hard-photon process}
\label{SEofph}

As mentioned before, in \PW-pair production the distinction between 
initial- and final-state 
radiation is not unique, unlike in the case of neutral particles such 
as the \PZ\ boson. In the matrix element, the universal leading 
logarithmic parts are easily separable, but the non-universal finite terms do 
not split naturally. An accurate calculation of the photon spectrum, beyond 
the leading logarithmic approximation, thus has to take into account 
initial-state radiation, final-state radiation, 
radiation off the \PW\ bosons, and various interference effects. These 
calculations have been performed for the final state consisting of 4 fermions 
and one photon \cite{Ae91}--\cite{KEKhardphoton}.  
However, the forward emission of many photons, described well by the structure
functions and parton-shower algorithms described in \refse{SEofrcpc}
and in \refapp{app}, is not included in these 
calculations. Recently efforts have been made 
to combine the two approaches \cite{GJhard,KEKphoton}. We give here the most 
salient points of these papers.

First we discuss the one-photon matrix element. This contains 
all graphs with two resonant \PW\ 
propagators, including radiation off the \PW\ bosons and the four-vector-boson 
interaction.
The resulting non-leading terms are negative and tend to decrease the energy 
lost in initial-state radiation. On top of that one can also include the 
non-doubly-resonant graphs. The universal ones that contribute for all 
channels are totally negligible \cite{Ae91,WWF}, just as in the non-radiative 
process, whereas the $t$-channel graphs show the expected collinear 
$\log(1\!-\!\cos\theta)$ behaviour.

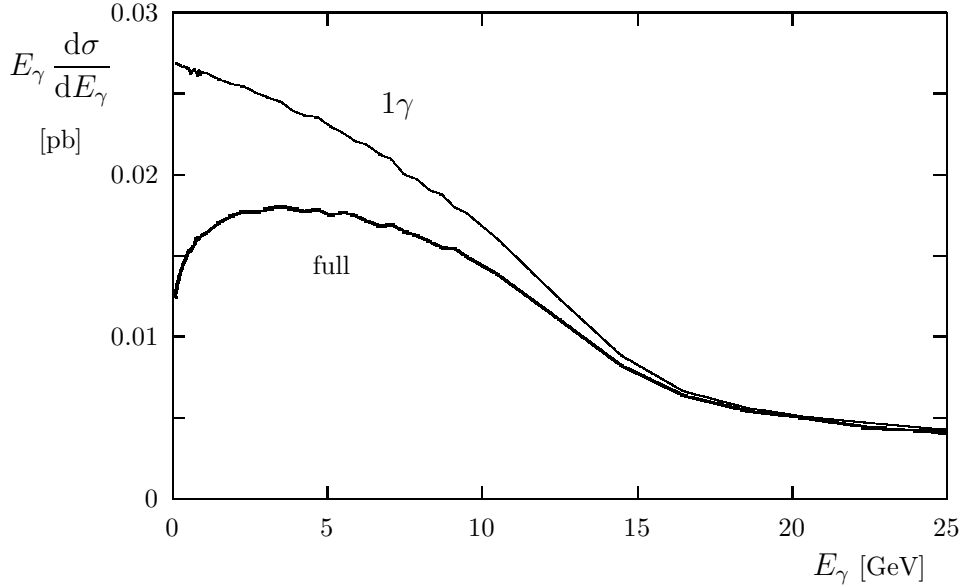
\begin{figure}
\setlength{\unitlength}{0.240900pt}
\ifx\plotpoint\undefined\newsavebox{\plotpoint}\fi
\sbox{\plotpoint}{\rule[-0.200pt]{0.400pt}{0.400pt}}%
\begin{picture}(1500,900)(-220,0)
\font\gnuplot=cmr10 at 10pt
\gnuplot
\sbox{\plotpoint}{\rule[-0.200pt]{0.400pt}{0.400pt}}%
\put(220.0,113.0){\rule[-0.200pt]{292.934pt}{0.400pt}}
\put(220.0,113.0){\rule[-0.200pt]{0.400pt}{184.048pt}}
\put(220.0,113.0){\rule[-0.200pt]{4.818pt}{0.400pt}}
\put(198,113){\makebox(0,0)[r]{0}}
\put(1416.0,113.0){\rule[-0.200pt]{4.818pt}{0.400pt}}
\put(220.0,240.0){\rule[-0.200pt]{4.818pt}{0.400pt}}
\put(1416.0,240.0){\rule[-0.200pt]{4.818pt}{0.400pt}}
\put(220.0,368.0){\rule[-0.200pt]{4.818pt}{0.400pt}}
\put(198,368){\makebox(0,0)[r]{0.01}}
\put(1416.0,368.0){\rule[-0.200pt]{4.818pt}{0.400pt}}
\put(220.0,495.0){\rule[-0.200pt]{4.818pt}{0.400pt}}
\put(1416.0,495.0){\rule[-0.200pt]{4.818pt}{0.400pt}}
\put(220.0,622.0){\rule[-0.200pt]{4.818pt}{0.400pt}}
\put(198,622){\makebox(0,0)[r]{0.02}}
\put(1416.0,622.0){\rule[-0.200pt]{4.818pt}{0.400pt}}
\put(220.0,750.0){\rule[-0.200pt]{4.818pt}{0.400pt}}
\put(1416.0,750.0){\rule[-0.200pt]{4.818pt}{0.400pt}}
\put(220.0,877.0){\rule[-0.200pt]{4.818pt}{0.400pt}}
\put(198,877){\makebox(0,0)[r]{0.03}}
\put(1416.0,877.0){\rule[-0.200pt]{4.818pt}{0.400pt}}
\put(220.0,113.0){\rule[-0.200pt]{0.400pt}{4.818pt}}
\put(220,68){\makebox(0,0){0}}
\put(220.0,857.0){\rule[-0.200pt]{0.400pt}{4.818pt}}
\put(463.0,113.0){\rule[-0.200pt]{0.400pt}{4.818pt}}
\put(463,68){\makebox(0,0){5}}
\put(463.0,857.0){\rule[-0.200pt]{0.400pt}{4.818pt}}
\put(706.0,113.0){\rule[-0.200pt]{0.400pt}{4.818pt}}
\put(706,68){\makebox(0,0){10}}
\put(706.0,857.0){\rule[-0.200pt]{0.400pt}{4.818pt}}
\put(950.0,113.0){\rule[-0.200pt]{0.400pt}{4.818pt}}
\put(950,68){\makebox(0,0){15}}
\put(950.0,857.0){\rule[-0.200pt]{0.400pt}{4.818pt}}
\put(1193.0,113.0){\rule[-0.200pt]{0.400pt}{4.818pt}}
\put(1193,68){\makebox(0,0){20}}
\put(1193.0,857.0){\rule[-0.200pt]{0.400pt}{4.818pt}}
\put(1436.0,113.0){\rule[-0.200pt]{0.400pt}{4.818pt}}
\put(1436,68){\makebox(0,0){25}}
\put(1436.0,857.0){\rule[-0.200pt]{0.400pt}{4.818pt}}
\put(220.0,113.0){\rule[-0.200pt]{292.934pt}{0.400pt}}
\put(1436.0,113.0){\rule[-0.200pt]{0.400pt}{184.048pt}}
\put(220.0,877.0){\rule[-0.200pt]{292.934pt}{0.400pt}}
\put(45,750){\makebox(0,0){\shortstack{\shortstack{$\displaystyle 
             E_\gamma \,\frac{\d\sigma}{\d E_\gamma}$\\[2mm]\relax[pb]}}}}
\put(1315,3){\makebox(0,0){$E_\gamma$ [GeV]}}
\put(220.0,113.0){\rule[-0.200pt]{0.400pt}{184.048pt}}
\sbox{\plotpoint}{\rule[-0.400pt]{0.800pt}{0.800pt}}%
\put(500,480){\makebox(0,0)[r]{full}}
\put(225,429){\usebox{\plotpoint}}
\put(224.34,429){\rule{0.800pt}{3.132pt}}
\multiput(223.34,429.00)(2.000,6.500){2}{\rule{0.800pt}{1.566pt}}
\put(226.34,442){\rule{0.800pt}{2.409pt}}
\multiput(225.34,442.00)(2.000,5.000){2}{\rule{0.800pt}{1.204pt}}
\put(228.34,452){\rule{0.800pt}{1.927pt}}
\multiput(227.34,452.00)(2.000,4.000){2}{\rule{0.800pt}{0.964pt}}
\put(230.34,460){\rule{0.800pt}{1.686pt}}
\multiput(229.34,460.00)(2.000,3.500){2}{\rule{0.800pt}{0.843pt}}
\put(232.34,467){\rule{0.800pt}{1.445pt}}
\multiput(231.34,467.00)(2.000,3.000){2}{\rule{0.800pt}{0.723pt}}
\put(234.34,473){\rule{0.800pt}{1.686pt}}
\multiput(233.34,473.00)(2.000,3.500){2}{\rule{0.800pt}{0.843pt}}
\put(236.34,480){\rule{0.800pt}{0.964pt}}
\multiput(235.34,480.00)(2.000,2.000){2}{\rule{0.800pt}{0.482pt}}
\put(238.34,484){\rule{0.800pt}{1.204pt}}
\multiput(237.34,484.00)(2.000,2.500){2}{\rule{0.800pt}{0.602pt}}
\put(240.34,489){\rule{0.800pt}{1.204pt}}
\multiput(239.34,489.00)(2.000,2.500){2}{\rule{0.800pt}{0.602pt}}
\put(242.34,494){\rule{0.800pt}{0.964pt}}
\multiput(241.34,494.00)(2.000,2.000){2}{\rule{0.800pt}{0.482pt}}
\put(244.34,498){\rule{0.800pt}{1.445pt}}
\multiput(243.34,498.00)(2.000,3.000){2}{\rule{0.800pt}{0.723pt}}
\put(247,501.34){\rule{0.482pt}{0.800pt}}
\multiput(247.00,502.34)(1.000,-2.000){2}{\rule{0.241pt}{0.800pt}}
\put(249,501.34){\rule{0.482pt}{0.800pt}}
\multiput(249.00,500.34)(1.000,2.000){2}{\rule{0.241pt}{0.800pt}}
\put(250.34,504){\rule{0.800pt}{1.686pt}}
\multiput(249.34,504.00)(2.000,3.500){2}{\rule{0.800pt}{0.843pt}}
\put(253,509.84){\rule{0.482pt}{0.800pt}}
\multiput(253.00,509.34)(1.000,1.000){2}{\rule{0.241pt}{0.800pt}}
\put(255,510.84){\rule{0.241pt}{0.800pt}}
\multiput(255.00,510.34)(0.500,1.000){2}{\rule{0.120pt}{0.800pt}}
\put(255.34,513){\rule{0.800pt}{2.409pt}}
\multiput(254.34,513.00)(2.000,5.000){2}{\rule{0.800pt}{1.204pt}}
\put(258,520.34){\rule{0.482pt}{0.800pt}}
\multiput(258.00,521.34)(1.000,-2.000){2}{\rule{0.241pt}{0.800pt}}
\put(261.34,521){\rule{0.800pt}{0.964pt}}
\multiput(260.34,521.00)(2.000,2.000){2}{\rule{0.800pt}{0.482pt}}
\put(264,523.84){\rule{0.482pt}{0.800pt}}
\multiput(264.00,523.34)(1.000,1.000){2}{\rule{0.241pt}{0.800pt}}
\put(266,526.34){\rule{1.600pt}{0.800pt}}
\multiput(266.00,524.34)(3.679,4.000){2}{\rule{0.800pt}{0.800pt}}
\multiput(273.00,531.41)(0.626,0.507){25}{\rule{1.200pt}{0.122pt}}
\multiput(273.00,528.34)(17.509,16.000){2}{\rule{0.600pt}{0.800pt}}
\multiput(293.00,547.40)(0.888,0.512){15}{\rule{1.582pt}{0.123pt}}
\multiput(293.00,544.34)(15.717,11.000){2}{\rule{0.791pt}{0.800pt}}
\multiput(312.00,558.40)(1.614,0.526){7}{\rule{2.486pt}{0.127pt}}
\multiput(312.00,555.34)(14.841,7.000){2}{\rule{1.243pt}{0.800pt}}
\put(332,561.84){\rule{4.577pt}{0.800pt}}
\multiput(332.00,562.34)(9.500,-1.000){2}{\rule{2.289pt}{0.800pt}}
\multiput(351.00,564.38)(2.943,0.560){3}{\rule{3.400pt}{0.135pt}}
\multiput(351.00,561.34)(12.943,5.000){2}{\rule{1.700pt}{0.800pt}}
\put(371,568.34){\rule{4.000pt}{0.800pt}}
\multiput(371.00,566.34)(10.698,4.000){2}{\rule{2.000pt}{0.800pt}}
\put(390,568.34){\rule{4.200pt}{0.800pt}}
\multiput(390.00,570.34)(11.283,-4.000){2}{\rule{2.100pt}{0.800pt}}
\put(410,564.34){\rule{4.000pt}{0.800pt}}
\multiput(410.00,566.34)(10.698,-4.000){2}{\rule{2.000pt}{0.800pt}}
\put(429,563.84){\rule{4.818pt}{0.800pt}}
\multiput(429.00,562.34)(10.000,3.000){2}{\rule{2.409pt}{0.800pt}}
\multiput(449.00,565.08)(1.116,-0.516){11}{\rule{1.889pt}{0.124pt}}
\multiput(449.00,565.34)(15.080,-9.000){2}{\rule{0.944pt}{0.800pt}}
\multiput(468.00,559.38)(2.943,0.560){3}{\rule{3.400pt}{0.135pt}}
\multiput(468.00,556.34)(12.943,5.000){2}{\rule{1.700pt}{0.800pt}}
\multiput(488.00,561.06)(2.775,-0.560){3}{\rule{3.240pt}{0.135pt}}
\multiput(488.00,561.34)(12.275,-5.000){2}{\rule{1.620pt}{0.800pt}}
\multiput(507.00,556.08)(1.116,-0.516){11}{\rule{1.889pt}{0.124pt}}
\multiput(507.00,556.34)(15.080,-9.000){2}{\rule{0.944pt}{0.800pt}}
\multiput(526.00,547.08)(1.358,-0.520){9}{\rule{2.200pt}{0.125pt}}
\multiput(526.00,547.34)(15.434,-8.000){2}{\rule{1.100pt}{0.800pt}}
\put(546,540.34){\rule{4.577pt}{0.800pt}}
\multiput(546.00,539.34)(9.500,2.000){2}{\rule{2.289pt}{0.800pt}}
\multiput(565.00,541.08)(0.938,-0.512){15}{\rule{1.655pt}{0.123pt}}
\multiput(565.00,541.34)(16.566,-11.000){2}{\rule{0.827pt}{0.800pt}}
\multiput(585.00,530.07)(1.913,-0.536){5}{\rule{2.733pt}{0.129pt}}
\multiput(585.00,530.34)(13.327,-6.000){2}{\rule{1.367pt}{0.800pt}}
\multiput(604.00,524.08)(1.179,-0.516){11}{\rule{1.978pt}{0.124pt}}
\multiput(604.00,524.34)(15.895,-9.000){2}{\rule{0.989pt}{0.800pt}}
\multiput(624.00,515.08)(0.988,-0.514){13}{\rule{1.720pt}{0.124pt}}
\multiput(624.00,515.34)(15.430,-10.000){2}{\rule{0.860pt}{0.800pt}}
\put(643,504.84){\rule{4.818pt}{0.800pt}}
\multiput(643.00,505.34)(10.000,-1.000){2}{\rule{2.409pt}{0.800pt}}
\multiput(663.00,504.08)(0.740,-0.509){19}{\rule{1.369pt}{0.123pt}}
\multiput(663.00,504.34)(16.158,-13.000){2}{\rule{0.685pt}{0.800pt}}
\multiput(682.00,491.09)(0.882,-0.504){49}{\rule{1.600pt}{0.121pt}}
\multiput(682.00,491.34)(45.679,-28.000){2}{\rule{0.800pt}{0.800pt}}
\multiput(731.00,463.09)(0.684,-0.501){135}{\rule{1.293pt}{0.121pt}}
\multiput(731.00,463.34)(94.316,-71.000){2}{\rule{0.646pt}{0.800pt}}
\multiput(828.00,392.09)(0.674,-0.501){137}{\rule{1.278pt}{0.121pt}}
\multiput(828.00,392.34)(94.348,-72.000){2}{\rule{0.639pt}{0.800pt}}
\multiput(925.00,320.09)(1.039,-0.502){87}{\rule{1.851pt}{0.121pt}}
\multiput(925.00,320.34)(93.158,-47.000){2}{\rule{0.926pt}{0.800pt}}
\multiput(1022.00,273.09)(2.090,-0.504){41}{\rule{3.467pt}{0.122pt}}
\multiput(1022.00,273.34)(90.805,-24.000){2}{\rule{1.733pt}{0.800pt}}
\multiput(1120.00,249.08)(4.333,-0.511){17}{\rule{6.667pt}{0.123pt}}
\multiput(1120.00,249.34)(83.163,-12.000){2}{\rule{3.333pt}{0.800pt}}
\multiput(1217.00,237.09)(3.395,-0.508){23}{\rule{5.373pt}{0.122pt}}
\multiput(1217.00,237.34)(85.847,-15.000){2}{\rule{2.687pt}{0.800pt}}
\multiput(1314.00,222.07)(10.732,-0.536){5}{\rule{13.267pt}{0.129pt}}
\multiput(1314.00,222.34)(70.464,-6.000){2}{\rule{6.633pt}{0.800pt}}
\put(1412,215.34){\rule{5.782pt}{0.800pt}}
\multiput(1412.00,216.34)(12.000,-2.000){2}{\rule{2.891pt}{0.800pt}}
\put(260.0,521.0){\usebox{\plotpoint}}
\sbox{\plotpoint}{\rule[-0.200pt]{0.400pt}{0.400pt}}%
\put(600,730){\makebox(0,0)[r]{$1\gamma$}}
\put(225,798){\usebox{\plotpoint}}
\put(225,796.67){\rule{0.482pt}{0.400pt}}
\multiput(225.00,797.17)(1.000,-1.000){2}{\rule{0.241pt}{0.400pt}}
\put(227,795.67){\rule{0.482pt}{0.400pt}}
\multiput(227.00,796.17)(1.000,-1.000){2}{\rule{0.241pt}{0.400pt}}
\put(229,794.67){\rule{0.482pt}{0.400pt}}
\multiput(229.00,795.17)(1.000,-1.000){2}{\rule{0.241pt}{0.400pt}}
\put(231,793.67){\rule{0.482pt}{0.400pt}}
\multiput(231.00,794.17)(1.000,-1.000){2}{\rule{0.241pt}{0.400pt}}
\put(233,792.17){\rule{0.482pt}{0.400pt}}
\multiput(233.00,793.17)(1.000,-2.000){2}{\rule{0.241pt}{0.400pt}}
\put(235,790.67){\rule{0.482pt}{0.400pt}}
\multiput(235.00,791.17)(1.000,-1.000){2}{\rule{0.241pt}{0.400pt}}
\put(237,791.17){\rule{0.482pt}{0.400pt}}
\multiput(237.00,790.17)(1.000,2.000){2}{\rule{0.241pt}{0.400pt}}
\put(239.17,790){\rule{0.400pt}{0.700pt}}
\multiput(238.17,791.55)(2.000,-1.547){2}{\rule{0.400pt}{0.350pt}}
\put(241,789.67){\rule{0.482pt}{0.400pt}}
\multiput(241.00,789.17)(1.000,1.000){2}{\rule{0.241pt}{0.400pt}}
\put(243.17,788){\rule{0.400pt}{0.700pt}}
\multiput(242.17,789.55)(2.000,-1.547){2}{\rule{0.400pt}{0.350pt}}
\put(245,787.67){\rule{0.482pt}{0.400pt}}
\multiput(245.00,787.17)(1.000,1.000){2}{\rule{0.241pt}{0.400pt}}
\put(247.17,779){\rule{0.400pt}{2.100pt}}
\multiput(246.17,784.64)(2.000,-5.641){2}{\rule{0.400pt}{1.050pt}}
\put(249.17,779){\rule{0.400pt}{1.100pt}}
\multiput(248.17,779.00)(2.000,2.717){2}{\rule{0.400pt}{0.550pt}}
\put(251,784.17){\rule{0.482pt}{0.400pt}}
\multiput(251.00,783.17)(1.000,2.000){2}{\rule{0.241pt}{0.400pt}}
\put(253,786.17){\rule{0.482pt}{0.400pt}}
\multiput(253.00,785.17)(1.000,2.000){2}{\rule{0.241pt}{0.400pt}}
\put(254.67,784){\rule{0.400pt}{0.964pt}}
\multiput(254.17,786.00)(1.000,-2.000){2}{\rule{0.400pt}{0.482pt}}
\put(258.17,776){\rule{0.400pt}{1.700pt}}
\multiput(257.17,780.47)(2.000,-4.472){2}{\rule{0.400pt}{0.850pt}}
\put(260.17,776){\rule{0.400pt}{2.300pt}}
\multiput(259.17,776.00)(2.000,6.226){2}{\rule{0.400pt}{1.150pt}}
\put(262.17,778){\rule{0.400pt}{1.900pt}}
\multiput(261.17,783.06)(2.000,-5.056){2}{\rule{0.400pt}{0.950pt}}
\put(264.17,778){\rule{0.400pt}{1.100pt}}
\multiput(263.17,778.00)(2.000,2.717){2}{\rule{0.400pt}{0.550pt}}
\put(266,781.67){\rule{1.686pt}{0.400pt}}
\multiput(266.00,782.17)(3.500,-1.000){2}{\rule{0.843pt}{0.400pt}}
\multiput(273.00,780.92)(1.017,-0.491){17}{\rule{0.900pt}{0.118pt}}
\multiput(273.00,781.17)(18.132,-10.000){2}{\rule{0.450pt}{0.400pt}}
\multiput(293.00,770.93)(1.408,-0.485){11}{\rule{1.186pt}{0.117pt}}
\multiput(293.00,771.17)(16.539,-7.000){2}{\rule{0.593pt}{0.400pt}}
\multiput(312.00,763.94)(2.821,-0.468){5}{\rule{2.100pt}{0.113pt}}
\multiput(312.00,764.17)(15.641,-4.000){2}{\rule{1.050pt}{0.400pt}}
\multiput(332.00,759.92)(0.964,-0.491){17}{\rule{0.860pt}{0.118pt}}
\multiput(332.00,760.17)(17.215,-10.000){2}{\rule{0.430pt}{0.400pt}}
\multiput(351.00,749.93)(1.286,-0.488){13}{\rule{1.100pt}{0.117pt}}
\multiput(351.00,750.17)(17.717,-8.000){2}{\rule{0.550pt}{0.400pt}}
\multiput(371.00,741.93)(1.666,-0.482){9}{\rule{1.367pt}{0.116pt}}
\multiput(371.00,742.17)(16.163,-6.000){2}{\rule{0.683pt}{0.400pt}}
\multiput(390.00,735.92)(0.668,-0.494){27}{\rule{0.633pt}{0.119pt}}
\multiput(390.00,736.17)(18.685,-15.000){2}{\rule{0.317pt}{0.400pt}}
\multiput(410.00,720.93)(1.408,-0.485){11}{\rule{1.186pt}{0.117pt}}
\multiput(410.00,721.17)(16.539,-7.000){2}{\rule{0.593pt}{0.400pt}}
\multiput(429.00,713.95)(4.258,-0.447){3}{\rule{2.767pt}{0.108pt}}
\multiput(429.00,714.17)(14.258,-3.000){2}{\rule{1.383pt}{0.400pt}}
\multiput(449.00,710.92)(0.680,-0.494){25}{\rule{0.643pt}{0.119pt}}
\multiput(449.00,711.17)(17.666,-14.000){2}{\rule{0.321pt}{0.400pt}}
\multiput(468.00,696.92)(1.017,-0.491){17}{\rule{0.900pt}{0.118pt}}
\multiput(468.00,697.17)(18.132,-10.000){2}{\rule{0.450pt}{0.400pt}}
\multiput(488.00,686.92)(0.734,-0.493){23}{\rule{0.685pt}{0.119pt}}
\multiput(488.00,687.17)(17.579,-13.000){2}{\rule{0.342pt}{0.400pt}}
\multiput(507.00,673.93)(1.666,-0.482){9}{\rule{1.367pt}{0.116pt}}
\multiput(507.00,674.17)(16.163,-6.000){2}{\rule{0.683pt}{0.400pt}}
\multiput(526.00,667.92)(0.668,-0.494){27}{\rule{0.633pt}{0.119pt}}
\multiput(526.00,668.17)(18.685,-15.000){2}{\rule{0.317pt}{0.400pt}}
\multiput(546.00,652.93)(1.220,-0.488){13}{\rule{1.050pt}{0.117pt}}
\multiput(546.00,653.17)(16.821,-8.000){2}{\rule{0.525pt}{0.400pt}}
\multiput(565.58,643.59)(0.496,-0.600){37}{\rule{0.119pt}{0.580pt}}
\multiput(564.17,644.80)(20.000,-22.796){2}{\rule{0.400pt}{0.290pt}}
\multiput(585.00,620.93)(1.408,-0.485){11}{\rule{1.186pt}{0.117pt}}
\multiput(585.00,621.17)(16.539,-7.000){2}{\rule{0.593pt}{0.400pt}}
\multiput(604.00,613.92)(0.588,-0.495){31}{\rule{0.571pt}{0.119pt}}
\multiput(604.00,614.17)(18.816,-17.000){2}{\rule{0.285pt}{0.400pt}}
\multiput(624.00,596.93)(1.408,-0.485){11}{\rule{1.186pt}{0.117pt}}
\multiput(624.00,597.17)(16.539,-7.000){2}{\rule{0.593pt}{0.400pt}}
\multiput(643.00,589.92)(0.498,-0.496){37}{\rule{0.500pt}{0.119pt}}
\multiput(643.00,590.17)(18.962,-20.000){2}{\rule{0.250pt}{0.400pt}}
\multiput(663.00,569.93)(1.077,-0.489){15}{\rule{0.944pt}{0.118pt}}
\multiput(663.00,570.17)(17.040,-9.000){2}{\rule{0.472pt}{0.400pt}}
\multiput(682.00,560.92)(0.597,-0.498){79}{\rule{0.578pt}{0.120pt}}
\multiput(682.00,561.17)(47.800,-41.000){2}{\rule{0.289pt}{0.400pt}}
\multiput(731.00,519.92)(0.516,-0.499){185}{\rule{0.513pt}{0.120pt}}
\multiput(731.00,520.17)(95.936,-94.000){2}{\rule{0.256pt}{0.400pt}}
\multiput(828.00,425.92)(0.545,-0.499){175}{\rule{0.536pt}{0.120pt}}
\multiput(828.00,426.17)(95.888,-89.000){2}{\rule{0.268pt}{0.400pt}}
\multiput(925.00,336.92)(0.868,-0.499){109}{\rule{0.793pt}{0.120pt}}
\multiput(925.00,337.17)(95.354,-56.000){2}{\rule{0.396pt}{0.400pt}}
\multiput(1022.00,280.92)(1.903,-0.497){49}{\rule{1.608pt}{0.120pt}}
\multiput(1022.00,281.17)(94.663,-26.000){2}{\rule{0.804pt}{0.400pt}}
\multiput(1120.00,254.92)(3.302,-0.494){27}{\rule{2.687pt}{0.119pt}}
\multiput(1120.00,255.17)(91.424,-15.000){2}{\rule{1.343pt}{0.400pt}}
\multiput(1217.00,239.93)(5.621,-0.489){15}{\rule{4.411pt}{0.118pt}}
\multiput(1217.00,240.17)(87.845,-9.000){2}{\rule{2.206pt}{0.400pt}}
\multiput(1314.00,230.93)(5.679,-0.489){15}{\rule{4.456pt}{0.118pt}}
\multiput(1314.00,231.17)(88.752,-9.000){2}{\rule{2.228pt}{0.400pt}}
\put(1412,221.67){\rule{5.782pt}{0.400pt}}
\multiput(1412.00,222.17)(12.000,-1.000){2}{\rule{2.891pt}{0.400pt}}
\put(256.0,784.0){\rule[-0.200pt]{0.482pt}{0.400pt}}
\end{picture}
\caption{Photon energy spectrum in the one-photon approximation ($1\ga$) and 
the same combined with resummed structure functions (full) for the 
reaction $\Pep\Pem\to \mu^+\Pnm\tau^-\bar{\nu}_\tau$.}
\label{fig:photenergy}
\end{figure}

The next step is to combine this hard matrix element, which is needed for 
large angles, with the resummed leading-log structure function \cite{GJhard} 
or parton shower \cite{KEKphoton}, which gives a good description of 
(multiple-) photon radiation at small angles. This is done by using the
exact matrix element (convoluted with initial-state radiation) outside a cone
\cite{GJhard} around the incoming electrons and positrons%
\footnote{Instead of a cone also a cut-off on the 
$\Pe^\pm_{\mathrm{in}}$--$\ga$ virtuality can be used to define the region 
of multiple-photon radiation \cite{KEKphoton}.}. 
Inside this cone one subtracts the leading logarithmic contribution from the 
hard matrix element, adds the tree-level matrix element, and subsequently
convolutes everything with initial-state radiation.
Of course the structure function or parton shower has to be restricted to 
radiate within the cone or virtuality cut-off. In the former case the scale
$Q^2=s\,(1\!-\!\cos\theta_c)/2$ is used, with $\theta_c$ defining the cone.

In order to compare this approach with the one-photon matrix element on the 
one hand, and a purely leading-log description on the other, we have to define 
some kinematical observables. The most relevant ones for the \PW-mass 
measurement are the observable and unobservable photon energies.  We define 
these with respect to the ADLO/TH set of canonical cuts defined in 
\cite{EGgroup}. A photon which passes all cuts is called `observable', 
and one that is combined with one of the beams is `unobservable'.  
Photons close to final-state particles are not counted either way. 
The unobservable photon energy is the quantity most relevant to 
the \PW-mass reconstruction without explicit photons, whereas the observable 
photon energy gives an indication how well the resummed leading-log and 
one-photon matrix elements describe large-angle radiation.

In order to get sensible results for these photon energies we will have to 
include an estimate of the unresummed soft and virtual corrections, which are 
not yet fully known. These corrections are only needed inside the cone as a
weight of the structure function. As such the corresponding uncertainty 
partially cancels out in the average photon energies 
$\Eg = \int\d E_\ga\, (\d\si/\d E_\ga)\,E_\ga/\si$. In the actual analysis
presented below we circumvent the problem by leaving out the virtual 
corrections and tuning the non-leading-log part
of the soft corrections (through the IR regulator mass $\la_{\mathrm{IR}}$) 
in such a way that it cancels in the
total \cs\ against the non-leading-log part of the hard-photon corrections.
Consequently, the resulting total \cs\ is simply given by the lowest-order 
\cs\ convoluted with leading logarithmic structure functions. As
the initial--final and final--final interference is expected to be of order 
$\alpha\GW/\MW$ (see \refse{SEofrcps}), this is a reasonable approximation. 

The results for the structure-function 
algorithm of \cite{GJhard} are given in \refta{tab:hardphoton} for 
leptonic, semi-leptonic, and hadronic final states%
\footnote{It should be noted that only the universal background diagrams
were taken into account.}. For the cone 
$\theta_c = 10^\circ$ was chosen, but the results do not depend strongly on 
this parameter ($\theta_c = 5^\circ$ only shifts the values by about 1\%).
\begin{table}
\begin{center}
$\displaystyle
\begin{array}{||r@{\;\;}|@{\;\;}c|c|c||}
\hline\hline
& \mbox{\ \ leptonic\ \ } & \mbox{\ semi-leptonic\ } & \mbox{\ \ hadronic\ \ }
\\ \hline
\sigma_0~[{\rm{pb}}] & 0.712 & 4.485 & 7.024 \\
\sigma_0^{\rm{isr}}~[{\rm{pb}}] & 0.594 & 3.732 & 5.845 \\
\sigma_{0+\rm{bkg}}^{\rm{isr}}~[{\rm{pb}}] & 0.595 & 3.736 & 5.853 \\
\hline
\sigma^{\rm{obs}}/\sigma_0^{\rm{isr}}& & & \\
\mbox{full} & 0.275 & 0.249 & 0.229 \\
1\gamma & 0.365 & 0.328 & 0.302 \\
\mbox{LL} & 0.286 & 0.260 & 0.232 \\
\mbox{with backgr} & 0.275 & 0.250 & 0.229 \\
\hline
\langle E_\gamma^{\rm{obs}}\rangle \mbox{ [GeV]} & & & \\
\mbox{full} & 1.227& 0.995 & 0.760 \\
1\gamma & 1.437 & 1.167 & 0.899 \\
\mbox{LL} & 1.238 & 1.028 & 0.801 \\
\mbox{with backgr} & 1.226 & 0.994 & 0.759 \\
\hline
\ \langle E_\gamma^{\rm{unobs}}\rangle \mbox{ [GeV]} & & & \\
\mbox{full} & 0.676 & 0.664 & 0.645 \\
1\gamma & 0.878 & 0.873 & 0.872 \\
\mbox{LL} & 0.666 & 0.667 & 0.664 \\
\mbox{with backgr} & 0.676 & 0.664 & 0.644 \\
\hline\hline
\end{array}$\end{center}
\caption[]{Cross-section for observable 
photons and the energy lost to observable and 
unobservable photons at $\sqrt{s} = 175$ GeV.}
\label{tab:hardphoton}
\end{table}
In \refta{tab:hardphoton} we first give the observable tree-level 
non-radiative doubly-resonant \cs\ ($\sigma_0$), the same 
convoluted with leading-log structure functions ($\sigma_0^{\rm{isr}}$), and 
with in addition the universal non-doubly-resonant (background) graphs 
($\sigma^{\rm{isr}}_{0+\rm{bkg}}$). For the next 
entries we consider the full calculation described above, the exact one-photon 
matrix element ($1\gamma$), and the leading-log result (LL). We also 
give the full result including the universal non-doubly-resonant diagrams 
(full$+$backgr). The statistical errors on the \css\ are 
${\cal O}(0.1\%)$, but the differences $(\mbox{LL} - \mbox{full})$ and 
$((\mbox{res}+\mbox{backgr}) - \mbox{res})$ were computed directly in this 
form and have relative errors of a few per cent and a few tens of per cent, 
respectively, on these differences.  
The statistical errors on the average photon energies are slightly larger,
0.3--0.5\%. Note that we introduced an upper cut-off on 
the photon energy to avoid the \PZ\ peak. This influences only the observable 
average energy.

One can see that an appreciable fraction (around one quarter) 
of the events will be accompanied by photons observable in the ADLO/TH 
set of cuts.  This is due to the excellent forward coverage ($\theta_\gamma > 
1^\circ$) and electromagnetic calorimeter ($E_\gamma>0.1$ GeV) assumed in the 
canonical cuts.  Reducing the angle to $10^\circ$ this fraction still is 
around 20\%, half of which also has an $E_\gamma > 1$~GeV. 
Neither the $1\gamma$ matrix 
element nor the leading-log approximation give a satisfactory description of 
the observable photons. By radiating only one photon one misses the
additional convolution with the structure functions, which scales down the
one-photon \cs\ as a result of the large, negative soft-photon effects. On the 
other hand, the leading-log approximation misses the negative effects from 
initial--final state interference and the radiation off 
the intermediate \PW\ bosons. Finally, given that the 
un-exponentiated large-angle contribution of the \cs\ still is 20\% 
of the total \cs, the $E_\ga$ spectrum in the region 
$E_\gamma < 1$ GeV should not be trusted 
to 1\%, even not in the full calculation. Most of the \cs\ here is, 
however, associated with the final state and hence does not influence the  
\PW-mass measurement.

The observable photon energy is dominated by final-state radiation and hence 
not very interesting. The unobservable energy spectrum is much more 
independent of the specific final state. It is not completely independent due 
to the possibility of observable jets in the beam pipe (there is no angular 
cut on jets in the canonical cuts). The initial-state radiation associated 
with these events is sometimes combined with the final state by the canonical 
cuts, thus lowering the average energy. As the radiation off jets is not 
modelled correctly anyway, a jet-angle cut will have to 
be imposed to exclude this contamination. One sees that analysing observable 
photons separately reduces the average energy loss, and hence the size of the 
theoretical corrections to be applied to the fitted \PW-mass.
For the unobservable radiation the leading-log approximation is, as 
expected, quite good. Like in the total \cs, the contribution 
of the universal non-doubly-resonant graphs is negligible.

We conclude that the one-photon matrix element does not describe observable 
photons or unobservable photons well, whereas a leading logarithmic 
description is not accurate enough for large-angle photons, as it does not 
include the negative contributions associated with interference terms and 
radiation off the \PW\ bosons.  
Furthermore, a separate analysis of the events with an observable photon 
(around one quarter of all events with the canonical cuts) reduces the 
average energy loss considerably, thus reducing the uncertainties coming from 
this theoretical input.

\subsubsection{General approach to radiative corrections within the
pole scheme}
\label{SEofrcps}

Finally, we discuss the way in which a reasonable approximation to the full 
radiative corrections to the process \eeffff\ can be computed.   
We do not consider QCD corrections. The perturbative QCD corrections to the 
final state are well-understood, as the main part is confined to the decay of 
a single \PW, which should be similar to the hadronic \PZ\ decays studied at 
LEP1. Perturbative QCD interference effects between the decay products of 
different \PW\ bosons are suppressed by a factor $\als^2/(N_{\mathrm{C}}^2-1)$.
The related non-perturbative interference effects associated with the 
hadronization are discussed in \refse{SEofrec}.

An electroweak \Oa\ calculation would consist of the traditional three parts: 
one-loop graphs, soft-photon bremsstrahlung, and hard-photon bremsstrahlung. 
Only the sum is IR finite. The separation between hard and soft radiation is, 
as usual, defined with respect to a photon-energy cut-off 
$E_{\mathrm{min}}^\ga$. Because of the finite \PW\ width one either has to use
$E_{\mathrm{min}}^\ga \ll \GW$ 
to ensure the validity of the soft-photon approximation in the 
soft-bremsstrahlung integrals (i.e., neglecting the effect of the photon 
momentum on the phase space and on the non-IR-singular parts of the matrix 
element), or one has to take into account finite photon energies in the 
Breit-Wigner resonances. Close to threshold, moreover, the limited amount of 
available phase space demands $E_{\mathrm{min}}^\ga \ll \MW\be^2$. 

The soft-bremsstrahlung integrals factorize into a simple 
multiplicative factor and the Born amplitude, and can easily be added, either 
using a `fixed-width scheme' or the `fermion-loop scheme'.
Hard-photon radiation (for an arbitrary cut-off) has already been addressed
in \refse{SEofph}. 

What remains are the virtual corrections. However, the full 
one-loop calculation of the virtual diagrams appears daunting.  For the most 
simple final state, $\Pep\Pem\to\mu^+\Pnm \bar{\Pu}\Pd$, there are 
3579 Standard-Model Feynman diagrams for massless fermions. This increases to 
7158 when one electron is included in the final state, and reaches 15948 for 
the most complicated final state $\Pep\Pem\Pne\bar{\Pne}$. However, not all
contributions are equally important. For instance the numerical significance 
of the non-doubly-resonant (background) diagrams is small in the Born and 
hard-photon calculations, especially if one requires that the event resembles 
\PW-pair production. So, we can assume that the one-loop 
corrections to these diagrams are even smaller. 
The problem then amounts to achieve a clean separation of the radiative 
corrections to \PW-pair production

One way to achieve this separation is the `pole scheme' introduced in 
\refse{SEoflogi} for tree-level matrix elements. There have been various 
attempts to define one-loop corrections in this scheme 
\cite{Ve63}--\cite{Ae94}, which differ considerably. Since sofar no actual 
calculation for \PW-pair production has been completed in such a scheme, we 
restrict ourselves to a few comments.
The idea behind the `pole scheme' is to include a minimal part of the 
higher-order corrections needed to generate a finite width.  By making a 
systematic expansion both in the coupling parameter $\alpha$ and in the 
width $\Gamma$ ($\propto\alpha$), one can identify gauge-invariant 
contributions of progressively smaller influence on the final result.  
We will roughly sketch the method for a single, neutral particle.
For factorizable diagrams, \ie diagrams that factorize into corrections to 
the production, propagation, and decay of the unstable particles, 
the amplitude to all orders can be Dyson resummed as follows \cite{St91,HVe92}
\begin{eqnarray}
\label{eq:Ainffact}
    \M^\infty_{\rm fact} & = & \frac{W(p^2,\theta)}{p^2-m^2}\,\sum_{n=0}^\infty
        \left(\frac{\Sigma(p^2)}{p^2-m^2}\right)^n           
                           = \frac{W(p^2,\theta)}{p^2-m^2-\Sigma(p^2)}
        \nonumber\\          
                        & = & \left[\frac{W(p^2,\theta)}{p^2-m^2-\Sigma(p^2)} 
        - \frac{W(M^2,\theta)}{p^2-M^2}\,\frac{1}{1-\Sigma'(M^2)} \right] 
        + \frac{W(M^2,\theta)}{p^2-M^2}\frac{1}{1-\Sigma'(M^2)},
\end{eqnarray}
where $m$ denotes the real mass, and $M$ the complex mass given by 
$M^2 - m^2 - \Sigma(M^2) = 0$. The corrections to production and decay are 
contained in the function $W$, and $\Sigma(p^2)$ is the 
one-particle-irreducible self-energy. The variable $\theta$ in the  
vertex-correction function $W$ stands for other variables, which have to be 
chosen sensibly (see \refse{SEoflogi}). 
The first term of \refeq{eq:Ainffact} does not have 
a pole and the residue at the pole $p^2=M^2$ of the last term is 
gauge-invariant. In \cite{Ae94} it has been shown how to derive the 
single-pole residue \,$W(M^2,\theta)/[1-\Sigma'(M^2)]$\, and the non-pole 
terms. The former consists essentially of the on-shell amplitude plus some 
terms of $\O(\Gamma)$. It should, however, be noted that this type of
decomposition only works when the on-shell limit exists. So, for instance
below the \PW-pair production threshold the `pole-scheme' method makes no 
sense, and as a result of that the procedure is not reliable just above 
threshold.  
 
When one turns to the production of charged unstable particles, additional
problems arise related to the infra-red divergences that occur in the
limit $\Gamma \to 0$. Again these problems can be overcome \cite{Ae94,BD94}.

So far the above discussion only took into account factorizable diagrams. 
There are also non-factorizable diagrams, \eg diagrams in which a photon 
connects the initial and final state or the decay products of different
unstable particles. Those non-factorizable diagrams can give rise to 
double-pole contributions when the virtual photon becomes soft. Combined with 
the related soft bremsstrahlung it has been shown that the non-factorizable 
double-pole contributions cancel up to order $\alpha\Gamma/m$ in the fully 
inclusive total \cs\ \cite{Fa93,Me95}. For sufficiently exclusive 
distributions this is in general not the case \cite{BD94,Me95,Me93}.

\subsection{Reconnection effects}
\label{SEofrec}
 
Nearly half of all \PW\ pairs decay hadronically. In the LEP2
energy range the average space--time distance between the \PWp\
and \PWm\ decay vertices is smaller than 0.1\,fm, \ie less than a
typical hadronic size of 1\,fm. Therefore the fragmentation of the
\PWp\ and the \PWm\ may not be independent, and this could
influence the \PW-mass reconstruction. Here a short summary
of the problem is given in a historical order.
More can be found in the report of the \PW-mass group \cite{Wmassgroup}.
 
The problem is related to two different physical effects:
colour reconnections and Bose--Einstein effects.
 
The colour-reconnection problem was first pointed out in
\cite{GuPeZe} using the string-model approximation.
Assuming that the \PW\ pair decays into two
quark-antiquark pairs $q\bar{q}$ and $Q\bar{Q}$, respectively,
it can either fragment into two strings stretched between $q\bar{q}$
and $Q\bar{Q}$, or into two strings stretched between $q\bar{Q}$ and
$\bar{q}Q$. In the three-colour world the probability for the second
configuration was taken to be 1/9 in a very crude approximation.
The properties of these two configurations
are very different. While the first one gives `classical' event
multiplicities and a flat rapidity distribution, the second one has
a small multiplicity and the particles are grouped at large rapidity
values \cite{GuPeZe}.
 
The fragmentation of the initial $q\bar{q}$ pairs into hadrons is
conventionally described in terms of a perturbative parton cascade
followed, at a later stage, by a non-perturbative hadronization
phase. In perturbative QCD (PQCD) the influence of one
W fragmentation on the other one is called `interconnection'.
Here the timing is very important \cite{Wmassgroup,KhSj}: the gluon-emission 
time of high-energy gluons, $\tau_g \sim 1/E_g$, is shorter than
the \PW\ lifetime, $\tau_\PW \sim 1/\GW \approx 0.5\GeV^{-1}$.
Therefore the gluons with $E_g\gg\GW$ are emitted independently from the 
$q\bar{q}$ and $Q\bar{Q}$ systems, while gluons with $E_g\leq\GW$ may feel 
four colour charges ($\Rightarrow$ PQCD interference) \cite{KhSj}. 
The non-perturbative hadronization (at distances 
$\sim$ 1\,fm) is NOT independent.
The influence of PQCD interference is shown to be not very important 
\cite{KhSj,Ac95}. It is suppressed by a colour factor
$1/(N^2_{\mathrm{C}}-1) = 1/8$ as compared to the total rate of double
primary gluon emission and, in addition, only gluons with
$E_g\leq\GW$ are concerned; thus the total suppression factor
is about 100 compared to a naive instantaneous reconnection scenario
with all events reconnected.
It is much more complicated to estimate the influence of the
fragmentation effects.
In the Lund string model analogies to two types of superconductors
have been used \cite{KhSj}. The reconnection probability is
taken to be either proportional
to the space--time volume over which the \PWp\ and \PWm\ strings
overlap (type I superconductor) or to the probability that vortex lines
cross (type II superconductor).
Using  these models the fragmentation error was estimated to be 30\MeV\
on the \PW-mass measurement. This error is increased to
40\MeV\ by adding the estimated error coming from the
perturbative effects as well
as from the interplay between perturbative and non-perturbative effects.
Also an extended version of this approach has been investigated 
\cite{Wmassgroup},
where additionally the space--time evolution of the parton cascade, multiple
reconnections, and a finite vortex-core radius have been taken into account.
 
In \cite{GuHa} the space--time picture is not used, except for the
timing of the gluon emission. Rather it is assumed that reconnections 
reducing the
total string length are preferred, as indicated by PQCD experience.
During fragmentation, in addition to the standard string connections
between quarks and gluons emitted from each \PW, also any other
possible interconnection (recoupling)
between partons emitted from different \PW\ bosons
is possible with a probability that may be very different from 1/9.
The reduced string length leads
to a reduction in multiplicity in the central rapidity region.
In this model the shift in the reconstructed \PW-mass
varies between 6 and 60\MeV\ (`shortest-string' version) or between 
13 and 130\MeV\ (`random-reconnection' version) if the
recoupling probability varies between 10 and 100\%.
Methods to measure this probability using data
from LEP2 have been suggested \cite{Wmassgroup,GuHa}.
A similar model has been introduced in the {\sc Ariadne} program
leading to similar mass shifts \cite{ARIANE}.
 
The preliminary estimate of the colour reconnection in {\sc Herwig},
based on a reduction of the space--time extent of clusters, gives mass shifts 
of the same order as the ones mentioned above \cite{Wmassgroup}.
 
The dependence of the shift in the reconstructed \PW-mass on the parameters of
different models
has been studied by the LEP experiments in more details \cite{Wmassgroup}.
It appears that uncertainties using experimental procedures are
not smaller than the ones obtained above. The issue whether diagnostic
signals for reconnection can be found is still open.
 
Thus the colour-reconnection effect may add a big
systematic error to the \PW-mass measurement in the four-quark
channel. The interesting aspect is that different approaches
give uncertainties of the same order.
 
It may be worthwhile to note that 
similar (but not identical) `unconventional' colour connections can also 
appear inside a single \PZ\ system, as is discussed in \cite{Wmassgroup}. 
The presence/absence of a signal in \PZ\ decays at LEP1 could give an
indication of the expected effects in the \PWp\PWm\ system.
 
Bose--Einstein correlations have been observed experimentally
as an enhancement in the two-particle correlation function for
identical bosons. In the case of a \PW\ pair decaying hadronically
a possibility exists of Bose--Einstein correlations
between particles that come from different \PW\ bosons \cite{LoSj}.
A test of this effect has been made in a model based on {\sc Pythia} and
{\sc Jetset} \cite{LoSj}.
The model gives large reconstructed
positive mass
shifts of the order of 100\MeV, rising with increasing c.m.s. energy and with
decreasing source radius. The real shift may be smaller due to various damping
factors not included in \cite{LoSj} (which is intended as a `worst-case'
scenario), but effects of the order of 50\MeV\ could be expected within
a large class of possible Bose--Einstein models.
 
In the coming years both colour (re)arrangement and Bose--Einstein effects
can and should be studied both theoretically
and experimentally. Also the large statistics collected already at LEP1 may
help to study both these effects.
 
\section{Concluding Remarks}
\label{CHconcl}

In this report the present status of the theoretical
knowledge of \PW-pair production and the related process of  
four-fermion production is reviewed.

In lowest-order complete evaluations for all four-fermion
final states exist, taking into account all 
Feynman diagrams. The theoretical problem how to
incorporate an energy-dependent width in the \PW\
and \PZ\ propagators without spoiling gauge invariance
has been solved for the lowest-order \cs.
The numerical relevance of imposing electromagnetic
gauge invariance is explicitly demonstrated.

As to the radiative corrections, the state of the art
depends on whether on- or off-shell \PW-pair production, 
or four-fermion production is considered. Complete
\Oa\ RC's only exist for on-shell \PW-pair
production. The dominant RC's that can be taken into account properly for all 
three types of processes comprise leading-log ISR, the leading weak effects 
related to \GF\ and $\al(s)$, and the Coulomb singularity. 
The \Oa\ RC's for four-fermion
production obviously do not exist, and can at best be calculated using 
approximative techniques. Only taking along   
the above dominant effects, the total and differential \css\ at 175 and 
190\GeV\ are estimated to be known at the 1--2\% level (based on on-shell 
experience). For the total \cs\ at 161\GeV\ an uncertainty of about 2\% is 
expected.
Part of the latter uncertainty is due to the possible
variation of the Higgs mass. This Higgs-mass dependence is largest for light
Higgs bosons and is most pronounced at threshold. 

Another question, relevant for the reconstruction of the
\PW\ mass, is the accuracy of the average emitted photon
energy \Eg. At 175 and 190\GeV\ a precision of
10--20\MeV\ seems feasible, provided a proper scale is used in the 
structure-function part of the calculation.

\appendix
 
\unitlength 1pt
 
\section{Various Methods of Calculating QED Corrections}
\label{app}

\subsection{The structure-function method}
\label{appSFM}
 
As pointed out in the previous sections, the virtual and real
corrections reveal the presence of large logarithmic QED effects of the
form $\alpha \LL/\pi \equiv (\alpha/\pi)\,\log(Q^2/\Me^2)$
with $Q^2 \gg \Me^2$. They arise when photons
or light fermions are radiated off in the direction of incoming or
outgoing light particles \cite{Na58,KLN}, provided the momentum of the
latter is kept fixed (exclusive) \cite{KLN}.
In a massless theory these large logarithms would
show up as collinear divergences (like in QCD), but in the SM the masses
of the particles act as natural cut-offs. The fact that they can
nevertheless give rise to sizeable corrections is due to the difference
in scale between the mass of the radiating particle and its energy.
This is illustrated by considering the propagator $P$ of a light
particle after photon emission \vspace*{20pt}
\newcommand{\longsim}{\begin{picture}(24,0)
                        \bezier{40}(0,0)( 6,3)(12,0)
                        \bezier{40}(12,0)(18,-3)(24,0)
                      \end{picture}}
\begin{displaymath}
  \begin{picture}(100,0)
    \savebox{\wigur}(12,12)[bl]
      {\bezier {20}(0.0,0.0)(0.0,6.0)(6.0,6.0)
       \bezier {20}(6.0,6.0)(12.0,6.0)(12.0,12.0)}
    \savebox{\Vur}(36,36)[bl]{\multiput(0,0)(12,12){3}{\usebox{\wigur}}}
    \put(0,3){\vector(1,0){27}}
    \put(27,3){\line(1,0){23}}
    \put(50,3){\line(1,-3){12}}
    \put(55.5,-13.5){\vector(1,-3){3}}
    \put(50,3){\circle*{4}}
    \put(50,3){\usebox{\Vur}}
    \multiput(52,3)(10,0){5}{\line(1,0){5}}
    \put(74,25.5){\vector(0,1){3}}
    \put(8,-10){\mbox{\footnotesize $q^2=m^2$}}
    \put(65.5,-18.5){\mbox{\footnotesize $q-k$}}
    \put(80,23){\mbox{\footnotesize $k$}}
    \put(22,10){\mbox{\footnotesize $q$}}
    \put(40,25){\mbox{\footnotesize $k^2=0$}}
    \bezier{40}(65,5)(65,12)(63.5,13.5)
    \put(70,8){\mbox{\footnotesize $\theta$}}
  \end{picture}
  \ \ \ \longrightarrow \ \ \
  P = \frac{1}{(q-k)^2-m^2} \stackrel{\raise 6pt
                                      \hbox{$\scriptstyle{q_0 \gg m}$}}
                                     {\raise 3pt \hbox{$\longsim$}}
  \frac{-1}{2 q_0 k_0\,[1 - (1-m^2/2q_0^2)\cos\theta]}.
  \vspace*{20pt}
\end{displaymath}
In the limit $m \to 0$ this propagator gives rise to a pole at
$\cos\theta = 1$. For finite $m$ it yields large logarithmic terms of
the form $\log(q_0^2/m^2)$ when the photon momentum is integrated over
(inclusive photon).
A consequence of the direct relation between the large QED logarithms
and collinear divergences is that they are controlled
by renormalization group equations and that they are universal, i.e.\
they are process-independent. They can be calculated using the
so-called structure-function method \cite{SFM}, taken over from QCD.
This procedure also allows the inclusion of soft-photon effects to all
orders by means of exponentiation.
 
The structure-function method is based on the
mass-factorization theorem (see \reffi{fig241}):
\beq
  s^4\,\frac{\d^4\sigma_{\Pep\Pem\to\PWp\PWm X}}{\d t_1\,\d u_1\,
  \d t_2\,\d u_2} =
  \int^1_0 \frac{\d x_{+}}{x_{+}^2} \int^1_0 \frac{\d x_{-}}{x_{-}^2}\,
  \Gamma_{i\Pep}(x_{+},Q^2)\,\Gamma_{j\Pem}(x_{-},Q^2)\,\hat{s}^4\,
  \frac{\d^4\hat\sigma_{ij\to\PWp\PWm X^\prime}}{\d\hat{t}_1\, \d\hat{u}_1\,
  \d\hat{t}_2\, \d\hat{u}_2}(Q^2), \label{massfac}
\eeq
which links the \cs\ $\sigma_{\Pep\Pem\to\PWp\PWm X}$, where
X indicates an arbitrary number of undetected particles, to the hard
scattering \cs\ $\hat{\sigma}_{ij\to\PWp\PWm X^\prime}$ which is free of
the large collinear logarithms\footnote{If the final state consists of
exclusively treated light particles (\eg \eeffff), large collinear
final-state QED logarithms appear. These large logs can be treated in
a way similar to the initial-state ones, provided proper account is
taken of the fact that now $\hat{p}=p/x$.}. The indices $i,j$ represent
all light-particle
transitions allowed by QED (e.g.\ $\gamma,e^\pm,\mu^\pm, \cdots$).
\bfi
\begin{center}
\begin{picture}(260,260)
  \savebox{\wigur}(12,12)[bl]
    {\bezier {20}(0.0,0.0)(0.0,6.0)(6.0,6.0)
     \bezier {20}(6.0,6.0)(12.0,6.0)(12.0,12.0)}
  \savebox{\Vur}(36,36)[bl]{\multiput(0,0)(12,12){3}{\usebox{\wigur}}}
  \savebox{\wigdr}(12,12)[bl]
    {\bezier {20}(0.0,0.0)(0.0,-6.0)(6.0,-6.0)
     \bezier {20}(6.0,-6.0)(12.0,-6.0)(12.0,-12.0)}
  \savebox{\Vdr}(36,36)[bl]{\multiput(0,0)(12,-12){3}{\usebox{\wigdr}}}
  \put(0,0){\line(1,1){40}}
  \put(54.1,54.1){\circle{40}}
  \put(68.3,68.3){\line(1,1){40}}
  \put(122.4,122.4){\circle{40}}
  \multiput(136.6,136.6)(36,36){2}{\usebox{\Vur}}
  \put(72.9,61){\line(5,3){47.1}}
  \put(73.8,57.6){\line(4,1){46.2}}
  \put(73.8,50.7){\line(4,-1){46.2}}
  \put(72.9,47.3){\line(5,-3){47.1}}
  \multiput(98,46.7)(0,6){3}{\line(0,1){3}}
  \put(0,244.8){\line(1,-1){40}}
  \put(54.1,190.7){\circle{40}}
  \put(68.3,176.5){\line(1,-1){40}}
  \multiput(136.6,72.2)(36,-36){2}{\usebox{\Vdr}}
  \put(72.9,197.6){\line(5,3){47.1}}
  \put(73.8,194.2){\line(4,1){46.2}}
  \put(73.8,187.3){\line(4,-1){46.2}}
  \put(72.9,183.9){\line(5,-3){47.1}}
  \multiput(98,183.3)(0,6){3}{\line(0,1){3}}
  \put(141.2,129.3){\line(5,3){47.1}}
  \put(142.1,125.9){\line(4,1){46.2}}
  \put(142.1,119){\line(4,-1){46.2}}
  \put(141.2,115.3){\line(5,-3){47.1}}
  \multiput(166.3,115)(0,6){3}{\line(0,1){3}}
  \put(46,51){\mbox{$\Gamma_{j\Pem}$}}
  \put(46,188){\mbox{$\Gamma_{i\Pep}$}}
  \put(119,119){\mbox{$\hat{\sigma}$}}
  \put(22,22){\vector(1,1){3}}
  \put(90,90){\vector(1,1){3}}
  \put(22,222.8){\vector(1,-1){3}}
  \put(90,154.8){\vector(1,-1){3}}
  \put(177,178.6){\vector(1,0){3}}
  \put(177,66.6){\vector(1,0){3}}
  \put(-12,26){\mbox{$\Pem,\,p_{-}$}}
  \put(61,94){\mbox{$j,\,\hat{p}_{-}$}}
  \put(187,169){\mbox{$\PWp,\,k_{+}$}}
  \put(-12,217){\mbox{$\Pep,\,p_{+}$}}
  \put(61,147){\mbox{$i,\,\hat{p}_{+}$}}
  \put(187,66){\mbox{$\PWm,\,k_{-}$}}
\end{picture}
\end{center}
\caption{Initial-state collinear radiation in $\eeWW$. The unlabeled
external lines represent an arbitrary number of undetected (inclusive)
particles.}
\label{fig241}
\efi
Furthermore we have defined
\beqar
  t_1 \equiv (p_{-}-k_{-})^2 - \MW^2,&\qquad&
  u_1 \equiv (p_{+}-k_{-})^2 - \MW^2, \nn \\
  t_2 \equiv (p_{+}-k_{+})^2 - \MW^2,&\qquad&
  u_2 \equiv (p_{-}-k_{+})^2 - \MW^2.
\eeqar
Analogously their hatted counterparts are given by the same
expressions with $p_{\pm}$ replaced by $\hat{p}_{\pm}=x_{\pm}p_{\pm}$.
It should be noted that (\ref{massfac}) allows the implementation
of various cuts on the energies and angles of the produced
\PW\ bosons, and that it can be used to extract more commonly used
distributions [like $\dsidctp$] by supplying the appropriate
Jacobians \cite{Be89}. The structure functions $\Gamma_{ij}$ describe
`mass singular' initial-state collinear radiation and represent the
probability of finding in the parent particle $j$ at the scale $Q^2$ a
particle $i$ with fraction $x$ of longitudinal momentum.
So $(1-x)$ denotes the fraction
of energy carried away by collinear radiation. All large collinear
logarithms are contained in the structure functions, which depend, just
like the hard scattering \cs, on the mass-factorization scale
$Q^2$. These structure functions can be decomposed into a part
containing the collinear logarithms and a non-log part according to
\beqar
  \Gamma_{ij}^{\mathrm{log}}(x,Q^2) &=& \delta_{ij}\,\delta(1-x)
     + \sum_{n=1}^{\infty} \left(\frac{\alpha}{\pi}\right)^n
       \sum_{m=1}^{n} a_{mn}^{ij}(x)\, \LL^m, \nn \\
  \Gamma_{ij}^{\mathrm{non}\mbox{-}\mathrm{log}}(x,Q^2) &=&
       \sum_{n=1}^{\infty} \left(\frac{\alpha}{\pi}\right)^n
       b_{n}^{ij}(x).
\eeqar
Analogously the hard scattering \cs\ can be decomposed
into the scale-indepen\-dent Born \cs\ plus non-log higher-order terms
\beqar
  \d\hat\si_{ij}(Q^2) &=& \d\si^\born_{ij} +
       \sum_{n=1}^{\infty} \left(\frac{\alpha}{\pi}\right)^n
       c_{n}^{ij}(Q^2).
\eeqar
For convenience we have dropped the explicit dependence on $x_\pm$ in
this decomposition. This dependence enters via the reduced momenta
$\hat{p}_{\pm}$.
 
The so-called leading-log (LL) approximation consists in only taking
along the terms $\propto (\alpha \LL/\pi)^n$, which automatically means
that the hard scattering \cs\ $\hat{\sigma}$ should be identical to the 
scale-independent Born \cs. In addition this Born \cs\ could be dressed by the
leading corrections that are not of the LL type, e.g.~running couplings or a
Coulomb factor, in order to improve the approximation. As these LL
contributions constitute the most important higher-order QED effects,
it is on most occasions sufficient to add the $\Oaa$ LL contributions
to the full \Oa\ results and to exponentiate the soft-photon
distribution. In this way the resulting \cs\ becomes
scale-dependent. A further simplification can be achieved by realizing that
in $\Pep\Pem$ collisions the bulk of the QED
corrections frequently originates from pure photon radiation,
especially when the soft-photon contributions are dominant%
\footnote{For more details concerning polarized structure functions and QED 
corrections that are not of the pure-photon-radiation type we refer to 
\cite{BD94}.}.

After these simplifications the relevant structure function takes on the
form \cite{Zshape89,epl}
\beqar
  \Gamma_{\mathrm{ee}}^{\mathrm{LL,exp}}(x,Q^2) &\equiv &
    \phi(\alpha,x,Q^2) =
    \frac{\exp(-\frac{1}{2}\,\gamma_E\,\bexp + \frac{3}{8}\,\bs)}
         {\Gamma(1+\frac{1}{2}\,\bexp)}\,\frac{\bexp}{2}\,
    (1-x)^{\bexp/2 - 1} - \frac{1}{4}\,\bh\,(1+x) \nn \\
                                                & & \hspace*{-2cm}
    -\, \frac{1}{4^2\,2!}\,\bh^2\,\left[ \frac{1+3x^2}{1-x}\,\log(x)
    + 4(1+x)\,\log(1-x) + 5 + x \right] \nn \\
                                                & & \hspace*{-2cm}
    -\, \frac{1}{4^3\,3!}\,\bh^3\,\left\{ \vphantom{\frac{1}{4}} 
    (1+x)\, \left[ 6\, \Li (x) + 12\,\log^2(1-x) - 3\pi^2 \right] 
    \right. \nn \\
                                                & & \hspace*{-2cm}
    + \left. \frac{1}{1-x}\,\left[\frac{3}{2}\,(1+8 x + 3 x^2)\,\log(x) 
    + 6\,(x+5)\,(1-x)\,\log(1-x)
    \right.\right. \nn \\
                                                & & \hspace*{-2cm}
    + \left.\left. 12\, (1+x^2)\,\log(x)\,\log(1-x) 
    - \frac{1}{2}\,(1+7 x^2)\,\log^2(x) + \frac{1}{4}\,(39 - 24 x - 15 x^2)
    \right] \right\}. \label{GeeLLexp}
\eeqar
Here $\Li(y)$ is the dilogarithm, $\gamma_E$ the Euler constant, and 
$\Gamma(y)$ the Gamma function (not to be confused with the structure 
functions).
In terms of this structure function the total \PW-pair production
\cs\ including exponentiated LL QED corrections can for example be written as
\beq
\sigma_{\mathrm{LL,exp}}(s,Q^2)=\int^{1}_{4\MW^2/s} \d z
    \,\phi(2\alpha,z,Q^2)\,\hat{\sigma}_{0}(zs),
\label{LLint}
\eeq
where $\hat{\sigma}_{0}(zs)$ denotes the (possibly improved) Born \cs\
at the reduced CM energy squared $zs$.

Note that some non-leading terms
can be incorporated, taking into account the fact that the residue of the
soft-photon pole is proportional to $\LL-1$ rather than $\LL$ for the
initial-state photon radiation. Using
\beq
  \eta = \frac{2\al}{\pi}\,\LL \qquad \mathrm{and} \qquad
  \beta =  \frac{2\al}{\pi}\,(\LL-1)
  \label{zetaalpha}
\eeq
we can identify a couple of popular options for including non-leading terms:
\begin{itemize}
\item $\bexp = \be$, $\bs = \bh = \eta$ \cite{BD94,bkp} ({\tt `MIXED'} choice)
\item $\bexp = \bs = \be$, $\bh = \eta$ \cite{mnpp} ({\tt `ETA'} choice); the 
      original Gribov--Lipatov form factor \cite{Gr72}
\item $\bexp = \bs = \bh = \be$ \cite{Ba93,wwgenpv} ({\tt `BETA'} choice).
\end{itemize}
The differences between these options are in the coefficient in front of the 
exponentiated soft-photon term and in the use of $\eta$ or $\be$ in the hard 
parts.

In \refta{taLL} we quantify for the {\tt CC3} process $\eeWW\to 4f$ the
uncertainty associated with these three different structure functions.
We give the total \cs\ without cuts, the radiative
energy loss \Eg, and the invariant-mass loss \Mg.
The radiative energy loss is defined as
\beq
\Eg \, = \, \frac{1}{\si} \, \int \, (2 - x_+ - x_-) E\, \d \si
\eeq
and the radiative invariant-mass loss as
\beq
\Mg \, = \, \frac{1}{\si} \, \int \, (1 - x_+ \, x_-) E\, \d \si.
\eeq

\btab
\begin{center}
\tabcolsep 6pt
\begin{tabular}{||c|c|c|c|c|c||}\hline\hline
$\sqrt{s}$ & & {\tt MIXED} & {\tt ETA} & {\tt BETA} & {\tt BETA} \\
$\, [\mathrm{GeV}]\, $ & & to $\O(\be^2)$ & to $\O(\be^2)$ & to $\O(\be^2)$ 
& to $\O(\be^3)$ \\ \hline
175 & $\si$ & 13.2218(11) &  13.1772(12) &  13.1828(13) &  13.1790(11) \\
    & \Eg   &  1.1120(3)  &   1.1115(3)  &   1.1147(3)  &   1.1146(3)  \\
    & \Mg   &  1.1093(3)  &   1.1090(3)  &   1.1120(3)  &   1.1119(3)  \\
190 & $\si$ & 16.3412(8)  &  16.2829(12) &  16.2969(13) &  16.2939(8)  \\
    & \Eg   &  2.1303(4)  &   2.1300(5)  &   2.1395(5)  &   2.1391(4)  \\
    & \Mg   &  2.1220(4)  &   2.1216(5)  &   2.1314(5)  &   2.1314(4)  \\
205 & $\si$ & 17.1302(17) &  17.0696(12) &  17.0897(17) &  17.0869(13) \\
    & \Eg   &  3.1823(9)  &   3.1817(9)  &   3.2011(9)  &   3.2019(7)  \\
    & \Mg   &  3.1666(10) &   3.1655(9)  &   3.1845(10) &   3.1854(9)  \\
\hline\hline
\end{tabular}
\end{center}
\caption{Effects of different structure functions on $\si$ (in \pba), 
         \Eg\ (in \GeV), and \Mg\ (in \GeV), for the {\tt CC3} process
         $\eeWW\to 4f$.}
\label{taLL}
\etab

As can be seen from the comparison between the third and fourth
column of \refta{taLL}, the effect on the total \cs\ due to the
different coefficient in front of the exponentiated soft-photon term 
in the {\tt MIXED} and {\tt ETA} structure functions is
of the order of 0.3--0.4\%. This different choice in the coefficient
does not affect \Eg\ and \Mg, 
since it is a factorized soft-photon contribution which
largely cancels out in the ratios. By comparing the fourth and fifth
column of \refta{taLL}, one can estimate the amount of the
so-called $\eta \to \beta$ effect in the hard part of the
structure function. The impact on the \cs\ is small, \ie 
of the order of 0.1\% (depending on the energy).
The effect on \Eg\ and \Mg\ 
is 3, 10, 19\MeV\ (at $\sqrt{s} = 175$, 190, 205\GeV, respectively)
and therefore not significant in view of the expected experimental accuracy at
LEP2. 
{}From the last two columns of \refta{taLL}, finally, we can infer that the
effects of the $\O(\be^3)$ hard part of the structure function is 
completely negligible.

At this point we would like to note that the leading effects related to
the production of undetected $\Pep\Pem$ pairs from conversion of a
virtual photon emitted from the initial state (+ corresponding loop
corrections) can be accounted for by replacing $\alpha$ in $\phi(\alpha,
x,Q^2)$ by $\alpha[1+\alpha \LL/(6\pi)]$.
This takes into account the QED $\beta$-function contributions appearing
in the renormalization group equations. Also other fermion pairs can be
included, but care has to be taken with the fact that the real
pair-production process requires the virtual photon to be sufficiently
hard.

Another source of uncertainties is related to the fact that the scale $Q^2$ 
is a free parameter%
\footnote{For a discussion of `appropriate' scale choices we refer to
\cite{BD94}.}.
As mentioned already in \refse{SEonrcho}, all `natural' scale choices are 
roughly equal close to threshold. Hence, when using the structure-function 
method to calculate higher-order corrections [beyond \Oa] the $Q^2$ dependence
is negligible. This is, of course, not the case for LL \Oa\ corrections,
as they are larger. So, in situations where the exact \Oa\ corrections 
are not known and one has to resort to a LL approximation instead, 
as is the case for \eeffff, the scale dependence is larger.
 
\subsection{The parton-shower method}
\label{appPS}

The basic assumption of the QED parton-shower method is that the 
structure function of an electron (or positron)
obeys the Altarelli--Parisi equation \cite{AP}, which can be
expressed by the integral equation
\begin{equation}
\Ga_{\Pe\Pe}(x,Q^2)= \Pi_{\Pe\Pe}(Q^2,Q_s^2)\,\Ga_{\Pe\Pe}(x,Q_s^2)
    + \frac{\al}{2\pi}\int\nolimits_{Q_s^2}^{Q^2}\frac{\d K^2}{K^2}\,
    \Pi_{\Pe\Pe}(Q^2,K^2)\int\nolimits_x^{1-\epsilon}\frac{\d y}{y}\,
           P_{\Pe\Pe}(y)\,\Ga_{\Pe\Pe}(x/y,K^2)      \label{eq:intform}
\end{equation}
in the leading-log approximation \cite{KEKll}.
Here $\epsilon$ is a small quantity specified later and
$P_{\Pe\Pe}(x)=(1+x^2)/(1-x)$. The function $\Pi_{\Pe\Pe}$, which is nothing
but the Sudakov factor, is given by
\beq
\Pi_{\Pe\Pe}(Q^2,{Q'}^2) = \exp\left[ - \frac{\al}{2\pi} \int_{{Q'}^2}^{Q^2}
   \frac{\d K^2}{K^2} \int_0^{1-\epsilon} \d x\, P_{\Pe\Pe}(x) \right],
\eeq
and denotes the probability that the electron (or positron) evolves from 
${Q'}^2$ to $Q^2$ without emitting a hard photon.
The scale $Q_s^2$ is, like in \refapp{appSFM}, a free
parameter (of order $\Me^2$). Often it is chosen such that, when the 
$K^2$ dependence is integrated out, the factor $\be$ emerges rather 
than $\eta$.

Equation~\refeqf{eq:intform} can be solved by iterating the
right-hand side in a successive way. Hence it is apparent that
the emission of $n$ photons corresponds to the \mbox{$n$-th} iteration. 
As such it is possible to regard the process as a stochastic one, suggesting
the following algorithm for the photon shower \cite{KEKps1}:
(a) Set  $x_b=1$. The variable $x_b$ is the fraction of the
    light-cone momentum of the virtual electron (or positron) that annihilates.
(b) Choose a random number $\xi$. If it is smaller than
    $\Pi_{\Pe\Pe}(Q^2,Q_s^2)$, then the evolution stops. If not, one proceeds 
    by finding the virtuality $K^2$ that satisfies 
    $\xi =\Pi_{\Pe\Pe}(K^2,Q_s^2)$. With this virtuality a branching takes 
    place.
(c) Fix $x$ according to the probability $P_{\Pe\Pe}(x)$ between 0 and
    $1-\epsilon$. Then $x_b$ is replaced by $x_b x$.
    Subsequently one should go to (b), replacing $Q_s^2$ by $K^2$, and
    repeat until the evolution stops.

Once an exclusive process is fixed by this algorithm, each branching
of a photon in the process is dealt with as a true process, that is,
an electron with $x,K^2$ decays as $\Pem(x,-K^2)\to
\Pem(xy,-{K'}^2)+\ga(x[1-y],Q_0^2)$.
Here $Q_0^2$ is a cut-off to avoid the IR divergence. It is unphysical, so any
physical observable should not depend on it. 
The momentum conservation at the branching gives\,
$-K^2=-{K'}^2/y+Q_0^2/(1-y)+{\bf k}_T^2/(y[1-y]),$
which in turn determines ${\bf k}_T^2$ from $y,K^2,{K'}^2$.
Hence the ${\bf k}_T^2$ distribution can be taken into account in
the simulation as well as the shape of $x$. This feature represents the 
essential difference between the structure-function method, which treats the 
photons inclusively, and the QED parton-shower method.
Keeping in mind that $y\leq 1-\epsilon$, the kinematical boundary 
$y(K^2+Q_0^2/[1-y])\le{K'}^2$, equivalent to ${\bf k}_T^2>0$, fixes 
$\epsilon$ to $\epsilon=Q_0^2/{K'}^2$,
since strong ordering ($K^2\ll {K'}^2$) is expected.

The above description of the algorithm represents the ``single-cascade 
scheme''.
This implies that only the \Pem\ or the \Pep\ is able to radiate photons
when the axial gauge vector is chosen along the momentum of the other
initial-state particle, namely \Pep\ or \Pem. 
For programming purposes, however, it is convenient to use a symmetrization
procedure (the so-called ``double-cascade scheme'') to ensure the symmetry 
of the radiation with respect to the electron 
and positron \cite{KEKps3,KEKdouble}. The so-obtained QED parton shower can 
then be combined with the matrix element of any hard
process initiated by \Pep\Pem\ annihilation. 

The Altarelli--Parisi equations can also be solved by a
different algorithm~\cite{Darmstadt-PS}, which yields equivalent
results for electromagnetic radiative corrections.

Starting from the observation that the most important corrections come
from multiple soft-photon emission, the photon-number distribution can
be approximated by a Poisson distribution with average photon number
\begin{equation}
  \bar n_\gamma
     = \frac{\al}{2\pi} \log\left(\frac{Q^2}{m_{\Pe}^2}\right)
       \int_0^{1-\epsilon} \d x\,P_{\Pe\Pe}(x).
\end{equation}
The technical infra-red cut-off $\epsilon$ will drop out of the physical
sum over soft photons.  After the number of photons has been chosen
from the Poisson distribution, their transverse and longitudinal
momenta can be generated properly, ordered according to the
$1/(p_\pm \cdot k)$ poles and splitting functions, respectively.

Using momentum conservation at each branching, the correct energy loss
and $p_T$~boost for the hard scattering is obtained. This algorithm
has been shown to yield results that are consistent with the
structure-function formalism and provides a realistic approximation for
the photonic ${\bf k}_T^2$ distributions.

\subsection{YFS exponentiation}
\label{appYFS}

Let us first answer the basic question: 
what the exclusive Yennie--Frautschi--Suura (YFS)
exponentiation {\em is} and what it {\em is not}?
YFS exponentiation {\em is}, in one word, a technique of summing up
all IR singularities for a given arbitrary process to infinite order.
The formal proof in \cite{YFS} is based on Feynman-diagram techniques
and goes `order by order'.
The YFS technique {\em is not} by any means bound 
to the leading-log summation technique and/or approximations.
It {\em is} applicable to arbitrary stable particles (with arbitrary mass and 
spin) in the initial and final state.
The YFS summation is inherently exclusive, i.e.\ all subtractions/summations
of IR-singular contributions are done {\em before} any phase-space 
integrations over virtual- or real-photon four-momenta are performed.
This means that the Monte Carlo technique can be used
to integrate over the multiple real-photon phase space.
The first practical solution of this kind was given in \cite{yfs2:1990}.
Here the QED matrix element that enters the YFS exponentiation consists of 
two parts. The \Oa\ part is taken to be exact, \ie from Feynman diagrams, 
whereas a LL approximation is used to write down an economic ansatz for the 
\Oaa\ part.
The YFS-exponentiation procedure, which involves a subtraction of the 
IR part of the matrix element, knows nothing
about the origin of the matrix element (ansatz or Feynman rules),
and it will fail if the ansatz has the wrong IR limit.
The above subtraction procedure is done on the fully differential
distribution {\em before} phase-space integration.
Hence it should be clear that a LL ansatz for the matrix element in which the 
emitted photons have zero transverse momenta {\em is not possible}.
Contrary to the LL techniques described in \refapps{appSFM} and 
\refappf{appPS}, in the YFS approach 
the parameter $\beta=2\,(\al/\pi)[\log(s/\Me^2)-1]$ results
from the integration over phase space and there is no freedom
to adjust it.
There is no discussion 
``do we have $-1$ in the definition of the leading logarithm or not''.
The $-1$ is mandatory, because otherwise the IR singularities
do not cancel.
The orthodox YFS exponentiation technique \cite{YFS} is
essentially `order by order' and as such it {\em does not}
sum up all the LL corrections to infinite order, something which is in 
principle possible for the LL techniques described in \refapps{appSFM} and 
\refappf{appPS}.

As the exact \Oa\ matrix element for \eeffff\ is not yet known, the present
implementation of YFS exponentiation for this process relies on a pure LL 
ansatz \cite{koralw:1995a}.
But this will, of course,  change as soon as an adequate approximative \Oa\ 
calculation becomes available.

\frenchspacing
 \newcommand{\zp}[3]{{\sl Z. Phys.} {\bf #1} (19#2) #3}
 \newcommand{\np}[3]{{\sl Nucl. Phys.} {\bf #1} (19#2)~#3}
 \newcommand{\pl}[3]{{\sl Phys. Lett.} {\bf #1} (19#2) #3}
 \newcommand{\pr}[3]{{\sl Phys. Rev.} {\bf #1} (19#2) #3}
 \newcommand{\prl}[3]{{\sl Phys. Rev. Lett.} {\bf #1} (19#2) #3}
 \newcommand{\fp}[3]{{\sl Fortschr. Phys.} {\bf #1} (19#2) #3}
 \newcommand{\nc}[3]{{\sl Nuovo Cimento} {\bf #1} (19#2) #3}
 \newcommand{\ijmp}[3]{{\sl Int. J. Mod. Phys.} {\bf #1} (19#2) #3}
 \newcommand{\ptp}[3]{{\sl Prog. Theo. Phys.} {\bf #1} (19#2) #3}
 \newcommand{\sjnp}[3]{{\sl Sov. J. Nucl. Phys.} {\bf #1} (19#2) #3}
 \newcommand{\cpc}[3]{{\sl Comp. Phys. Commun.} {\bf #1} (19#2) #3}
 \newcommand{\mpl}[3]{{\sl Mod. Phys. Lett.} {\bf #1} (19#2) #3}
 \newcommand{\cmp}[3]{{\sl Commun. Math. Phys.} {\bf #1} (19#2) #3}
 \newcommand{\jmp}[3]{{\sl J. Math. Phys.} {\bf #1} (19#2) #3}
 \newcommand{\nim}[3]{{\sl Nucl. Instr. Meth.} {\bf #1} (19#2) #3}
 \newcommand{\el}[3]{{\sl Europhysics Letters} {\bf #1} (19#2) #3}
 \newcommand{\ap}[3]{{\sl Ann. of Phys.} {\bf #1} (19#2) #3}
 \newcommand{\jetp}[3]{{\sl JETP} {\bf #1} (19#2) #3}
 \newcommand{\acpp}[3]{{\sl Acta Physica Polonica} {\bf #1} (19#2) #3}
 \newcommand{\vj}[4]{{\sl #1~}{\bf #2} (19#3) #4}
 \newcommand{\ej}[3]{{\bf #1} (19#2) #3}
 \newcommand{\vjs}[2]{{\sl #1~}{\bf #2}}
\def\ch#1{\smallskip\item[]\hskip-\labelwidth {\bf\boldmath#1}
                                  \nopagebreak}
\def\comment#1{#1\\*}
\def\ch#1{\relax}
\def\comment#1{\relax}

\end{document}